\documentclass[aps,prc,twocolumn,superscriptaddress,showpacs,nofootinbib,floatfix]{revtex4}

\usepackage{graphicx}
\usepackage{dcolumn}
\usepackage{bm}

\newcommand{\pp} {\mbox{$p+p$}}
\newcommand{\pt} {\mbox{$p_T$}}
\newcommand{\mt} {\mbox{$m_T$}}
\newcommand{\snn} {\mbox{$\sqrt{s_{NN}}$}}
\newcommand{\meanpt} {\mbox{$\langle p_{T} \rangle$}}
\newcommand{\piz} {\mbox{$\pi^{0}$}}
\newcommand{\pbar} {\mbox{$\overline{p}$}}
\newcommand{\lbar} {\mbox{$\overline{\Lambda}$}}
\newcommand{\npart} {\mbox{$N_{part}$}}
\newcommand{\npartav} {\mbox{$\langle N_{part} \rangle$}}
\newcommand{\ncoll} {\mbox{$N_{coll}$}}
\newcommand{\ncollav} {\mbox{$\langle N_{coll} \rangle$}}
\newcommand{\ut} {\mbox{$\langle u_t \rangle$}}

\def\simge{\stackrel{>}{\sim} }

\def\Journal#1#2#3#4{#1 {\bf #2}, #3 (#4)}
\def\JournalAPPEAR#1#2{#1 {\bf #2}}

\def\NIMA{Nucl. Instrum. Methods~A}
\def\NPA{Nucl. Phys.~A}
\def\NPB{Nucl. Phys.~B}
\def\PLB{Phys. Lett.~B}

\def\PRL{Phys. Rev. Lett.}
\def\PRD{Phys. Rev.~D}
\def\PRC{Phys. Rev.~C}
\def\ZPC{Z. Phys.~C}
\def\APPB{Acta Phys. Polon.~B}

\begin{document}
\title{Identified Charged Particle Spectra and Yields\\
in Au+Au Collisions at $\snn$~=~200~GeV}

\newcommand{\abilene}{Abilene Christian University, Abilene, TX 79699, USA}
\newcommand{\acadsin}{Institute of Physics, Academia Sinica, Taipei 11529, Taiwan}
\newcommand{\banaras}{Department of Physics, Banaras Hindu University, Varanasi 221005, India}
\newcommand{\barc}{Bhabha Atomic Research Centre, Bombay 400 085, India}
\newcommand{\bnl}{Brookhaven National Laboratory, Upton, NY 11973-5000, USA}
\newcommand{\caucr}{University of California - Riverside, Riverside, CA 92521, USA}
\newcommand{\ciae}{China Institute of Atomic Energy (CIAE), Beijing, People's Republic of China}
\newcommand{\cns}{Center for Nuclear Study, Graduate School of Science, University of Tokyo, 7-3-1 Hongo, Bunkyo, Tokyo 113-0033, Japan}
\newcommand{\columbia}{Columbia University, New York, NY 10027 and Nevis Laboratories, Irvington, NY 10533, USA}
\newcommand{\dapnia}{Dapnia, CEA Saclay, F-91191, Gif-sur-Yvette, France}
\newcommand{\debrecen}{Debrecen University, H-4010 Debrecen, Egyetem t{\'e}r 1, Hungary}
\newcommand{\fsu}{Florida State University, Tallahassee, FL 32306, USA}
\newcommand{\gsu}{Georgia State University, Atlanta, GA 30303, USA}
\newcommand{\hiroshima}{Hiroshima University, Kagamiyama, Higashi-Hiroshima 739-8526, Japan}
\newcommand{\ihepprot}{Institute for High Energy Physics (IHEP), Protvino, Russia}
\newcommand{\isu}{Iowa State University, Ames, IA 50011, USA}
\newcommand{\jinrdubna}{Joint Institute for Nuclear Research, 141980 Dubna, Moscow Region, Russia}
\newcommand{\kaeri}{KAERI, Cyclotron Application Laboratory, Seoul, South Korea}
\newcommand{\kangnung}{Kangnung National University, Kangnung 210-702, South Korea}
\newcommand{\kek}{KEK, High Energy Accelerator Research Organization, Tsukuba-shi, Ibaraki-ken 305-0801, Japan}
\newcommand{\kfki}{KFKI Research Institute for Particle and Nuclear Physics (RMKI), H-1525 Budapest 114, POBox 49, Hungary}
\newcommand{\korea}{Korea University, Seoul, 136-701, Korea}
\newcommand{\kurchatov}{Russian Research Center ``Kurchatov Institute", Moscow, Russia}
\newcommand{\kyoto}{Kyoto University, Kyoto 606, Japan}
\newcommand{\labllr}{Laboratoire Leprince-Ringuet, Ecole Polytechnique, CNRS-IN2P3, Route de Saclay, F-91128, Palaiseau, France}
\newcommand{\lawllnl}{Lawrence Livermore National Laboratory, Livermore, CA 94550, USA}
\newcommand{\losalamos}{Los Alamos National Laboratory, Los Alamos, NM 87545, USA}
\newcommand{\lpc}{LPC, Universit{\'e} Blaise Pascal, CNRS-IN2P3, Clermont-Fd, 63177 Aubiere Cedex, France}
\newcommand{\lund}{Department of Physics, Lund University, Box 118, SE-221 00 Lund, Sweden}
\newcommand{\muenster}{Institut fuer Kernphysik, University of Muenster, D-48149 Muenster, Germany}
\newcommand{\myongji}{Myongji University, Yongin, Kyonggido 449-728, Korea}
\newcommand{\nagasaki}{Nagasaki Institute of Applied Science, Nagasaki-shi, Nagasaki 851-0193, Japan}
\newcommand{\newmex}{University of New Mexico, Albuquerque, NM, USA}
\newcommand{\nmsu}{New Mexico State University, Las Cruces, NM 88003, USA}
\newcommand{\ornl}{Oak Ridge National Laboratory, Oak Ridge, TN 37831, USA}
\newcommand{\orsay}{IPN-Orsay, Universite Paris Sud, CNRS-IN2P3, BP1, F-91406, Orsay, France}
\newcommand{\pnpi}{PNPI, Petersburg Nuclear Physics Institute, Gatchina, Russia}
\newcommand{\riken}{RIKEN (The Institute of Physical and Chemical Research), Wako, Saitama 351-0198, JAPAN}
\newcommand{\rkrbrc}{RIKEN BNL Research Center, Brookhaven National Laboratory, Upton, NY 11973-5000, USA}
\newcommand{\saispbstu}{St. Petersburg State Technical University, St. Petersburg, Russia}
\newcommand{\saopaulo}{Universidade de S{\~a}o Paulo, Instituto de F\'{\i}sica, Caixa Postal 66318, S{\~a}o Paulo CEP05315-970, Brazil}
\newcommand{\seoulnat}{System Electronics Laboratory, Seoul National University, Seoul, South Korea}
\newcommand{\stonybrkc}{Chemistry Department, Stony Brook University, SUNY, Stony Brook, NY 11794-3400, USA}
\newcommand{\stonycrkp}{Department of Physics and Astronomy, Stony Brook University, SUNY, Stony Brook, NY 11794, USA}
\newcommand{\subatech}{SUBATECH (Ecole des Mines de Nantes, CNRS-IN2P3, Universit{\'e} de Nantes) BP 20722 - 44307, Nantes, France}
\newcommand{\tenn}{University of Tennessee, Knoxville, TN 37996, USA}
\newcommand{\titech}{Department of Physics, Tokyo Institute of Technology, Tokyo, 152-8551, Japan}
\newcommand{\tsukuba}{Institute of Physics, University of Tsukuba, Tsukuba, Ibaraki 305, Japan}
\newcommand{\vandy}{Vanderbilt University, Nashville, TN 37235, USA}
\newcommand{\waseda}{Waseda University, Advanced Research Institute for Science and Engineering, 17 Kikui-cho, Shinjuku-ku, Tokyo 162-0044, Japan}
\newcommand{\weizmann}{Weizmann Institute, Rehovot 76100, Israel}
\newcommand{\yonsei}{Yonsei University, IPAP, Seoul 120-749, Korea}
\affiliation{\abilene}
\affiliation{\acadsin}
\affiliation{\banaras}
\affiliation{\barc}
\affiliation{\bnl}
\affiliation{\caucr}
\affiliation{\ciae}
\affiliation{\cns}
\affiliation{\columbia}
\affiliation{\dapnia}
\affiliation{\debrecen}
\affiliation{\fsu}
\affiliation{\gsu}
\affiliation{\hiroshima}
\affiliation{\ihepprot}
\affiliation{\isu}
\affiliation{\jinrdubna}
\affiliation{\kaeri}
\affiliation{\kangnung}
\affiliation{\kek}
\affiliation{\kfki}
\affiliation{\korea}
\affiliation{\kurchatov}
\affiliation{\kyoto}
\affiliation{\labllr}
\affiliation{\lawllnl}
\affiliation{\losalamos}
\affiliation{\lpc}
\affiliation{\lund}
\affiliation{\muenster}
\affiliation{\myongji}
\affiliation{\nagasaki}
\affiliation{\newmex}
\affiliation{\nmsu}
\affiliation{\ornl}
\affiliation{\orsay}
\affiliation{\pnpi}
\affiliation{\riken}
\affiliation{\rkrbrc}
\affiliation{\saispbstu}
\affiliation{\saopaulo}
\affiliation{\seoulnat}
\affiliation{\stonybrkc}
\affiliation{\stonycrkp}
\affiliation{\subatech}
\affiliation{\tenn}
\affiliation{\titech}
\affiliation{\tsukuba}
\affiliation{\vandy}
\affiliation{\waseda}
\affiliation{\weizmann}
\affiliation{\yonsei}
\author{S.S.~Adler}	\affiliation{\bnl}
\author{S.~Afanasiev}	\affiliation{\jinrdubna}
\author{C.~Aidala}	\affiliation{\bnl}
\author{N.N.~Ajitanand}	\affiliation{\stonybrkc}
\author{Y.~Akiba}	\affiliation{\kek} \affiliation{\riken}
\author{J.~Alexander}	\affiliation{\stonybrkc}
\author{R.~Amirikas}	\affiliation{\fsu}
\author{L.~Aphecetche}	\affiliation{\subatech}
\author{S.H.~Aronson}	\affiliation{\bnl}
\author{R.~Averbeck}	\affiliation{\stonycrkp}
% ...
\collaboration{PHENIX Collaboration} \noaffiliation

\date{\today}        
\begin{abstract}
The centrality dependence of transverse momentum distributions and yields for 
$\pi^{\pm}$, $K^{\pm}$, $p$ and $\pbar$ in Au+Au collisions at $\snn$~=~200~GeV 
at mid-rapidity are measured by the PHENIX experiment at RHIC. We observe 
a clear particle mass dependence of the shapes of transverse momentum spectra 
in central collisions below $\sim$~2~GeV/$c$ in $\pt$. Both mean transverse 
momenta and particle yields per participant pair increase from peripheral to 
mid-central and saturate at the most central collisions for all particle species. 
We also measure particle ratios of $\pi^{-}/\pi^{+}$, $K^{-}/K^{+}$, $\pbar/p$, 
$K/\pi$, $p/\pi$ and $\pbar/\pi$ as a function of $\pt$ and collision centrality. 
The ratios of equal mass particle yields are independent of $\pt$ and 
centrality within the experimental uncertainties. In central collisions at 
intermediate transverse momenta $\sim$~1.5 -- 4.5~GeV/$c$, proton and anti-proton 
yields constitute a significant fraction of the charged hadron production and 
show a scaling behavior different from that of pions. 

\end{abstract}
\maketitle

%%%%%%%%%%%%%%%%%%%%
% (1) Introduction %
%%%%%%%%%%%%%%%%%%%%
\section{INTRODUCTION}
\label{sec:intro}
The motivation for ultra-relativistic heavy-ion experiments at the Relativistic 
Heavy Ion Collider (RHIC) at Brookhaven National Laboratory is the study of nuclear 
matter at extremely high temperature and energy density with the hope of creating and 
detecting deconfined matter consisting of quarks and gluons -- the quark gluon plasma (QGP). 
Lattice QCD calculations~\cite{lattice} predict that the transition to a deconfined 
state occurs at a critical temperature $T_{c} \approx$~170~MeV and an energy density 
$\epsilon \approx$ 2~GeV/${\rm fm^3}$. Based on the Bjorken estimation~\cite{bjorken} 
and the measurement of transverse energy ($E_{T}$) in Au+Au collisions at 
$\snn$~=~130~GeV~\cite{PPG002} and 200~GeV, the spatial energy density in central 
Au+Au collisions at RHIC is believed to be high enough to create such deconfined 
matter in a laboratory~\cite{PPG002}. 

The hot and dense matter produced in relativistic heavy ion collisions may evolve 
through the following scenario: pre-equilibrium, thermal (or chemical) equilibrium 
of partons, possible formation of QGP or a QGP-hadron gas mixed state, a gas of hot 
interacting hadrons, and finally, a freeze-out state when the produced hadrons 
no longer strongly interact with each other. Since produced hadrons carry information 
about the collision dynamics and the entire space-time evolution of the system from 
the initial to the final stage of collisions, a precise measure of the transverse 
momentum ($\pt$) distributions and yields of identified hadrons as a function of 
collision geometry is essential for the understanding of the dynamics and properties 
of the created matter. 

In the low $\pt$ region ($<$~2~GeV/$c$), hydrodynamic models~\cite{hydro_1,hydro_2} 
that include radial flow successfully describe the measured $\pt$ distributions in 
Au+Au collisions at $\snn$~=~130~GeV~\cite{PPG006,STAR_pbar_spectra,PPG009}. The $\pt$ spectra 
of identified charged hadrons below $\pt \approx$~2~GeV/$c$ in central collisions 
have been well reproduced by two simple parameters: transverse flow velocity $\beta_{T}$ 
and freeze-out temperature $T_{fo}$~\cite{PPG009} under the assumption of thermalization
with longitudinal and transverse flow~\cite{hydro_1}. The particle production in this 
$\pt$ region is considered to be dominated by secondary interactions among produced 
hadrons and participating nucleons in the reaction zone. Another model which successfully 
describes the particle abundances at low $\pt$ is the statistical thermal 
model~\cite{thermal_becattini}. Particle ratios have been shown to be well reproduced 
by two parameters: a baryon chemical potential $\mu_{B}$ and a chemical freeze-out 
temperature $T_{ch}$. It is found that there is an overall good agreement between 
measured particle ratios at $\snn$~=~130~GeV Au+Au and the thermal model 
calculations~\cite{thermal_1,thermal_2}. 

On the other hand, at high $\pt$ ($\ge$~4~GeV/$c$) the dominant 
particle production mechanism is the hard scattering described by
perturbative Quantum Chromodynamics (pQCD), which produces particles 
from the fragmentation of energetic partons. 
One of the most interesting observations at RHIC is that the yield 
of high $\pt$ neutral pions and non-identified charged hadrons in central Au+Au 
collisions at RHIC are below the expectation of the scaling with the number 
of nucleon-nucleon binary collisions, $\ncoll$~\cite{PPG003,PPG013,PPG014}.  
This effect could be a consequence of the energy 
loss suffered by partons moving through deconfined matter~\cite{quench_effect,
quenching_theory}. It has also been observed that the yield of neutral pions is 
more strongly suppressed than that for non-identified charged hadrons~\cite{PPG003}
in central Au+Au collisions at RHIC. Another interesting feature is that the proton and 
anti-proton yields in central events are comparable to that of pions at 
$\pt \approx$ 2~GeV/$c$~\cite{PPG006}, differing from the expectation of pQCD. 
These observations suggest that a detailed study of particle composition at 
intermediate $\pt$ ($\approx$~2 -- 4~GeV/$c$) is very important to understand hadron 
production and collision dynamics at RHIC.

The PHENIX experiment~\cite{PHENIX_overview} has a unique hadron identification 
capability in a broad momentum range. Pions and kaons are 
identified up to 3~GeV/$c$ and 2~GeV/$c$ in $\pt$, respectively, and protons and 
anti-protons can be identified up to 4.5~GeV/$c$ by using a high resolution 
time-of-flight detector~\cite{PHENIX_PID}. Neutral pions are reconstructed 
via $\piz \rightarrow \gamma\gamma$  up to $\pt \approx$~10~GeV/$c$ through 
an invariant mass analysis of $\gamma$ pairs detected in an electro-magnetic 
calorimeter (EMCal)~\cite{PHENIX_EMC} with wide azimuthal coverage.
During the measurements of Au+Au collisions at $\snn$~=~200~GeV in year 2001 at 
RHIC, the PHENIX experiment accumulated enough events to address the above issues 
at intermediate $\pt$ as well as the particle production at low $\pt$ with 
precise centrality dependences. In this paper, we present the centrality 
dependence of $\pt$ spectra, $\meanpt$,  yields, and ratios for $\pi^{\pm}$, $K^{\pm}$, 
$p$ and $\pbar$ in Au+Au collisions at $\snn$~=~200~GeV at mid-rapidity measured 
by the PHENIX experiment. We also present results on the scaling behavior 
of charged hadrons compared with results of $\piz$ measurements~\cite{PPG014}, which have
been published separately.  

%%% paper road map %%%
The paper is organized as follows. Section~\ref{sec:phenix} describes 
the PHENIX detector used in this analysis. In Section~\ref{sec:analysis} 
the analysis details including event selection, track selection, particle 
identification, and corrections applied to the data are described. 
The systematic errors on the measurements are also discussed in this section. 
For the experimental results, centrality dependence of $\pt$ spectra for 
identified charged particles are presented in Section~\ref{sec:pt_spectra}, 
and transverse mass spectra are given in Section~\ref{sec:mt_spectra}. 
Particle yields and mean transverse momenta as a function of centrality 
are presented in Section~\ref{sec:meanpt_dndy}. In Section~\ref{sec:ratio} 
the systematic study of particle ratios as a function $\pt$ and centrality 
are presented. Section~\ref{sec:ncoll} studies the scaling behavior 
of identified charged hadrons. A summary is given in Section~\ref{sec:summary}.

%%%%%%%%%%%%%%%%%%%%%%%%%
% (2) PHENIX Detector %
%%%%%%%%%%%%%%%%%%%%%%%%%
\section{PHENIX DETECTOR}
\label{sec:phenix}
The PHENIX experiment is composed of two central arms, two forward 
muon arms, and three global detectors. The east and west central arms 
are placed at zero rapidity and designed to detect electrons, 
photons and charged hadrons. The north and south forward muon arms have 
full azimuthal coverage and are designed to detect muons. The 
global detectors measure the start time, vertex, and multiplicity of 
the interactions.  The following sections describe the parts of the
detector that are used in the present analysis. A detailed description 
of the complete detector can be found elsewhere~\cite{PHENIX_overview,
PHENIX_PID,PHENIX_EMC,PHENIX_inner,PHENIX_ZDC,PHENIX_tracking}.

\subsection{Global Detectors}
In order to characterize the centrality of Au+Au collisions, zero-degree 
calorimeters (ZDC)~\cite{PHENIX_ZDC} and beam-beam counters (BBC)~\cite{PHENIX_inner} 
are employed. The zero-degree calorimeters are small hadronic calorimeters 
which measure the energy carried by spectator neutrons. They are 
placed 18~m up- and downstream of the interaction point along the beam line. 
Each ZDC consists of three modules. Each module has a depth of 2 hadronic 
interaction lengths and is read out by a single photo-multiplier tube (PMT). 
Both time and amplitude are digitized for each PMT along with the analog sum 
of the three PMT signals for each ZDC.

Two sets of beam-beam counters are placed 1.44 m from the nominal interaction 
point along the beam line (one on each side). Each counter consists of 64 
$\check{\rm C}$erenkov telescopes, arranged radially around the beam line.
The BBC measures the number of charged particles in the pseudo-rapidity 
region 3.0~$<$~$|\eta|$~$<$~3.9. The correlation between BBC charge sum and 
ZDC total energy is used for centrality determination. The BBC also 
provides a collision vertex position and start time information for 
time-of-flight measurement. 

\subsection{Central Arm Detectors}
Charged particles are tracked using the central arm 
spectrometers~\cite{PHENIX_tracking}. The spectrometer on 
the east side of the PHENIX detector (east arm) contains the 
following subsystems used in this analysis: drift chamber (DC), 
pad chamber (PC) and time-of-flight (TOF).

The drift chambers are the closest tracking detectors to the beam line --
at a radial distance of 2.2~m. They measure charged particle trajectories 
in the azimuthal direction to determine the transverse momentum of each particle. 
By combining the polar angle information from the first layer of the PC with the
transverse momentum, the total momentum $p$ is determined. 
The momentum resolution is $\delta p/p \simeq 0.7\% \oplus 1.0\%\times p$ (GeV/$c$), 
where the first term is due to the multiple scattering before the DC and the 
second term is the angular resolution of the DC. The momentum scale is known
to 0.7\%, from the reconstructed proton mass using the TOF. 

The pad chambers are multi-wire proportional chambers that 
form three separate layers of the central tracking system.
The first pad chamber layer (PC1) is located at the radial 
outer edge of each drift chamber at a distance of 2.49~m, while 
the third layer (PC3) is 4.98~m from the interaction point. 
The second layer (PC2) is located at a radial distance of 
4.19~m in the west arm only. PC1 and the DC, along with the vertex 
position measured by the BBC, are used in the global track reconstruction 
to determine the polar angle of each charged track.

The time-of-flight detector serves as the primary particle 
identification device for charged hadrons by measuring the stop 
time. The start time is given by the BBC. The TOF wall is located 
at a radial distance of 5.06~m from the interaction point in the east 
central arm. This contains 960 scintillator slats oriented along 
the azimuthal direction. It is designed to cover $|\eta|< 0.35$ and 
$\Delta\phi=45^{o}$ in azimuthal angle. The intrinsic timing resolution 
is $\sigma \simeq 115$ ps, which allows for a 
3$\sigma$ $\pi/K$ separation up to $\pt \simeq 2.5$~GeV/$c$, and 
3$\sigma$ $K/p$ separation up to $\pt \simeq 4$~GeV/$c$.  

%%%%%%%%%%%%%%%%%%%%%
% (3) Data Analysis %
%%%%%%%%%%%%%%%%%%%%%
\section{DATA ANALYSIS}
\label{sec:analysis}
In this section, we describe the event and track selection, charged 
particle identification and various corrections, including geometrical 
acceptance, particle decay, multiple scattering and absorption effects, 
detector occupancy corrections and weak decay contributions from 
$\Lambda$ and $\lbar$ to proton and anti-proton spectra. The estimations 
of systematic uncertainties on the measurements are addressed at the 
end of this section.  

%%%%%%%%%%%%%%%%%%%
% Event Selection %
%%%%%%%%%%%%%%%%%%%
\subsection{Event Selection}
\label{sec:event_selection}
%%%%%%%%%%%%%%%%%%%%%%%%%%%%%%%%%%%%%%%% Figure 1.
\begin{figure}[t]
\includegraphics[width=1.0\linewidth]{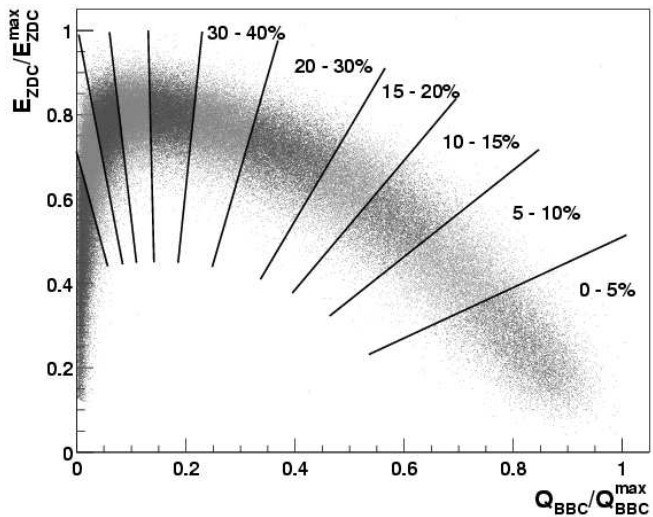}
\caption{BBC versus ZDC analog response. 
The lines represent the centrality cut boundaries.}
\label{fig:bbc_zdc} 
\end{figure} 

For the present analysis, we use the PHENIX minimum bias trigger events, 
which are determined by a coincidence between north and south BBC 
signals. We also require a collision vertex within $\pm$ 30~cm from the 
center of the spectrometer. The collision vertex resolution determined by 
the BBC is about 6~mm in Au+Au collisions in minimum bias events~\cite{PHENIX_inner}. 
The PHENIX minimum bias trigger events include $92.2^{+2.5}_{-3.0}$\% of 
the 6.9 barn Au+Au total inelastic cross section~\cite{PPG014}. 
Figure~\ref{fig:bbc_zdc} shows the correlation between the BBC charge sum 
and ZDC total energy for Au+Au at $\snn$~=~200~GeV. The lines on the plot 
indicate the centrality definition in the analysis. For the centrality 
determination, these events are subdivided into 11 bins using the BBC and 
ZDC correlation: 0--5\%, 5--10\%, 10--15\%, 15--20\%, 20--30\%, ..., 70--80\% 
and 80--92\%. Due to the statistical limitations in the peripheral events, 
we also use the 60--92\% centrality bin as the most peripheral bin. After 
event selection, we analyze 2.02$\times 10^{7}$ minimum bias events, which 
represents $\sim$ 140 times more events than used in our published Au+Au data 
at 130~GeV~\cite{PPG006,PPG009}. Based on a Glauber model calculation~\cite{PPG009,PPG014}
we use two global quantities to characterize the event centrality: the average 
number of participants $\npartav$ and the average number of collisions 
$\ncollav$ associated with each centrality bin (Table~\ref{tab:Ncoll}).

%%%%%%%%%%%%%%%%%%%%%%
% Ncoll, Npart table %
%%%%%%%%%%%%%%%%%%%%%%
\begin{table}
\caption{The average nuclear overlap function ($\langle T_{\rm AuAu}\rangle$),
the number of nucleon-nucleon binary collisions ($\ncollav$), and the number
of participant nucleons ($\npartav$) obtained from a Glauber
Monte Carlo~\cite{PPG009,PPG014} correlated with the BBC and ZDC response for Au+Au
at $\snn$~=~200~GeV as a function of centrality. Centrality is expressed as
percentiles of $\sigma_{\rm AuAu}$ = 6.9 barn with 0\% representing the most
central collisions. The last line refers to minimum bias collisions.}
\begin{ruledtabular}\begin{tabular}{cccc}
Centrality & $\langle T_{\rm AuAu}\rangle$ (mb$^{-1}$) & $\ncollav$ & $\npartav$ \\ \hline 
 0- 5\%    &  25.37  $\pm$  1.77  & 1065.4  $\pm$  105.3  &  351.4  $\pm$   2.9 \\
 0-10\%    &  22.75  $\pm$  1.56  &  955.4  $\pm$   93.6  &  325.2  $\pm$   3.3 \\
 5-10\%    &  20.13  $\pm$  1.36  &  845.4  $\pm$   82.1  &  299.0  $\pm$   3.8 \\
10-15\%    &  16.01  $\pm$  1.15  &  672.4  $\pm$   66.8  &  253.9  $\pm$   4.3 \\
10-20\%    &  14.35  $\pm$  1.00  &  602.6  $\pm$   59.3  &  234.6  $\pm$   4.7 \\
15-20\%    &  12.68  $\pm$  0.86  &  532.7  $\pm$   52.1  &  215.3  $\pm$   5.3 \\
20-30\%    &   8.90  $\pm$  0.72  &  373.8  $\pm$   39.6  &  166.6  $\pm$   5.4 \\
30-40\%    &   5.23  $\pm$  0.44  &  219.8  $\pm$   22.6  &  114.2  $\pm$   4.4 \\
40-50\%    &   2.86  $\pm$  0.28  &  120.3  $\pm$   13.7  &   74.4  $\pm$   3.8 \\
50-60\%    &   1.45  $\pm$  0.23  &   61.0  $\pm$    9.9  &   45.5  $\pm$   3.3 \\
60-70\%    &   0.68  $\pm$  0.18  &   28.5  $\pm$    7.6  &   25.7  $\pm$   3.8 \\
60-80\%    &   0.49  $\pm$  0.14  &   20.4  $\pm$    5.9  &   19.5  $\pm$   3.3 \\
60-92\%    &   0.35  $\pm$  0.10  &   14.5  $\pm$    4.0  &   14.5  $\pm$   2.5 \\
70-80\%    &   0.30  $\pm$  0.10  &   12.4  $\pm$    4.2  &   13.4  $\pm$   3.0 \\
70-92\%    &   0.20  $\pm$  0.06  &    8.3  $\pm$    2.4  &    9.5  $\pm$   1.9 \\
80-92\%    &   0.12  $\pm$  0.03  &    4.9  $\pm$    1.2  &    6.3  $\pm$   1.2 \\
min. bias  &   6.14  $\pm$  0.45  &  257.8  $\pm$   25.4  &  109.1  $\pm$   4.1 \\ 
\end{tabular}\end{ruledtabular}
\label{tab:Ncoll}
\end{table}

%%%%%%%%%%%%%%%%%%%
% Track Selection %
%%%%%%%%%%%%%%%%%%%
\subsection{Track Selection}
\label{sec:track}
Charged particle tracks are reconstructed by the DC based on a combinatorial 
Hough transform~\cite{PHENIX_recoNIM} -- which gives the angle of the track 
in the main bend plane. The main bend plane is perpendicular to the beam axis 
(azimuthal direction). PC1 is used to measure the position of the hit in the 
longitudinal direction (along the beam axis). When combined with the location
of the collision vertex along the beam axis (from the BBC), the PC1
hit gives the polar angle of the track. Only tracks with valid 
information from both the DC and PC1 are used in the analysis. 
In order to associate a track with a hit on the TOF, 
the track is projected to its expected hit location on the TOF.
Tracks are required to have a hit on the TOF within $\pm$2$\sigma$
of the expected hit location in both the azimuthal and beam directions.
Finally, a cut on the energy loss in the TOF scintillator
is applied to each track. This $\beta$-dependent energy loss cut 
is based on a parameterization of the Bethe-Bloch formula, 
i.e. $dE/dx \approx \beta^{-5/3}$, where $\beta = L/(c \cdot t_{\rm TOF})$,
$L$ is the path-length of the track trajectory from the collision
vertex to the hit position of the TOF wall, $t_{\rm TOF}$ is the 
time-of-flight, and $c$ is the speed of light. The flight path-length
is calculated from a fit to the reconstructed track trajectory. 
The background due to random association of DC/PC1 tracks with
TOF hits is reduced to a negligible level when the mass cut used for 
particle identification is applied (described in the next section). 

%%%%%%%
% PID %
%%%%%%%
\subsection{Particle Identification}
\label{sec:pid}
%%%%%%%%%%%%%%%%%%%%%%%%%%%%%%%%%%%%%%%% Figure 2.
\begin{figure}[t]
\includegraphics[width=1.0\linewidth]{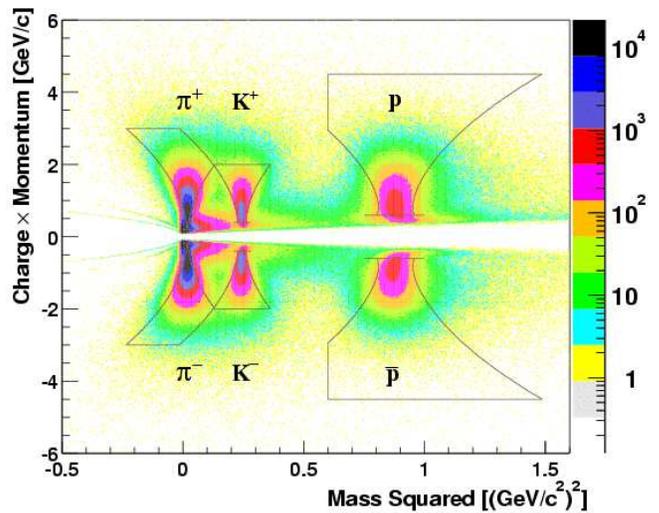}
\caption{Mass squared versus momentum multiplied by charge distribution
in Au+Au collisions at $\snn$~=~200~GeV. The lines indicate the PID 
cut boundaries for pions, kaons, and protons (anti-protons) from 
left to right, respectively.}
\label{fig:PID}
\end{figure}   

The charged particle identification (PID) is performed by using
the combination of three measurements: time-of-flight from the BBC 
and TOF, momentum from the DC, and flight path-length from the collision
vertex point to the hit position on the TOF wall. The square of the mass 
is derived from the following formula, 

\begin{equation}
m^{2} = \frac{p^{2}}{c^{2}} \Bigl[ {\Bigl( \frac{t_{\rm TOF}}{L/c} \Bigr)}^{2} -1 \Bigr], 
\label{eq:m2}
\end{equation}
where $p$ is the momentum, $t_{\rm TOF}$ is the time-of-flight, $L$ is 
a flight path-length, and $c$ is the speed of light. The charged 
particle identification is performed using cuts in $m^{2}$ and momentum space. 

In Figure~\ref{fig:PID}, a plot of $m^{2}$ versus momentum multiplied by 
charge is shown together with applied PID cuts as solid curves. We use 
2$\sigma$ standard deviation PID cuts in $m^{2}$ and momentum space for 
each particle species. The PID cut is based on a parameterization of 
the measured $m^{2}$ width as a function of momentum,

\begin{eqnarray}
{\sigma^2_{m^2}} &=& \frac{\sigma_{\alpha}^2} {K_{1}^{2}} (4m^{4}p^{2}) + 
                     \frac{\sigma_{ms}^2} {K_{1}^{2}} 
                           \Bigl[ 4m^{4} \Bigl( 1+\frac{m^2}{p^2} \Bigr) \Bigl] \nonumber \\
                 &+& \frac{\sigma_{t}^2 c^2} {L^2}  
                           \bigl[ 4p^{2} \bigl( m^2 + p^2 \bigr) \bigr],
\label{eq:pid}
\end{eqnarray}
where $\sigma_\alpha$ is the angular resolution, $\sigma_{ms}$ 
is the multiple scattering term, $\sigma_{t}$ is the overall time-of-flight 
resolution, $m$ is the centroid of $m^{2}$ distribution for each
particle species, and $K_1$ is a magnetic field integral constant term of 87.0~mrad$\cdot$GeV. 
The parameters for PID are, $\sigma_\alpha = 0.835$~mrad, $\sigma_{ms} = 0.86$~mrad$\cdot$GeV 
and $\sigma_{t} = 120$~ps. Through improvements in alignment and calibrations, 
the momentum resolution is improved over the 130~GeV data~\cite{PPG009}.
The centrality dependence of the width and the mean position of the $m^{2}$ distribution has also 
been checked. There is no clear difference seen between central and peripheral 
collisions. For pion identification above 2~GeV/$c$, we apply an asymmetric PID 
cut to reduce kaon contamination of the pions. As shown by the lines 
in Figure~\ref{fig:PID}, the overlap region which is within the 2$\sigma$
cuts for both pions and kaons is excluded. For kaons, the upper momentum
cut-off is 2~GeV/$c$ since the pion contamination level for kaons is 
$\approx$ 10\% at that momentum. The upper momentum cut-off on the pions
is $\pt = $ 3~GeV/$c$ -- where the kaon contamination reaches $\approx$ 10\%.
The contamination of protons by kaons reaches about 5\% at 4~GeV/$c$.
Electron (positron) and decay muon background at very low $\pt$ ($<$ 0.3~GeV/$c$)
are well separated from the pion mass-squared peak. The contamination background
on each particle species is not subtracted in the analysis. For protons, 
the upper momentum cut-off is set at 4.5~GeV/$c$ due to statistical limitations and 
background at high $\pt$. An additional cut on $m^{2}$ for protons and anti-protons, 
$m^{2} >$ 0.6 $({\rm GeV}/c^{2})^{2}$, is introduced to reduce background. 
The lower momentum cut-offs are 0.2~GeV/$c$ for pions, 0.4~GeV/$c$ for kaons, 
and 0.6~GeV/$c$ for $p$ and $\pbar$. This cut-off value for $p$ and $\pbar$ 
is larger than those for pions and kaons due to the large energy loss effect.

%%%%%%%%%%%%%%%%%%%%%%%%%%%%%%%%%%%%%%%% Figure 3.
\begin{figure}[t]
\includegraphics[width=1.0\linewidth]{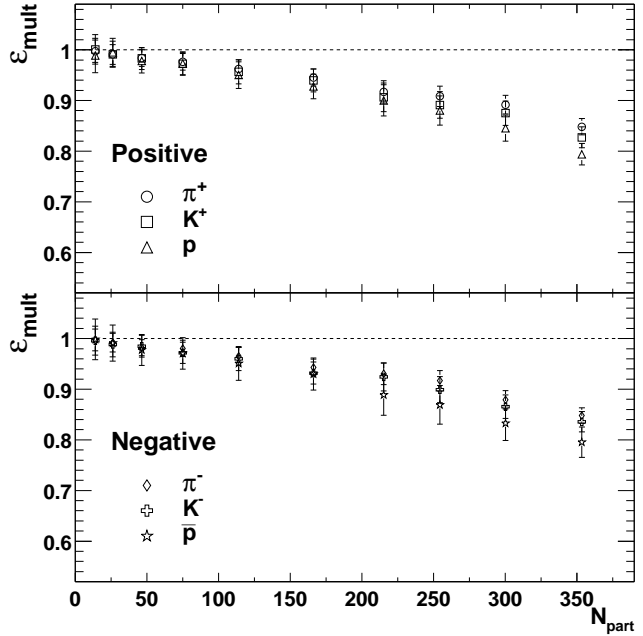}
\caption{Track reconstruction efficiency ($\epsilon_{\rm mult}$)
as a function of centrality. The error bars on the plot represent 
the systematic errors.}
\label{fig:embed}
\end{figure} 

%%%%%%%%%%%%%%%%%
% MC simulation %
%%%%%%%%%%%%%%%%%
\subsection{Acceptance, Decay and Multiple Scattering Corrections}
\label{sec:MC}
In order to correct for 1) the geometrical acceptance, 2) in-flight 
decay for pions and kaons, 3) the effect of multiple scattering, 
and 4) nuclear interactions with materials in the detector (including
anti-proton absorption), we use PISA (PHENIX Integrated Simulation Application),
a GEANT~\cite{GEANT} based Monte Carlo (MC) simulation program of the PHENIX detector.
The single particle tracks are passed from GEANT through the PHENIX 
event reconstruction software~\cite{PHENIX_recoNIM}. 
In this simulation, the BBC, TOF, and DC detector responses are tuned 
to match the real data. For example, dead areas of DC and TOF are 
included, and momentum and time-of-flight resolution are tuned. 
The track association to TOF in both azimuth ($\phi$) and along 
the beam axis ($z$) as a function of momentum and the PID cut 
boundaries are parameterized to match the real data. 
A fiducial cut is applied to choose identical 
active areas on the TOF in both the simulation and data. We generate 
1$\times 10^{7}$ single particle events for each particle species 
($\pi^{\pm}$, $K^{\pm}$, $p$ and $\pbar$) with low $\pt$ enhanced
($<$ 2 GeV/$c$) + flat $\pt$ distributions for high $\pt$ 
(2 -- 4 GeV/$c$ for pions and kaons, 2 -- 8 GeV/$c$ for $p$ 
and $\pbar$)~\footnote{Due to the good momentum resolution at the high $\pt$ 
region, the momentum smearing effect for a steeply falling spectrum is $<$1\% 
at $\pt$~=~5 GeV/$c$. The flat $\pt$ distribution up to 5 GeV/$c$ can 
be used to obtain the correction factors.}.
The efficiencies are determined in each $\pt$ bin by dividing
the reconstructed output by the generated input as expressed as follows:

\begin{equation}
\epsilon_{\rm acc}(j,\pt) = \frac{\mbox{\rm \# of reconstructed MC tracks}}
                                 {\mbox{\rm \# of generated MC tracks}},
\label{eq:acc}
\end{equation}
where $j$ is the particle species. The resulting correction factors 
(1/$\epsilon_{\rm acc}$) are applied to the data in each $\pt$ bin and for 
each individual particle species.  

%%%%%%%%%%%%%%%%%%%%%%%%%%%%%%%%%
% Detector Occupancy Correction %
%%%%%%%%%%%%%%%%%%%%%%%%%%%%%%%%%
\subsection{Detector Occupancy Correction}
\label{sec:occ_corr}

Due to the high multiplicity environment in heavy ion collisions,
which causes high occupancy and multiple hits on a detector cell 
such as scintillator slats of the TOF, it is expected that the 
track reconstruction efficiency in central events is lower than that 
in peripheral events. The typical occupancy at TOF is 
less than 10\% in the most central Au+Au collisions. To correct for 
this effect, we merge single particle simulated events with real 
events and calculate the track reconstruction efficiency for each 
simulated track as follows:

\begin{equation}
\epsilon_{\rm mult}(i,j) = \frac{\mbox{\rm \# of reconstructed embedded tracks}}
                                {\mbox{\rm \# of embedded tracks}},
\label{eq:track_eff}
\end{equation}
where $i$ is the centrality bins and $j$ is the particle species. 
This study has been performed for each particle species and each centrality bin.
The track reconstruction efficiencies are factorized (into independent terms
depending on centrality and $\pt$) for $\pt > 0.4$~GeV/$c$,
since there is no $\pt$ dependence in the efficiencies above that $\pt$.
Figure~\ref{fig:embed} shows the dependence of track reconstruction efficiency
for $\pi^{\pm}$, $K^{\pm}$, $p$ and $\pbar$ as a function of centrality expressed 
as $\npart$. The efficiency in the most central 0--5\% events is about 80\% for 
protons ($\pbar$), 83\% for kaons and 85\% for pions. Slower particles are more 
likely lost due to high occupancy in the TOF because the system responds to the 
earliest hit. For the most peripheral 80--92\% events, the efficiency for detector 
occupancy effect is $\approx$ 99\% for all particle species. The factors are 
applied to the spectra for each particle species and centrality bin. Systematic 
uncertainties on detector occupancy corrections (1/$\epsilon_{\rm mult}$) are 
less than 3\%.

%%%%%%%%%%%%%%%%%%%%%%%%
% Feed-down correction %
%%%%%%%%%%%%%%%%%%%%%%%%
\subsection{Weak Decay Correction}
\label{sec:feed-down}
%%%%%%%%%%%%%%%%%%%%%%%%%%%%%%%%%%%%%%%% Figure 4.
\begin{figure}[t]
\includegraphics[width=1.0\linewidth]{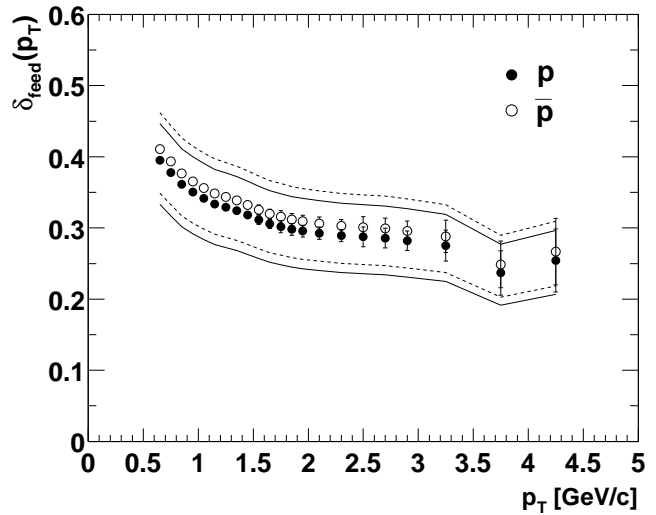}
\caption{The fractional contribution of protons ($\pbar$) from $\Lambda$ ($\lbar$) 
decays in all measured protons ($\pbar$), $\delta_{\rm feed}(\pt)$, as a function of
$\pt$. The solid (dashed) lines represent the systematic errors for protons ($\pbar$).
The error bars are statistical errors.}
\label{fig:feeddown}
\end{figure}
%%%%%%%%%%%%%%%%%%%%%%%%%%%%%%%%%%%%%%%% Figure 5.
\begin{figure*}[t]
\includegraphics[width=1.0\linewidth]{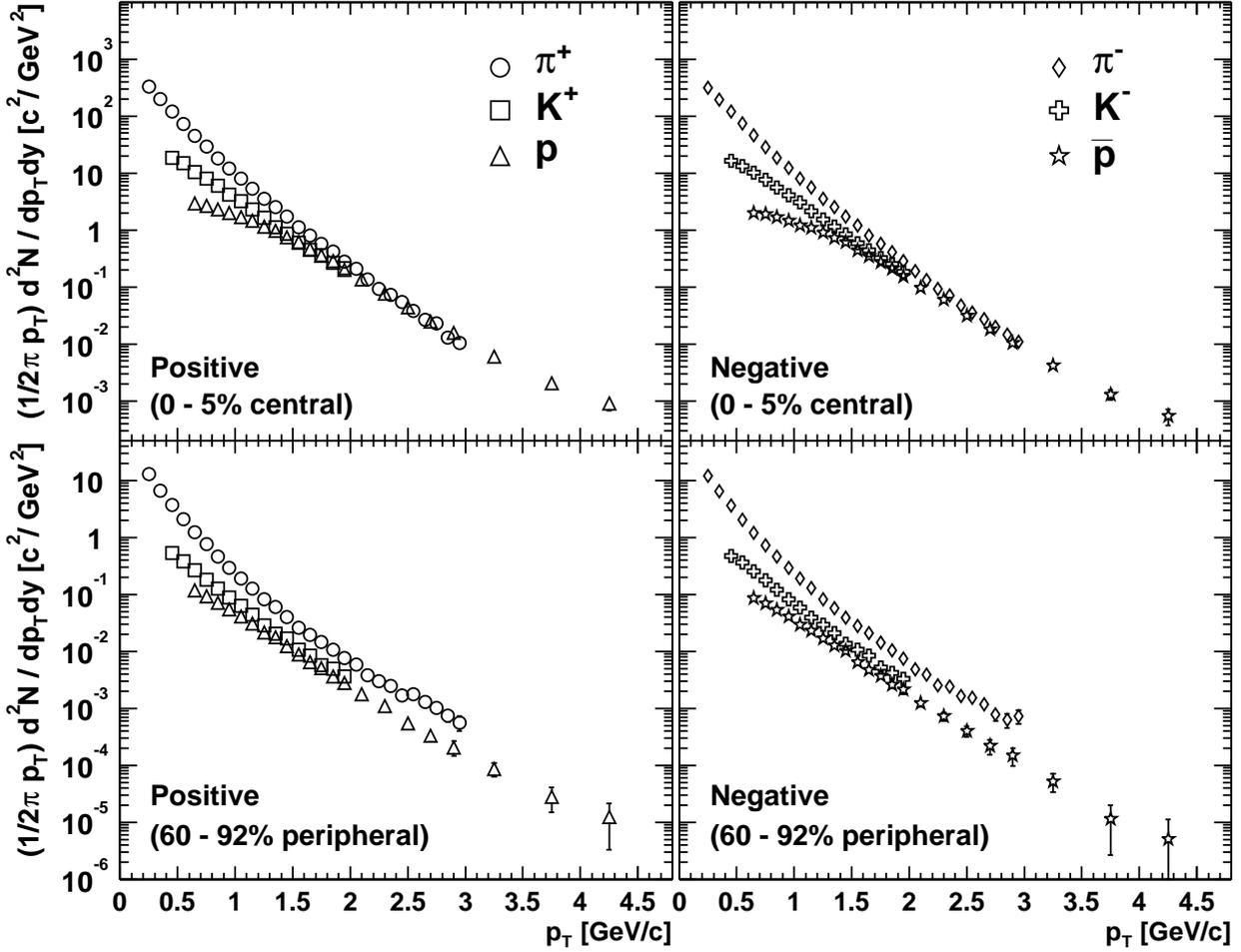}
\caption{Transverse momentum distributions for pions, kaons, protons
and anti-protons in Au+Au collisions at $\sqrt{s_{NN}}$~=~200~GeV. 
The top two figures show $\pt$ spectra for the most central 0--5\% collisions.
The bottom two are for the most peripheral 60--92\% collisions. The error bars 
are statistical only. The $\Lambda$ ($\lbar$) feed-down corrections for protons 
(anti-protons) have been applied.}
\label{fig:pt_spectra_all}
\end{figure*} 

Protons and anti-protons from weak decays (e.g. from $\Lambda$ and $\lbar$) can be 
reconstructed as tracks in the PHENIX spectrometer. The proton and anti-proton 
spectra are corrected to remove the feed-down contribution from weak decays using 
a HIJING~\cite{HIJING} simulation. HIJING output has been tuned to reproduce the 
measured particle ratios of $\Lambda/p$ and $\lbar/\pbar$ along with their $\pt$ 
dependencies in $\sqrt{s_{NN}}$~=~130~GeV Au+Au collisions~\cite{PPG012} which include
contribution from $\Xi$ and $\Sigma^{0}$. Corrections for feed-down from $\Sigma^{\pm}$
are not applied, as these yields were not measured. About 2$\times 10^{6}$ central 
HIJING events (impact parameter $b = 0 - 3$ fm) covering the TOF acceptance have been 
generated and processed through the PHENIX reconstruction software. 
To calculate the feed-down corrections, the $\pbar/p$ and $\lbar/\Lambda$ yield 
ratios were assumed to be independent of $p_{T}$ and centrality. 
The systematic error due to the feed-down correction is estimated
at 6\% by varying the $\Lambda/p$ and $\lbar/\pbar$ ratios within
the systematic errors of the $\snn = $130~GeV Au+Au measurement~\cite{PPG012} 
($\pm$24\%) and assuming $\mt$-scaling at high $\pt$. This uncertainty
could be larger if the $\Lambda/p$ and $\lbar/\pbar$ ratios change
significantly with $\pt$ and beam energy. 
The fractional contribution to the $p$ ($\pbar$) yield from $\Lambda (\lbar)$, 
$\delta_{\rm feed}(\pt)$, is shown in Figure~\ref{fig:feeddown}. The solid (dashed) 
lines represent the systematic errors for protons ($\pbar$). The obtained factor 
is about 40\% below 1~GeV/$c$ and 30\% at 4~GeV/$c$. We multiply the proton and 
anti-proton spectra by the factor, $C_{\rm feed}$, for all centrality 
bins as a function of $\pt$: 

\begin{equation}
C_{\rm feed}(j,\pt) = 1-\delta_{\rm feed}(j,\pt),
\label{eq:feed}
\end{equation}
where $j = p, \pbar$. 

%%%%%%%%%%%%%%%%%%%%
% Final pt spectra %
%%%%%%%%%%%%%%%%%%%%
\subsection{Invariant Yield}
\label{sec:final_spectra}

Applying the data cuts and corrections discussed above, 
the final invariant yield for each particle species
and centrality bin are derived using the following equation.

\begin{equation}
\frac{1}{2\pi \pt} \frac{d^{2}N}{dp_T dy} 
  = \frac{1}{2\pi \pt} \cdot 
    \frac{1}{N_{evt}(i)} \cdot
    C_{ij}(\pt) \cdot
    \frac{N_{j}(i,\pt)}{\Delta \pt \Delta y},
\label{eq:final_spectra}
\end{equation}
where $y$ is rapidity, $N_{evt}(i)$ is the number of events in 
each centrality bin $i$, $C_{ij}(\pt)$ is the total 
correction factor and $N_{j}(i,\pt)$ is the number of counts 
in each centrality bin $i$, particle species $j$, and $\pt$. 
The total correction factor is composed of:  
\begin{equation}
C_{ij}(\pt) = 
    \frac{1}{\epsilon_{\rm acc}(j,\pt)} \cdot
    \frac{1}{\epsilon_{\rm mult}(i,j)} \cdot
    C_{\rm feed}(j,\pt).
\label{eq:total_eff}
\end{equation}

%%%%%%%%%%%%%%%%%%%%%%%%%%%%%%%%%%%%%%%% Figure 6.
\begin{figure*}
\includegraphics[width=1.0\linewidth]{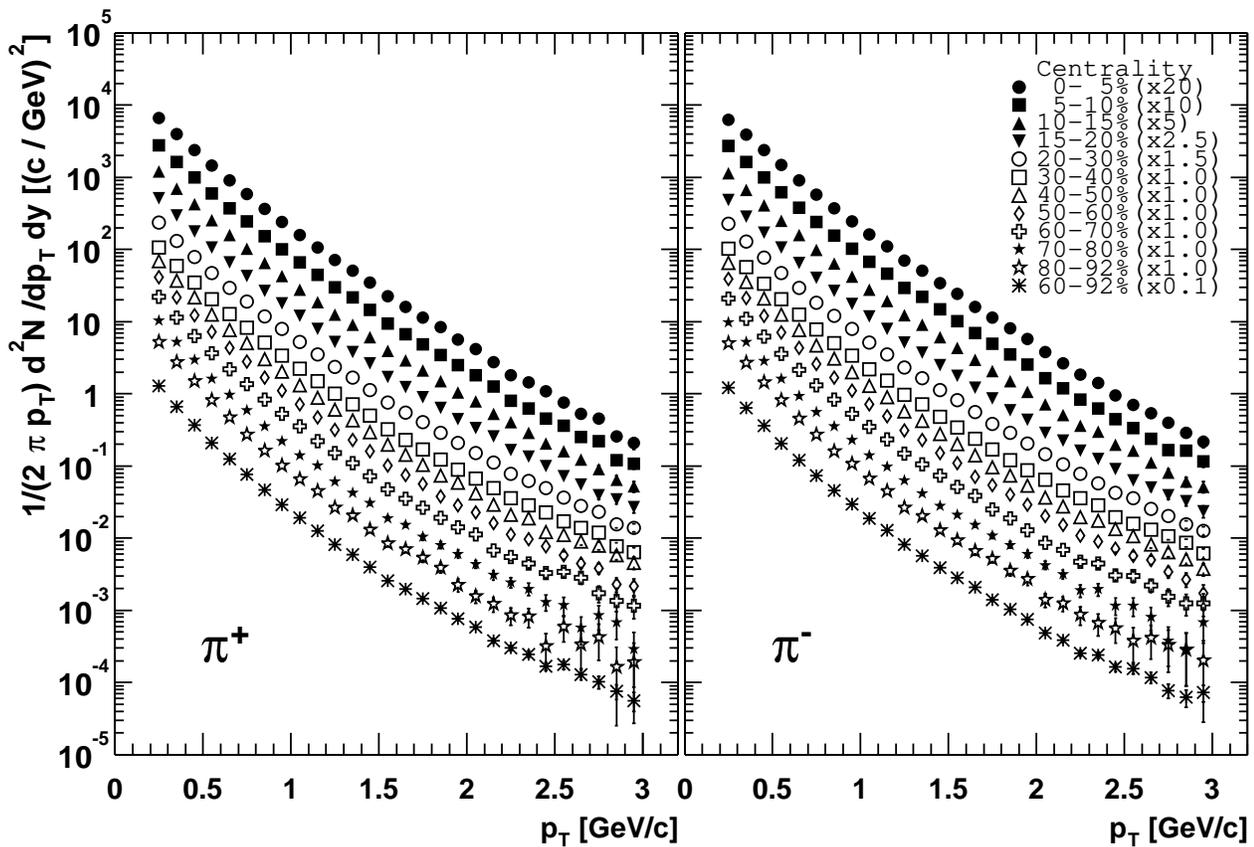}
\caption{Centrality dependence of the $\pt$ distribution for $\pi^{+}$ 
(left) and $\pi^{-}$ (right) in Au+Au collisions at $\sqrt{s_{NN}}$ 
= 200~GeV. The different symbols correspond to different centrality bins. 
The error bars are statistical only. For clarity, the data points are scaled 
vertically as quoted in the figure.}
\label{fig:pt_spectra_pion}
\end{figure*}   
%%%%%%%%%%%%%%%%%%%%%%%%%%%%%%%%%%%%%%%% Figure 7.
\begin{figure*}
\includegraphics[width=1.0\linewidth]{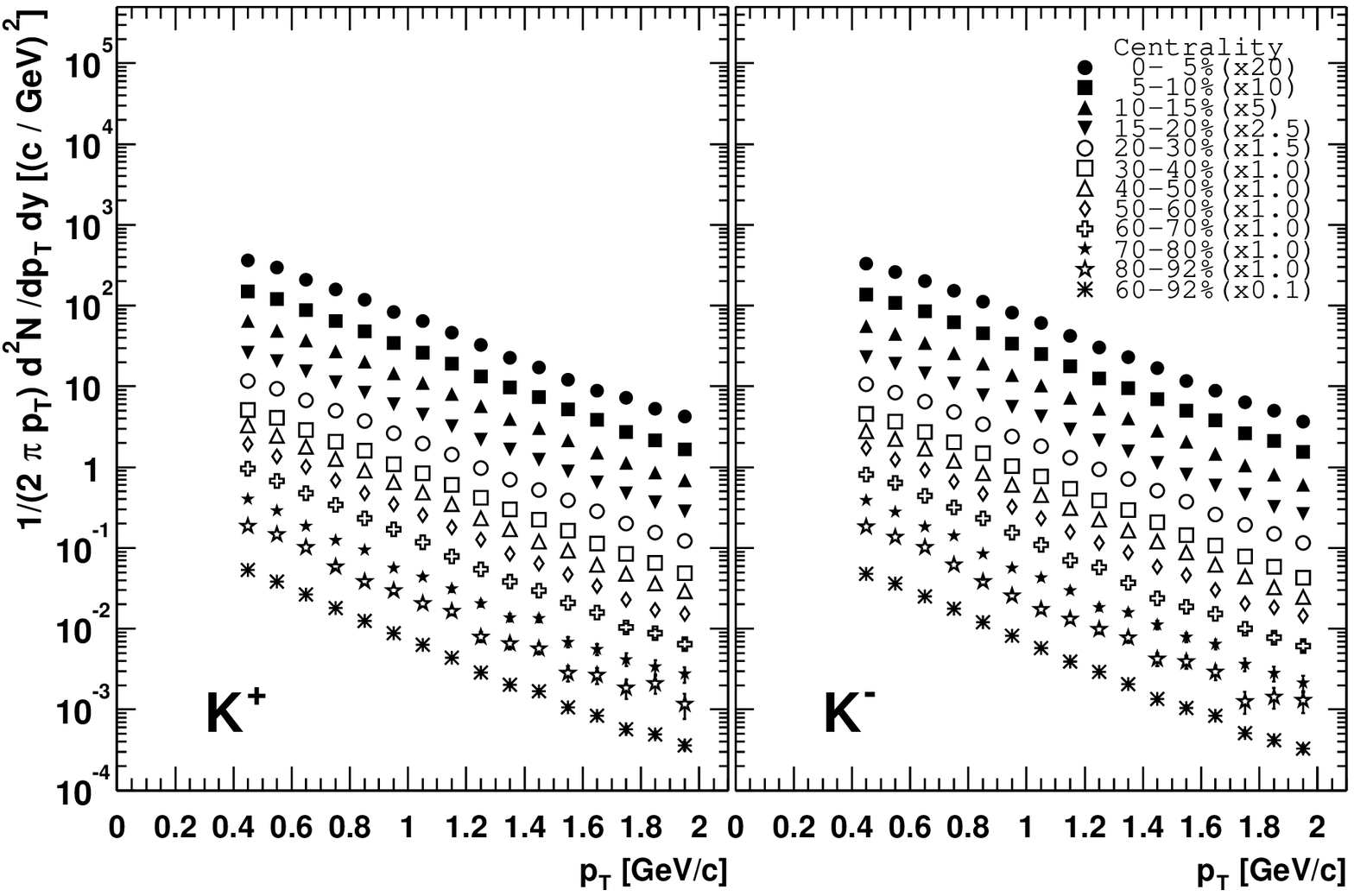}
\caption{Centrality dependence of the $\pt$ distribution for $K^{+}$ 
(left) and $K^{-}$ (right) in Au+Au collisions at $\sqrt{s_{NN}}$~=~200~GeV. 
The different symbols correspond to different centrality bins. 
The error bars are statistical only. For clarity, the data points are scaled 
vertically as quoted in the figure.}
\label{fig:pt_spectra_kaon}
\end{figure*}   

%%%%%%%%%%%%%%%%%%%%
% Systematic Error %
%%%%%%%%%%%%%%%%%%%%
\subsection{Systematic Uncertainties}
\label{sec:sys_error}
To estimate systematic uncertainties on the $\pt$ distribution 
and particle ratios, various sets of $\pt$ spectra and particle 
ratios were made by changing the cut parameters including the 
fiducial cut, PID cut, and track association windows slightly 
from what was used in the analysis. For each of these spectra 
and ratios using modified cuts, the same changes in the cuts 
were made in the Monte Carlo analysis. The absolutely normalized 
spectra with different cut conditions are divided by the spectra 
with the baseline cut conditions, resulting in uncertainties 
associated with each cut condition as a function of $\pt$. 
The various uncertainties are added in quadrature. Three different 
centrality bins  (minimum bias, central 0--5\%, and peripheral 60--92\%) 
are used to study the centrality dependence of systematic errors. 
The same procedure has been applied for the following particle ratios: 
$\pi^{-}/\pi^{+}$, $K^{-}/K^{+}$, $\pbar/p$, $K/\pi$, $p/\pi^{+}$, 
and $\pbar/\pi^{-}$.
 
Table~\ref{tab:sys_spectra_cent} shows the systematic errors of the 
$\pt$ spectra for central collisions. The systematic uncertainty on 
the absolute value of momentum (momentum scale) are estimated as 3\% 
in the measured $\pt$ range by comparing the known proton mass
to the value measured as protons in real data. It is found that the 
total systematic error on the $\pt$ spectra is 8--14\% in both central 
and peripheral collisions. For the particle ratios, the typical systematic 
error is about 6\% for all particle species. The dominant source of 
uncertainties on the central-to-peripheral ratio scaled by 
$N_{coll}$ ($R_{CP}$) are the systematic errors on the nuclear overlap 
function, $T_{\rm AuAu}$ (see Table~\ref{tab:sys_rcp_ncoll}). 
The systematic errors on $dN/dy$ and $\meanpt$ are discussed in 
Section~\ref{sec:meanpt_dndy} together with the procedure for the 
determination of these quantities.

%%%%%%%%%%%%%%%
% (4) Results %
%%%%%%%%%%%%%%%
\section{RESULTS}
\label{sec:results}
In this section, the $\pt$ and transverse mass spectra and yields 
of identified charged hadrons as a function of centrality are shown. Also a  
systematic study of particle ratios in Au+Au collisions at 
$\snn$~=~200~GeV at mid-rapidity is presented.

%%%%%%%%%%%%%%%%%%%%%%%%%%%%%%%%%%%%%%%% Figure 8.
\begin{figure*}
\includegraphics[width=1.0\linewidth]{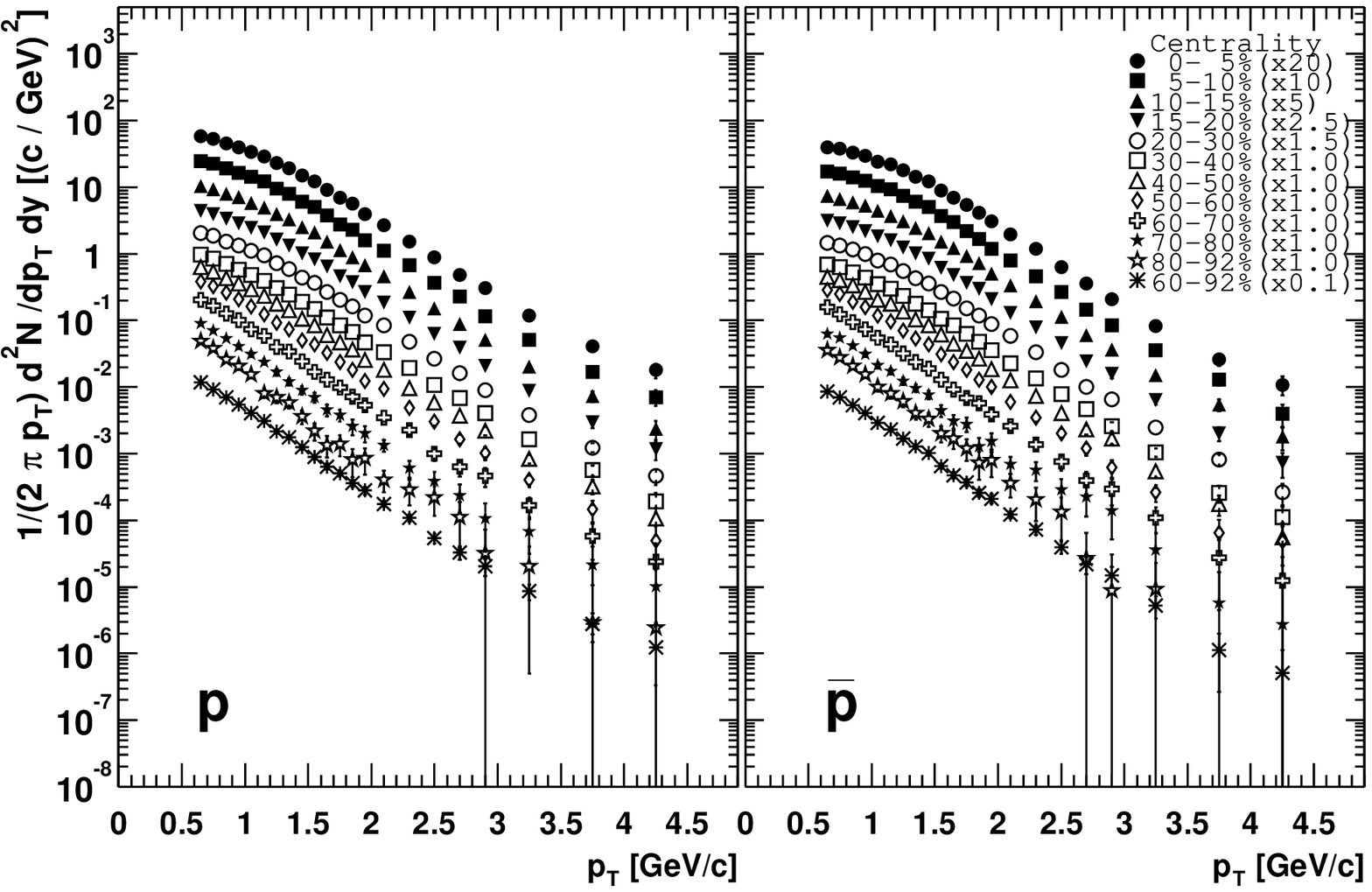}
\caption{Centrality dependence of the $\pt$ distribution for protons
(left) and anti-protons (right) in Au+Au collisions at $\sqrt{s_{NN}}$~=~200~GeV. 
The different symbols correspond to different centrality bins. 
The error bars are statistical only. Feed-down corrections for $\Lambda$ 
($\lbar$) decaying into proton ($\pbar$) have been applied. For clarity, 
the data points are scaled vertically as quoted in the figure.}
\label{fig:pt_spectra_proton}
\end{figure*} 

%%%%%%%%%%%%%%%%%%
% 4.1 pt spectra %
%%%%%%%%%%%%%%%%%%
\subsection{Transverse Momentum Distributions}
\label{sec:pt_spectra}

Figure~\ref{fig:pt_spectra_all} shows the $\pt$ distributions for pions, kaons, 
protons, and anti-protons. The top two plots are for the most 
central 0--5\% collisions, and the bottom two are for the most peripheral 
60--92\% collisions. The spectra for positive particles are presented on 
the left, and those for negative particles on the right. For $\pt <$ 1.5 
GeV/$c$ in central events, the data show a clear mass dependence in the 
shapes of the spectra. The $p$ and $\pbar$ spectra have a shoulder-arm shape,
the pion spectra have a concave shape, and the kaons fall exponentially. 
On the other hand, in the peripheral events, the mass dependences of the 
$\pt$ spectra are less pronounced and the $\pt$ spectra are more nearly parallel 
to each other. Another notable observation is that at $\pt$ above $\approx$ 
2.0~GeV/$c$ in central events, the $p$ and $\pbar$ yields become comparable 
to the pion yields, which is also observed in 130~GeV Au+Au collisions~\cite{PPG006}. 
This observation shows that a significant fraction of the total particle 
yield at $\pt \approx$ 2.0 -- 4.5~GeV/$c$ in Au+Au central collisions 
consists of $p$ and $\pbar$.

These high statistics Au+Au data at $\snn$~=~200~GeV allow us to perform a detailed
study of the centrality dependence of the $\pt$ spectra. In this analysis, 
we use the eleven centrality bins described in Section~\ref{sec:event_selection}
as well as the combined peripheral bin (60--92\%) for each particle species. 
Figure~\ref{fig:pt_spectra_pion} shows the centrality dependence of 
the $\pt$ spectrum for $\pi^{+}$ (left) and $\pi^{-}$ (right).
For clarity, the data points are scaled vertically as quoted in the figures. 
The error bars are statistical only. The pion spectra show an approximately 
power-law shape for all centrality bins. The spectra become steeper
(fall faster with increasing $\pt$) for more peripheral collisions.

%%%%%%%%%%%%%%%%%%%%%%%%%%%%%%%%%%%%%%%% Figure 9.
\begin{figure*}
\includegraphics[width=1.0\linewidth]{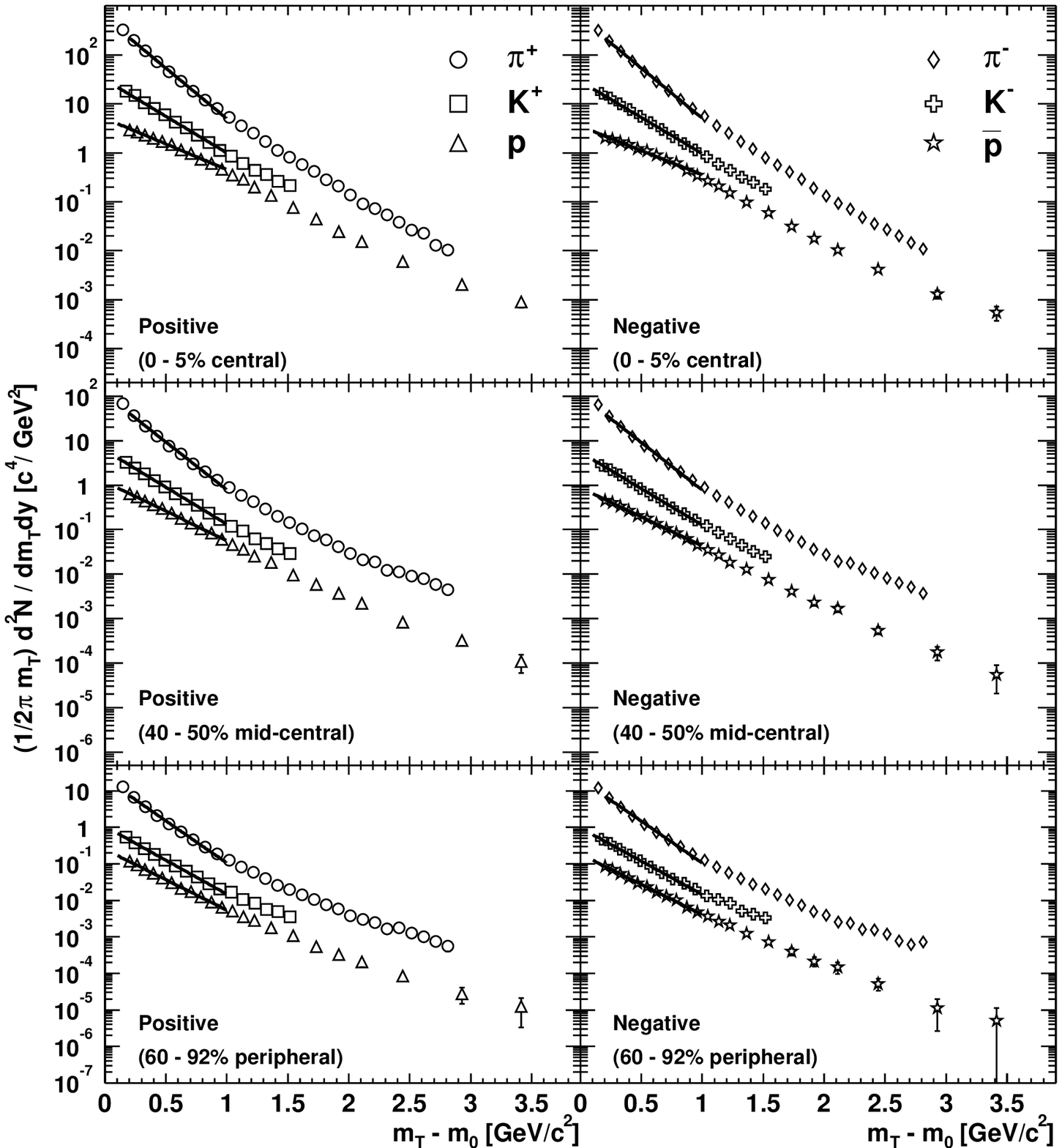}
\caption{Transverse mass distributions for $\pi^{\pm}$, $K^{\pm}$,
protons and anti-protons for central 0--5\% (top panels), mid-central 40--50\%
(middle panels) and peripheral 60--92\% (bottom panels) in Au+Au collisions 
at $\sqrt{s_{NN}}$~=~200~GeV. The lines on each spectra are the fitted 
results using $m_T$ exponential function. The fit ranges are 0.2 -- 1.0~GeV/$c^{2}$ 
for pions and 0.1 -- 1.0~GeV/$c^{2}$ for kaons, protons, and anti-protons in 
$m_T - m_0$. The error bars are statistical errors only.}
\label{fig:mt_spectra_all}
\end{figure*} 

%%%%%%%%%%%%%%%%%%%%%%%%%%%%%%%%%%%
% Systematic errors on pT spectra %
%%%%%%%%%%%%%%%%%%%%%%%%%%%%%%%%%%%
\begin{table*}
\caption{Systematic errors on the $\pt$ spectra for central events.
All errors are given in percent.}
\begin{ruledtabular}\begin{tabular}{ccccccccc}
                        & $\pi^{+}$ & $\pi^{-}$ & $K^{+}$   & $K^{-}$    & \multicolumn{2}{c}{$p$} & \multicolumn{2}{c}{$\pbar$}   \\ \hline
$\pt$ range (GeV/$c$)   & 0.2 - 3.0 & 0.2 - 3.0 & 0.4 - 2.0 & 0.4 - 2.0  & 0.6 - 3.0 & 3.0 - 4.5    & 0.6 - 3.0 & 3.0 - 4.5  \\ \hline
Cuts                    & 6.2       & 6.2       & 11.2      & 9.5        & 6.6       & 11.6         & 6.6       & 11.6       \\
Momentum scale          & 3         & 3         & 3         & 3          & 3         & 3            & 3         & 3          \\
Occupancy correction    & 2         & 2         & 3         & 3          & 3         & 3            & 3         & 3          \\
Feed-down correction    & -         & -         & -         & -          & 6.0       & 6.0          & 6.0       & 6.0        \\ \hline
Total                   & 7.2       & 7.2       & 12.0      & 10.4       & 9.9       & 13.7         & 9.9       & 9.9       \\ 
\end{tabular}\end{ruledtabular}
\label{tab:sys_spectra_cent}
\end{table*}

%%%%%%%%%%%%%%%%%%%%%%%%%%%%%
% Systematic errors on R_CP %
%%%%%%%%%%%%%%%%%%%%%%%%%%%%%
\begin{table*}
\caption{Systematic errors on Central-to-Peripheral ratio ($R_{CP}$). 
All errors are given in percent.}
\begin{ruledtabular}\begin{tabular}{cccc}
Source                                   & $(\pi^{+} + \pi^{-})/2$ & $(K^{+} + K^{-})/2$   & $(p + \pbar)/2$ \\ \hline
Occupancy correction (central)           & 2                       & 3                     & 3               \\ 
Occupancy correction (peripheral)        & 2                       & 3                     & 3               \\ 
$\langle T_{\rm AuAu}\rangle$ (0--10\%)  & 6.9                     & 6.9                   & 6.9             \\ 
$\langle T_{\rm AuAu}\rangle$ (60--92\%) & 28.6                    & 28.6                  & 28.6            \\ \hline 
Total                                    & 29.5                    & 29.7                  & 29.7            \\ 

\end{tabular}\end{ruledtabular}
\label{tab:sys_rcp_ncoll}
\end{table*}

Figure~\ref{fig:pt_spectra_kaon} shows similar plots for kaons. 
The data can be well approximated by an exponential function in 
$\pt$ for all centralities. Finally, the centrality dependence of 
the $\pt$ spectra for protons (left) and anti-protons (right) is shown in 
Figure~\ref{fig:pt_spectra_proton}. As in Figure~\ref{fig:pt_spectra_all}, 
both $p$ and $\pbar$ spectra show a strong centrality dependence below 1.5~GeV/$c$, 
i.e. they develop a shoulder at low $\pt$ and the spectra flatten (fall more 
slowly with increasing $\pt$) with increasing collision centrality.  

%%% [Hydro models for pT spectra] %%%
Up to $\pt = 1.5 \sim 2$~GeV/$c$,  it has been found that hydrodynamic models 
can reproduce the data well for $\pi^{\pm}$, $K^{\pm}$, $p$ and $\pbar$ spectra 
at 130~GeV~\cite{PPG009}, and also the preliminary data at 200~GeV in Au+Au collisions 
(e.g.~\cite{hydro_2,hydro_200gev}). These models assume thermal equilibrium and 
that the created particles are affected by a common transverse flow velocity 
$\beta_T$ and freeze-out (stop interacting) at a temperature $T_{fo}$ with a 
fixed initial condition governed by the equation of state (EOS) of matter. 
There are several types of hydrodynamic calculations, e.g., (1) a conventional 
hydrodynamic fit to the experimental data with two free parameters, $\beta_T$ 
and $T_{fo}$~\cite{Schnedermann}, (2) a combination of hydrodynamics and a 
hadronic cascade model~\cite{hydro_2}, (3) transverse and longitudinal flow 
with simultaneous chemical and thermal freeze-outs within the statistical 
thermal model~\cite{hydro_3}, (4) requiring the early thermalization with a 
QGP type EOS~\cite{Heinz}. Despite the differences between the hydrodynamic models, 
all models are in qualitative agreement with the identified single particle 
spectra in central collisions at low $\pt$ as seen in reference~\cite{PPG009}. 
However, they fail to reproduce the peripheral spectra above $\pt \simeq 1$~GeV/$c$ 
and their applicability in the high $\pt$ region ($>$ 2~GeV/$c$) is limited. 
Comparison with the detailed centrality dependence of hadron spectra presented 
here would shed light on further understanding of the EOS, chemical properties 
in the model, and the freeze-out conditions at RHIC. 

%%%%%%%%%%%%%%%%%%
% 4.2 mt spectra %
%%%%%%%%%%%%%%%%%%
\subsection{Transverse Mass Distributions}
\label{sec:mt_spectra}
%%%%%%%%%%%%%%%%%%%%%%%%%%%%%%%%%%%%%%%% Figure 10.
\begin{figure}[t]
\includegraphics[width=1.0\linewidth]{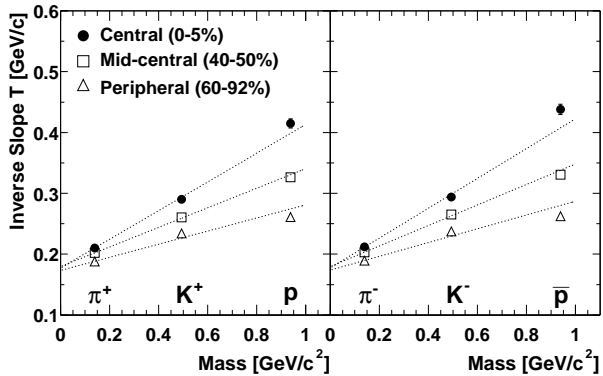}
\caption{Mass and centrality dependence of inverse slope parameters $T$
in $m_T$ spectra for positive (left) and negative (right) particles 
in Au+Au collisions at $\sqrt{s_{NN}}$~=~200~GeV. The fit ranges are
0.2 -- 1.0~GeV/$c^{2}$ for pions and 0.1 -- 1.0~GeV/$c^{2}$ for kaons, 
protons, and anti-protons in $m_T - m_0$. The dotted lines represent 
a linear fit of the results from each centrality bin as a function of mass
using Eq.~\ref{eq:flow}.}
\label{fig:slope_mass}
\end{figure}  

In order to quantify the observed particle mass dependence
of the $\pt$ spectra shape and their centrality dependence, 
the transverse mass spectra for identified charged hadrons
are presented here. From former studies at lower beam energies,
it is known that the invariant differential cross sections in 
$p+p$, $p+A$, and $A+A$ collisions generally show a shape of 
an exponential in $m_{T}-m_{0}$, where $m_{0}$ is particle 
mass, and $m_T = \sqrt{p_{T}^{2} + m_{0}^{2}}$ is transverse 
mass. For an $\mt$ spectrum with an exponential shape, one can 
parameterize it as follows:

\begin{equation}
	\frac{d^{2}N}{2\pi m_{T}dm_{T}dy} 
	= \frac{1}{2\pi T(T+m_{0})} \cdot A
	\cdot \exp{(-\frac{m_{T}-m_{0}}{T})}, 
\label{eq:mt_exp}
\end{equation}
where $T$ is referred to as the inverse slope parameter,
and $A$ is a normalization parameter which contains information 
on $dN/dy$. In Figure~\ref{fig:mt_spectra_all}, $m_T$ distributions 
for $\pi^{\pm}$, $K^{\pm}$, $p$ and $\pbar$ for central 0--5\% (top panels), 
mid-central 40--50\% (middle panels) and peripheral 60--92\% (bottom panels) 
collisions are shown. The spectra for positive particles are on the left and for 
negative particles are on the right. The solid lines overlaid on each spectra are the fit 
results using Eq.~\ref{eq:mt_exp}.  The error bars are statistical only. 
As seen in Figure~\ref{fig:mt_spectra_all}, all the $\mt$ spectra display an 
exponential shape in the low $\mt$ region. However, at higher $\mt$, the spectra 
become less steep, which corresponds to a power-law behavior in $\pt$. Thus, the inverse
slope parameter in Eq.~\ref{eq:mt_exp} depends on the fitting range. 
In this analysis, the fits cover the range 0.2 -- 1.0~GeV/$c^{2}$ 
for pions and 0.1 -- 1.0~GeV/$c^{2}$ for kaons, protons, 
and anti-protons in $m_T - m_0$. The low $\mt$ region ($m_T - m_0 <$ 0.2~GeV/$c^2$) 
for pions is excluded from the fit to eliminate the contributions from 
resonance decays. The inverse slope parameters for each particle species 
in the three centrality bins are summarized in Figure~\ref{fig:slope_mass} and in
Table~\ref{tab:mt_slope}. The inverse slope parameters increase with increasing 
particle mass in all centrality bins. This increase for central collisions is more 
rapid for heavier particles.

Such a behavior was derived, under certain conditions, by 
E.~Schnedermann~{\it et al.}~\cite{Schnedermann1994} for central collisions and 
by T. Cs\"org\H{o}~{\it et al.}~\cite{Csorgo} for non-central heavy ion collisions: 

\begin{equation}
T = T_{0} + m \ut^{2}.
\label{eq:flow}
\end{equation}
Here $T_0$ is a freeze-out temperature and $\ut$ is a measure of the strength 
of the (average radial) transverse flow. The dotted lines in Figure~\ref{fig:slope_mass} 
represent a linear fit of the results from each centrality bin as a function of mass 
using Eq.~\ref{eq:flow}. The fit parameters for positive and negative particles
are shown in Table~\ref{tab:mt_slope}. It indicates, that the linear extrapolation 
of the slope parameter $T(m)$ to zero mass has the same intercept parameters $T_0$ in all 
the centrality classes, indicating that the freeze-out temperature is approximately 
independent of the centrality. On the other hand, $\ut$, the strength of the average 
transverse flow is increasing with increasing centrality, supporting the 
hydrodynamic picture.

Motivated by the idea of a Color Glass Condensate, the authors of reference~\cite{mt_scaling} 
argued that the $\mt$ spectra (not $\mt - m_{0}$) of identified hadrons at RHIC energy 
follow a generalized scaling law for all centrality classes when the proton (kaon)
spectrum is multiplied by a factor of 0.5 (2.0). The 200 GeV Au+Au pion and kaon
spectra seem to follow this $\mt$ scaling, but proton and anti-proton spectra
are below it by a factor of $\sim$ 2 for all centralities. Since $p$ and $\pbar$ 
spectra presented here are corrected for weak decays from $\Lambda$ and 
$\overline{\Lambda}$, the model also needs to study the feed-down effect to 
conclude that a universal $\mt$ scaling law is seen at RHIC.

%%%%%%%%%%%%%%%%%%%%%%%
% Inverse Slope Table %
%%%%%%%%%%%%%%%%%%%%%%%
\begin{table}
\caption{(Top) Inverse slope parameters for $\pi$, $K$, $p$ and $\pbar$ for the 0--5\%, 40--50\% and
60--92\% centrality bins, in units of MeV/$c^{2}$. The errors are statistical only. 
(Bottom) The extracted fit parameters of the freeze-out temperature ($T_0$) in units 
of MeV/$c^{2}$ and the measure of the strength of the average radial transverse flow 
($\ut$) using Eq.~\ref{eq:flow}. The fit results shown here are for positive and 
negative particles, as denoted in the superscripts, and for three different centrality bins.}
\begin{ruledtabular}\begin{tabular}{cccc}
\bf {Particle} &  0--5\%              &  40--50\%            &  60--92\%           \\ \hline 
$\pi^{+}$  &  210.2  $\pm$  0.8   &  201.9  $\pm$  0.8   &  187.8  $\pm$  0.7  \\
$\pi^{-}$  &  211.9  $\pm$  0.7   &  203.0  $\pm$  0.7   &  189.2  $\pm$  0.7  \\
$K^{+}$    &  290.2  $\pm$  2.2   &  260.6  $\pm$  2.4   &  233.9  $\pm$  2.6  \\
$K^{-}$    &  293.8  $\pm$  2.2   &  265.1  $\pm$  2.3   &  237.4  $\pm$  2.6  \\
$p$        &  414.8  $\pm$  7.5   &  326.3  $\pm$  5.9   &  260.7  $\pm$  5.4  \\
$\pbar$    &  437.9  $\pm$  8.5   &  330.5  $\pm$  6.4   &  262.1  $\pm$  5.9  \\ \hline \hline
\bf {Fit parameter}  &  0--5\%           &  40--50\%          &  60--92\%         \\ \hline
$T_{0}^{(+)}$         & 177.0 $\pm$ 1.2   &  179.5 $\pm$ 1.2   &  173.1 $\pm$ 1.2  \\
$T_{0}^{(-)}$         & 177.3 $\pm$ 1.2   &  179.6 $\pm$ 1.2   &  173.7 $\pm$ 1.1  \\
$\ut^{(+)}$           & 0.48  $\pm$ 0.07  &  0.40  $\pm$ 0.07  &  0.32  $\pm$ 0.07 \\
$\ut^{(-)}$           & 0.49  $\pm$ 0.07  &  0.41  $\pm$ 0.07  &  0.33  $\pm$ 0.07 \\ 
\end{tabular}\end{ruledtabular}
\label{tab:mt_slope}
\end{table}

%%%%%%%%%%%%%%%%%%%%%%%%%
% 4.3 mean pT and dN/dy %
%%%%%%%%%%%%%%%%%%%%%%%%%
\subsection{Mean Transverse Momentum and Particle Yields versus $\npart$}
\label{sec:meanpt_dndy}
%%%%%%%%%%%%%%%%%%%%%%%%%%%%%%%%%%%%%%%% Figure 11.
\begin{figure}[t]
\includegraphics[width=1.0\linewidth]{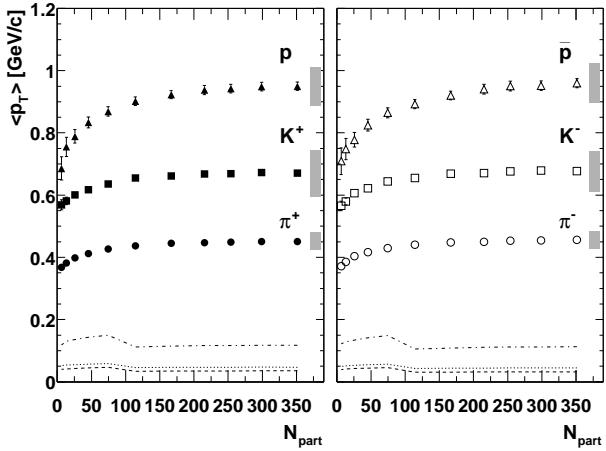}
\caption{Mean transverse momentum as a function of $\npart$ for 
pions, kaons, protons and anti-protons in Au+Au collisions at $\sqrt{s_{NN}}$ 
= 200~GeV. The left (right) panel shows the $\meanpt$ for positive (negative) 
particles. The error bars are statistical errors. The systematic errors from 
cuts conditions are shown as shaded boxes on the right for each particle species. 
The systematic errors from extrapolations, which are scaled by a factor of 2 for 
clarity, are shown in the bottom for protons and anti-protons (dashed-dot lines), 
kaons (dotted lines), and pions (dashed lines).}
\label{fig:meanpt_Npart}
\end{figure}   
%%%%%%%%%%%%%%%%%%%%%%%%%%%%%%%%%%%%%%%% Figure 12.
\begin{figure}[t]
\includegraphics[width=1.0\linewidth]{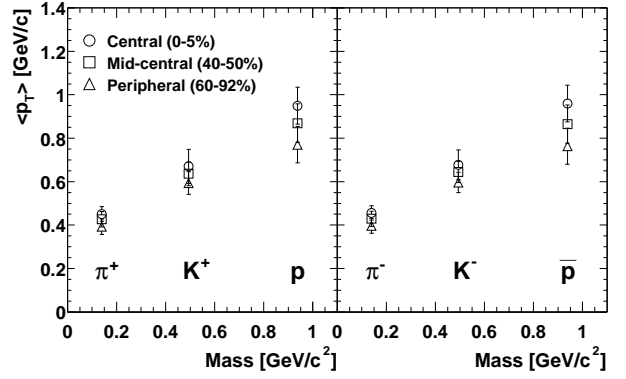}
\caption{Mean transverse momentum versus particle mass for 
central 0--5\%, mid-central 40--50\% and peripheral 60--92\%
in Au+Au collisions at $\sqrt{s_{NN}}$~=~200~GeV. 
The left (right) panel shows the $\meanpt$ for positive (negative) 
particles. The error bars represent the total systematic errors.
The statistical errors are negligible.}
\label{fig:meanpt_mass}
\end{figure}   
%%%%%%%%%%%%%%%%%%%%%%%%%%%%%%%%%%%%%%%% Figure 13.
\begin{figure}[ht]
\includegraphics[width=1.0\linewidth]{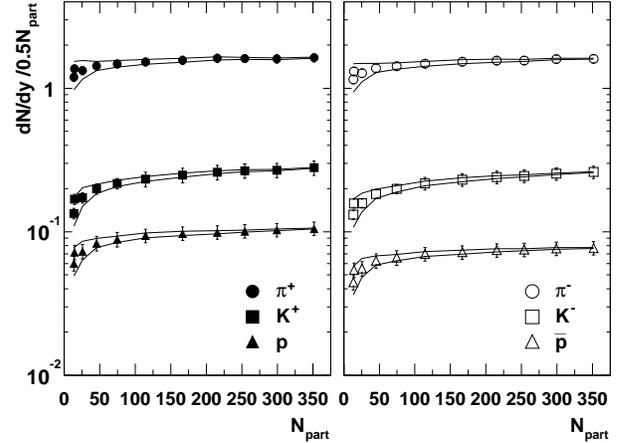}
\caption{Particle yield per unit rapidity ($dN/dy$) per participant 
pair (0.5 $N_{part}$) as a function of $\npart$ for pions, kaons, protons and 
anti-protons in Au+Au collisions at $\sqrt{s_{NN}}$~=~200~GeV. The left (right) 
panel shows the $dN/dy$ for positive (negative) particles. The error bars represent
the quadratic sum of statistical errors and systematic errors from cut conditions. 
The lines represent the effect of the systematic error 
on $\npart$ which affects all curves in the same way.} 
\label{fig:dndy_Npart}
\end{figure}   

By integrating a measured $\pt$ spectrum over $\pt$, one can determine
the mean transverse momentum, $\meanpt$, and particle yield per unit
rapidity, $dN/dy$, for each particle species. The procedure to determine 
the mean $\pt$ and $dN/dy$ is described below: 
(1) Determine $dN/dy$ and $\meanpt$ by integrating 
    over the measured $\pt$ range from the data. 
(2) Fit several appropriate functional forms (detailed below) to the $\pt$ spectra.
    Note that all of the fits are reasonable approximations to the data.  
    Integrate from zero to the first data point and from the last data 
    point to infinity.
(3) Sum the data yield and the two functional yield pieces together
    to get $dN/dy$ and $\meanpt$ in each functional form.
(4) Take the average between the upper and lower bounds from the different 
    functional forms to obtain the final $dN/dy$ and $\meanpt$. The statistical 
    uncertainties are determined from the data. The systematic errors from 
    the extrapolation of yield are defined as half of the
    difference between the upper and lower bounds.
(5) Determine the final systematic errors on $dN/dy$ and $\meanpt$ for each
    centrality bin by taking the quadrature sum of the extrapolation errors, 
    errors associated with cuts, detector occupancy corrections (for $dN/dy$)
    and feed-down corrections (for $p$ and $\pbar$).

For the extrapolation of $dN/dy$ and $\meanpt$, the following functional 
forms are used for different particle species: a power-law function and 
a $\pt$ exponential for pions, a $\pt$ exponential and  an $\mt$ exponential
for kaons, and a Boltzmann function, $\pt$ exponential, and $\mt$ exponential
for protons and anti-protons. The effects of contamination background at high
$\pt$ region for both $dN/dy$ and $\meanpt$ are estimated as less than 1\% 
for all particle species. 
The overall systematic uncertainties on both $dN/dy$ and $\meanpt$ are 
about 10--15\%. See Table~\ref{tab:sys_dndy} for the systematic errors 
of $dN/dy$ and Table~\ref{tab:sys_meanPt} for those of $\meanpt$.

%%%%%%%%%%%%%%%%%%%%%%%%%%%%%%
% Systematic errors on dN/dy %
%%%%%%%%%%%%%%%%%%%%%%%%%%%%%%
\begin{table}
\caption{Systematic errors on $dN/dy$ for central 0--5\% (top) and 
peripheral 60--92\% (bottom) collisions. All errors are given in 
percent.}
\begin{ruledtabular}\begin{tabular}{ccccccc}
Source                      & $\pi^{+}$  & $\pi^{-}$  & $K^{+}$   & $K^{-}$  & $p$   & $\pbar$  \\ \hline
 & \multicolumn{6}{c}{Central 0--5\%} \\ \hline
Cuts + occupancy            & 6.5        & 6.5        & 11.6      & 10.0     & 7.2   & 7.2  \\ 
Extrapolation               & 5.4        & 4.8        & 5.7       & 5.6      & 9.6   & 9.2  \\
Contamination background    & $<$1       & $<$1       & $<$1      & $<$1     & $<$1  & $<$1 \\
Feed-down                   & -          & -          & -         & -        & 8.0   & 8.0  \\ \hline
Total                       & 8.4        & 8.0        & 12.9      & 11.4     & 14.4  & 14.4 \\ \hline \hline
 & \multicolumn{6}{c}{Peripheral 60--92\%}  \\ \hline
Cuts + occupancy            & 6.5        & 6.5        & 8.3       & 7.2      & 8.3   & 8.3  \\
Extrapolation               & 8.4        & 8.0        & 7.4       & 7.5      & 13.6  & 13.6 \\
Contamination background    & $<$1       & $<$1       & $<$1      & $<$1     & $<$1  & $<$1 \\
Feed-down                   & -          & -          & -         & -        & 8.0   & 8.0  \\ \hline
Total                       & 10.6       & 10.3       & 11.1      & 10.3     & 17.8  & 17.8 \\ 
\end{tabular}\end{ruledtabular}
\label{tab:sys_dndy}
\end{table}

%%%%%%%%%%%%%%%%%%%%%%%%%%%%%
% Systematic errors on <pT> %
%%%%%%%%%%%%%%%%%%%%%%%%%%%%%
\begin{table}
\caption{Systematic errors on $\meanpt$ for central 0--5\% (top) and 
peripheral 60--92\% (bottom) collisions. All errors are given in 
percent.}
\begin{ruledtabular}\begin{tabular}{ccccccc}
Source                     & $\pi^{+}$  & $\pi^{-}$  & $K^{+}$   & $K^{-}$  & $p$    & $\pbar$ \\ \hline
 & \multicolumn{6}{c}{Central 0--5\%} \\ \hline
Cuts                       & 6.2        & 6.2        & 11.2      & 9.5      & 6.6    & 6.6     \\
Extrapolation              & 3.9        & 3.5        & 3.5       & 3.3      & 6.2    & 5.9     \\
Contamination background   & $<$1       & $<$1       & $<$1      & $<$1     & $<$1   & $<$1    \\
Feed-down                  & -          & -          & -         & -        & 1.0    & 1.0     \\ \hline
Total                      & 7.3        & 7.1        & 13.5      & 10.0     & 9.1    & 8.9     \\ \hline \hline
 & \multicolumn{6}{c}{Peripheral 60--92\%}  \\ \hline
Cuts                       & 6.2        & 6.2        & 7.7       & 6.6      & 7.7    & 7.7     \\
Extrapolation              & 5.4        & 5.3        & 4.6       & 4.4      & 8.6    & 8.6     \\
Contamination background   & $<$1       & $<$1       & $<$1      & $<$1     & $<$1   & $<$1    \\
Feed-down                  & -          & -          & -         & -        & 1.0    & 1.0     \\ \hline
Total                      & 8.2        & 8.1        & 8.9       & 7.9      & 11.5   & 11.5    \\ 
\end{tabular}\end{ruledtabular}
\label{tab:sys_meanPt}
\end{table}

%%%%%%%%%%%%%%%%%%%%%
% mean pt vs. Npart %
%%%%%%%%%%%%%%%%%%%%%
\begin{table*}
\caption{Centrality dependence of $\meanpt$ for $\pi^{\pm}$, $K^{\pm}$, $p$ and $\pbar$ in MeV/$c$.
The errors are systematic only. The statistical errors are negligible.}
\begin{ruledtabular}\begin{tabular}{rcccccc}
$N_{part}$ &  $\pi^{+}$   &  $\pi^{-}$  &  $K^{+}$   &  $K^{-}$  &  $p$  & $\pbar$  \\ \hline
351.4      &  451  $\pm$   33  &  455 $\pm$ 32  &  670  $\pm$  78 &  677  $\pm$  68  &  949  $\pm$  85  &  959  $\pm$  84 \\
299.0      &  450  $\pm$   33  &  454 $\pm$ 33  &  672  $\pm$  78 &  679  $\pm$  68  &  948  $\pm$  84  &  951  $\pm$  83 \\
253.9      &  448  $\pm$   33  &  453 $\pm$ 33  &  668  $\pm$  78 &  676  $\pm$  68  &  942  $\pm$  84  &  950  $\pm$  83 \\
215.3      &  447  $\pm$   34  &  449 $\pm$ 33  &  667  $\pm$  78 &  670  $\pm$  67  &  937  $\pm$  84  &  940  $\pm$  83 \\
166.6      &  444  $\pm$   35  &  447 $\pm$ 34  &  661  $\pm$  77 &  668  $\pm$  67  &  923  $\pm$  85  &  920  $\pm$  83 \\
114.2      &  436  $\pm$   35  &  440 $\pm$ 35  &  655  $\pm$  77 &  654  $\pm$  66  &  901  $\pm$  83  &  892  $\pm$  82  \\
74.4       &  426  $\pm$   35  &  429 $\pm$ 35  &  636  $\pm$  54 &  644  $\pm$  48  &  868  $\pm$  88  &  864  $\pm$  88 \\
45.5       &  412  $\pm$   35  &  416 $\pm$ 34  &  617  $\pm$  53 &  621  $\pm$  47  &  833  $\pm$  86  &  824  $\pm$  86  \\
25.7       &  398  $\pm$   34  &  403 $\pm$ 33  &  600  $\pm$  52 &  606  $\pm$  46  &  788  $\pm$  84  &  777  $\pm$  83  \\
13.4       &  381  $\pm$   32  &  385 $\pm$ 32  &  581  $\pm$  51 &  579  $\pm$  46  &  755  $\pm$  82  &  747  $\pm$  80  \\
6.3        &  367  $\pm$   30  &  371 $\pm$ 30  &  568  $\pm$  51 &  565  $\pm$  45  &  685  $\pm$  78  &  708  $\pm$  81  \\ 
\end{tabular}\end{ruledtabular}
\label{tab:meanpt_npart}
\end{table*}

%%%%%%%%%%%%%%%%%%%
% dN/dy vs. Npart %
%%%%%%%%%%%%%%%%%%%
\begin{table*}
\caption{Centrality dependence of $dN/dy$ for $\pi^{\pm}$, 
$K^{\pm}$, $p$ and $\pbar$. The errors are systematic only. 
The statistical errors are negligible.}
\begin{ruledtabular}\begin{tabular}{rcccccc}
$N_{part}$ &  $\pi^{+}$          &  $\pi^{-}$          &  $K^{+}$          &  $K^{-}$          &   $p$              &  $\pbar$  \\ \hline
351.4      &  286.4  $\pm$  24.2   &  281.8  $\pm$  22.8   &  48.9  $\pm$  6.3   &  45.7  $\pm$  5.2   &   18.4  $\pm$  2.6   &  13.5  $\pm$  1.8  \\  
299.0      &  239.6  $\pm$  20.5   &  238.9  $\pm$  19.8   &  40.1  $\pm$  5.1   &  37.8  $\pm$  4.3   &   15.3  $\pm$  2.1   &  11.4  $\pm$  1.5  \\  
253.9      &  204.6  $\pm$  18.0   &  198.2  $\pm$  16.7   &  33.7  $\pm$  4.3   &  31.1  $\pm$  3.5   &   12.8  $\pm$  1.8   &   9.5  $\pm$  1.3  \\
215.3      &  173.8  $\pm$  15.6   &  167.4  $\pm$  14.4   &  27.9  $\pm$  3.6   &  25.8  $\pm$  2.9   &   10.6  $\pm$  1.5   &   7.9  $\pm$  1.1  \\
166.6      &  130.3  $\pm$  12.4   &  127.3  $\pm$  11.6   &  20.6  $\pm$  2.6   &  19.1  $\pm$  2.2   &    8.1  $\pm$  1.1   &   5.9  $\pm$  0.8  \\ 
114.2      &   87.0  $\pm$  8.6    &   84.4  $\pm$  8.0    &  13.2  $\pm$  1.7   &  12.3  $\pm$  1.4   &    5.3  $\pm$  0.7   &   3.9  $\pm$  0.5  \\ 
74.4       &   54.9  $\pm$  5.6    &   52.9  $\pm$  5.2    &   8.0  $\pm$  0.8   &   7.4  $\pm$  0.6   &    3.2  $\pm$  0.5   &   2.4  $\pm$  0.3  \\ 
45.5       &   32.4  $\pm$  3.4    &   31.3  $\pm$  3.1    &   4.5  $\pm$  0.4   &   4.1  $\pm$  0.4   &    1.8  $\pm$  0.3   &   1.4  $\pm$  0.2  \\
25.7       &   17.0  $\pm$  1.8    &   16.3  $\pm$  1.6    &   2.2  $\pm$  0.2   &   2.0  $\pm$  0.1   &   0.93  $\pm$  0.15  &  0.71  $\pm$  0.12 \\  
13.4       &   7.9   $\pm$  0.8    &    7.7  $\pm$  0.7    &  0.89  $\pm$  0.09  &  0.88  $\pm$  0.09  &   0.40  $\pm$  0.07  &  0.29  $\pm$  0.05 \\ 
6.3        &   4.0   $\pm$  0.4    &    3.9  $\pm$  0.3    &  0.44  $\pm$  0.04  &  0.42  $\pm$  0.04  &   0.21  $\pm$  0.04  &  0.15  $\pm$  0.02 \\ 
\end{tabular}\end{ruledtabular}
\label{tab:dndy_npart}
\end{table*}

In Figure~\ref{fig:meanpt_Npart}, the centrality dependence of $\meanpt$ 
for $\pi^{\pm}$, $K^{\pm}$, $p$ and $\pbar$ is shown. The error bars in the 
figure represent the statistical errors. The systematic errors from cuts conditions 
are shown as shaded boxes on the right for each particle species. The systematic errors
from extrapolations, which are scaled by a factor of 2 for clarity, are shown in the 
bottom for each particle species. The data are also summarized in 
Table~\ref{tab:meanpt_npart}. It is found that $\meanpt$ for all particle species 
increases from the most peripheral to mid-central collisions, and appears to saturate 
from the mid-central to central collisions (although the $\meanpt$ values for $p$ and 
$\pbar$ may continue to rise). It should be noted that while the total systematic errors 
on $\meanpt$ listed in Table~\ref{tab:sys_meanPt} is large, the trend shown in the 
figure is significant. One of the main sources of the uncertainty is the yield extrapolation 
in unmeasured $\pt$ range (e.g. $\pt <$ 0.6~GeV/$c$ for protons and anti-protons). 
These systematic errors are correlated, and therefore move the curve up and down simultaneously. 
In Figure~\ref{fig:meanpt_mass}, the particle mass and centrality dependence of $\meanpt$ 
are shown. The data presented here are the $\meanpt$ for the 0--5\%, 40--50\% and 60--92\% 
centrality bins. Figure~\ref{fig:meanpt_mass} is similar to Figure~\ref{fig:slope_mass}, 
which shows the inverse slope parameters, in that the $\meanpt$ increases with particle 
mass and with centrality. This is qualitatively consistent with the hydrodynamic expansion 
picture~\cite{hydro_200gev,Schnedermann1994,Csorgo}. 

Figure~\ref{fig:dndy_Npart} shows the centrality dependence of $dN/dy$ per participant 
pair (0.5 $N_{part}$). The data are summarized in Table~\ref{tab:dndy_npart}. The error 
bars on each point represent the quadratic sum of the statistical errors and systematic 
errors from cut conditions. The statistical errors are negligible. The lines represent 
the effect of the systematic error on $\npart$ which affects 
all curves in the same way. The data indicate that $dN/dy$ per participant pair increases 
for all particle species with $\npart$ up to $\approx$ 100, and saturates from the mid-central 
to the most central collisions. From $dN/dy$ for protons and anti-protons, we obtain the 
net proton number at mid-rapidity for the most central 0--5\% collisions, 
$dN/dy|_{p} - dN/dy|_{\pbar} = 18.47 - 13.52 = 4.95 \pm 2.74$, 
which is consistent with the preliminary result at 200~GeV Au+Au (mid-rapidity) 
reported by the BRAHMS collaboration~\cite{BRAHMS_200_netproton}. 

%%%%%%%%%%%%%%%%%%%%%%
% 4.4 Particle Ratio %
%%%%%%%%%%%%%%%%%%%%%%
\subsection{Particle Ratios}
\label{sec:ratio}
The ratios of $\pi^{-}/\pi^{+}$, $K^{-}/K^{+}$, $p/\pbar$, $K/\pi$, 
$p/\pi$ and $\pbar/\pi$ measured as a function of $\pt$ and centrality
at $\snn=$200~GeV in Au+Au collisions are presented here. 

\subsubsection{Particle Ratios versus $\pt$}
\label{sec:ratio_pt}
\begin{figure}[t]
\includegraphics[width=1.0\linewidth]{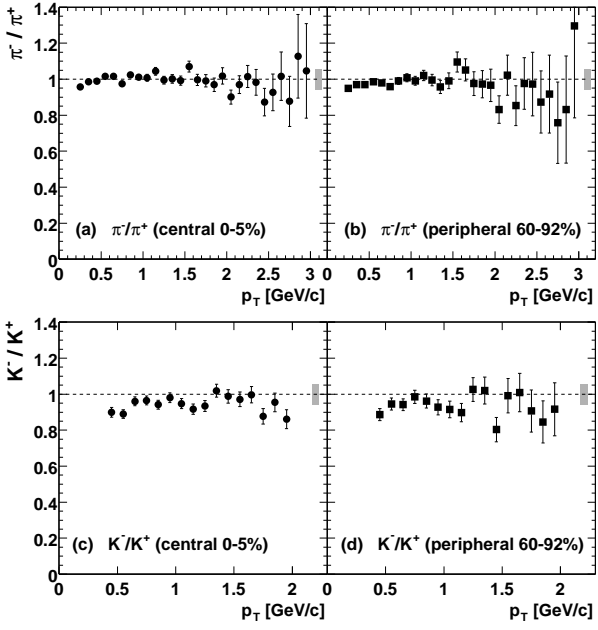}
\caption{Particle ratios of (a) $\pi^{-}/\pi^{+}$ for central 0--5\%, 
(b) $\pi^{-}/\pi^{+}$ for peripheral 60--92\%, (c) $K^{-}/K^{+}$ 
for central 0--5\%, and (d) $K^{-}/K^{+}$ for peripheral 60--92\%
in Au+Au collisions at $\sqrt{s_{NN}}$~=~200~GeV. The error bars indicate 
the statistical errors and shaded boxes around unity on each panel indicate 
the systematic errors.}
\label{fig:ratio_pi_and_k}
\end{figure}

\begin{figure}[t]
\includegraphics[width=1.0\linewidth]{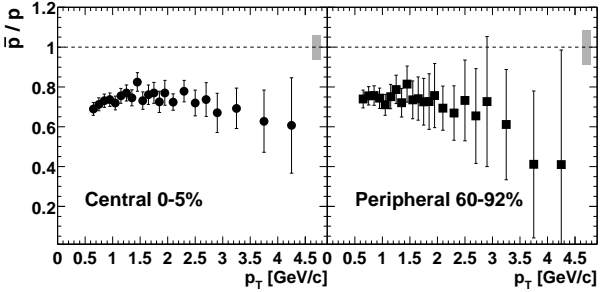}
\caption{Ratio of $\pbar/p$ as a function of $\pt$ for central
0--5\% (left) and peripheral 60--92\% (right) in Au+Au collisions at 
$\sqrt{s_{NN}}$~=~200~GeV. The error bars indicate the statistical 
errors and shaded boxes around unity on each panel indicate the systematic 
errors.}
\label{fig:ratio_pbarp}
\end{figure}

%%%%%%%%%%%%%%%%%%%%%%%%%%%%%%%%%
% pi-/pi+, K+/K-, pbar/p vs. pT %
%%%%%%%%%%%%%%%%%%%%%%%%%%%%%%%%%
Figure~\ref{fig:ratio_pi_and_k} shows the particle ratios of (a) 
$\pi^{-}/\pi^{+}$ for central 0--5\%, (b) $\pi^{-}/\pi^{+}$ for
peripheral 60--92\%, (c) $K^{-}/K^{+}$  for central 0--5\%, and 
(d) $K^{-}/K^{+}$  for peripheral 60--92\%. Similar plots for the 
$\pbar/p$ ratios are shown in Figure~\ref{fig:ratio_pbarp}. 
The error bars represent statistical errors and the shaded boxes on each 
panel represent the systematic errors.
For each of these particle species and centralities, the particle ratios 
are constant within the experimental errors over the measured $\pt$ range. 
Similar centrality and $\pt$ dependences are observed in 130~GeV Au+Au 
data~\cite{PPG009,STAR_pbarp_130,STAR_pbarp_erratum,STAR_kaon,STAR_strange,BRAHMS_pbarp_130,PHOBOS_ratio_130} 
and previously published 200~GeV Au+Au data~\cite{BRAHMS_ratio_200,PHOBOS_ratio_200}. 

\begin{figure}[ht]
\includegraphics[width=1.0\linewidth]{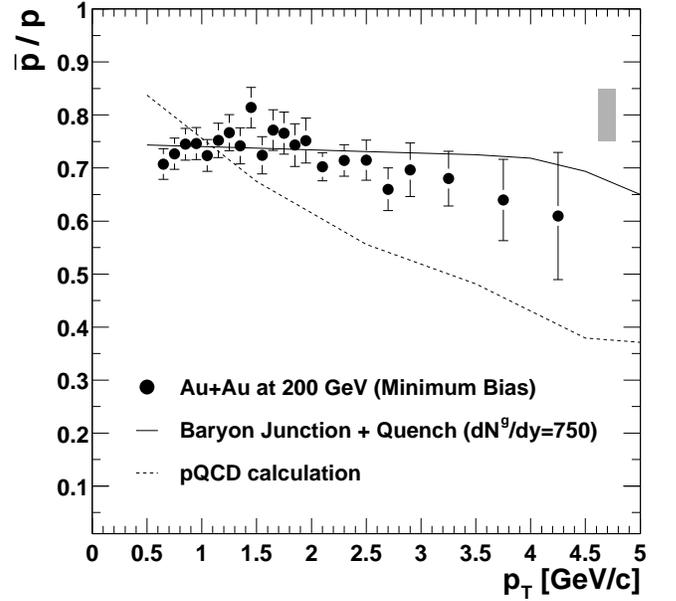}
\caption{$\pbar/p$ ratios as a function of $\pt$ for 
minimum bias events in  Au+Au at $\sqrt{s_{NN}}$~=~200~GeV. 
The error bars indicate the statistical errors and shaded 
box on the right indicates the systematic errors.
Two theoretical calculations are shown: baryon junction
model (solid line) and pQCD calculation (dashed line) taken
from reference~\cite{bj_pbarp}.}
\label{fig:ratio_pbarp_mb}
\end{figure}

To investigate the $\pt$ dependence of the $\pbar/p$ ratio in detail, 
it is shown in Figure~\ref{fig:ratio_pbarp_mb} for minimum bias events with 
two theoretical calculations: a pQCD calculation (dashed line), and a baryon 
junction model with jet-quenching~\cite{bj_pbarp} (solid line). The baryon junction 
calculation agrees well with the measured $\pbar/p$ ratio over the measured $\pt$ 
range within the experimental uncertainties, while the pQCD calculation does not 
explain the constant $\pbar/p$ ratio over the wide $\pt$ range. The statistical thermal 
model (discussed in more detail later in this section) predicted~\cite{thermal_1} 
a baryon chemical potential of $\mu_{B} = 29$ MeV and a freeze-out temperature of 
$T_{ch} = 177$ MeV for central Au+Au collisions at 200~GeV. From these, the 
expected $\pbar/p$ ratio is $e^{-2\mu_{B}/T_{ch}} = 0.72$, which agrees with our 
data (0.73). The parton recombination model~\cite{recombination} also reproduces 
the $\pbar/p$ ratio and its flat $\pt$ dependence. The $\pbar/p$ ratio in this 
model is 0.72 since the statistical thermal model is used.

\begin{figure}[t]
\includegraphics[width=1.0\linewidth]{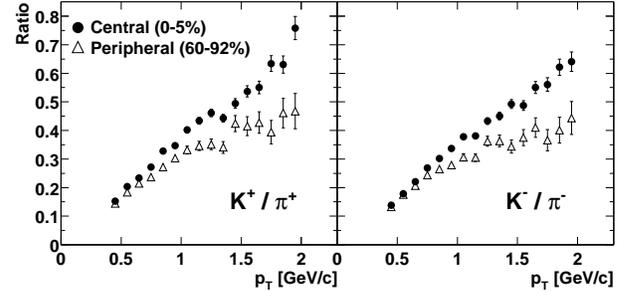}
\caption{$K/\pi$ ratios as a function of $\pt$ for central
0--5\% and peripheral 60--92\% in Au+Au collisions at 
$\snn$~=~200~GeV. The left is for $K^{+}/\pi^{+}$ and the right is for 
$K^{-}/\pi^{-}$. The error bars indicate the statistical errors.}
\label{fig:kpi_ratio}
\end{figure}

%%%%%%%%%%%%%%%
% K/pi vs. pT %
%%%%%%%%%%%%%%%
In Figure~\ref{fig:kpi_ratio}, the $\pt$ dependence of the $K/\pi$ ratio 
is shown for the most central 0--5\% and the most peripheral 60--92\% 
centrality bins. The $K^{+}/\pi^{+}$ ($K^{-}/\pi^{-}$) ratios are shown 
on the left (right). Both ratios increase with $\pt$ and the increase is
faster in central collisions than in peripheral ones. 

%%%%%%%%%%%%%%%%%%%%%%%%%%%%%%%%%%%%%%%% Figure 14.
%%%%%%%%%%%%%%%%%%%%%%%%%%%%%%%%%%%%%%%% Figure 15.
%%%%%%%%%%%%%%%%%%%%%%%%%%%%%%%%%%%%%%%% Figure 16.

\begin{figure}[ht]
\includegraphics[width=1.0\linewidth]{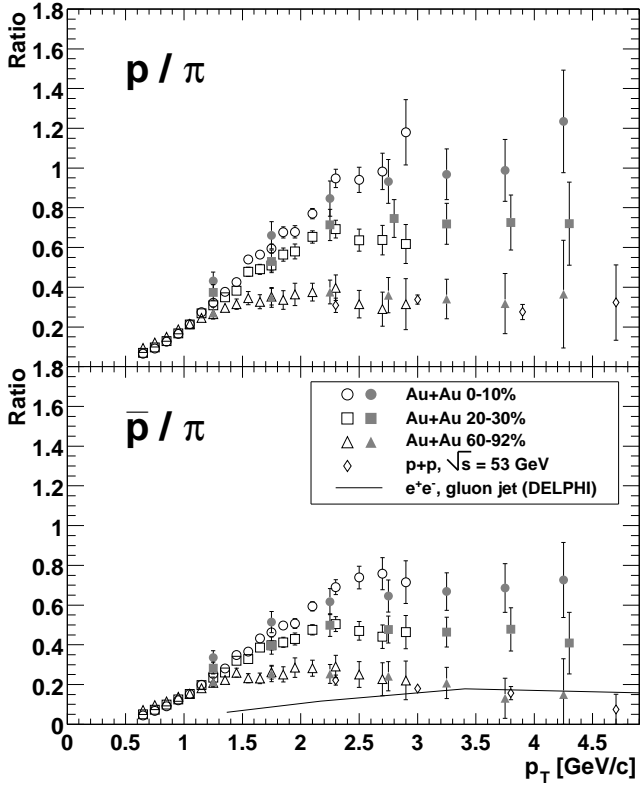}
\caption{Proton/pion (top) and anti-proton/pion (bottom) ratios for 
central 0--10\%, mid-central 20--30\% and peripheral 60--92\% in 
Au+Au collisions at $\snn$~=~200~GeV. Open (filled) points are for 
charged (neutral) pions. The data at $\sqrt{s} = 53 $~GeV $\pp$ 
collisions~\cite{ISR} are also shown. The solid line is the 
$(\pbar + p)/(\pi^{+} + \pi^{-})$ ratio measured in gluon jets~\cite{DELPHI}.}
\label{fig:ppi_ratio}
\end{figure}

%%%%%%%%%%%%%%
% p/pi vs pT %
%%%%%%%%%%%%%%
In Figure~\ref{fig:ppi_ratio}, the $p/\pi$ and $\pbar/\pi$ ratios are shown as 
a function of $\pt$ for the 0--10\%, 20--30\% and 60--92\% centrality bins. 
In this figure, the results of $p/\piz$ and $\pbar/\piz$~\cite{PPG014} are 
presented above 1.5~GeV/$c$ and overlaid on the results 
of $p/\pi^{+}$ and $\pbar/\pi^{-}$, respectively. The
absolutely normalized $\pt$ spectra of charged and neutral pions 
agree within 5--15\%. The error bars on the PHENIX data points in the 
figure show the quadratic sum of the statistical errors and the point-to-point 
systematic errors. There is an additional normalization uncertainty of 8\%
for $p/\pi^{+}$, $\pbar/\pi^{-}$ and 12\% for $p/\piz$, $\pbar/\piz$  
(the quadratic sum of the systematic errors on $p$ (or $\pbar$) normalization
and $\pt$ independent systematic errors from $\piz$~\cite{PPG015}), 
which may shift the data up or down for all three centrality bins together, 
but does not affect their shape.  The ratios increase rapidly at low $\pt$, 
but saturate at different values of $\pt$ which increase from peripheral 
to central collisions. In central collisions, the yields of both protons 
and anti-protons are comparable to that of pions for $\pt > 2$~GeV/$c$. 
For comparison, the corresponding ratios for $\pt >$ 2~GeV/$c$ observed in 
$\pp$ collisions at lower energies~\cite{ISR}, and in gluon jets produced in 
$e^{+}+e^{-}$ collisions~\cite{DELPHI}, are also shown. 
Within the uncertainties those ratios are compatible with the 
peripheral Au+Au results. In hard-scattering processes described by pQCD, 
the $p/\pi$ and $\pbar/\pi$ ratios at high $\pt$ are determined by the 
fragmentation of energetic partons, independent of the initial colliding 
system, which is seen as agreement between $p+p$ and $e^{+}+e^{-}$ collisions. 
Thus, the clear increase in the $p/\pi$ ($\pbar/\pi$) ratios at high $\pt$ 
from $p+p$ and peripheral to the mid-central and to the central Au+Au collisions 
requires ingredients other than pQCD. 

%% [p/pi theory] %%
The first observation of the enhancement of protons and anti-protons
compared to pions in the intermediate $\pt$ region was in the
130~GeV Au+Au data~\cite{PPG006}. The data inspired several new 
theoretical interpretations and models. Hydrodynamics calculations~\cite{Heinz} 
predict that the $\pbar/\pi$ ratio at high $\pt$ exceeds unity for central collisions.
The expected $\pbar/\pi$ ratio in the thermal model at fixed and sufficiently
large $\pt$ is determined by $2 e^{-\mu_{B}/T_{ch}}$ $\approx$ 1.7 using $T_{ch}= 177$ 
MeV and $\mu_{B}=29$ MeV~\cite{thermal_1} for 200~GeV Au+Au central collisions. 
Due to the strong radial flow effect at RHIC at relativistic transverse 
momenta ($\pt \gg m$), all hadron spectra have a similar shape.
The hydrodynamic model thus explains the excess of $\pbar/\pi$ in central collisions at 
intermediate $\pt$. However, the hydrodynamic model~\cite{Cleymans} predicts 
no or very little dependence on the centrality, which clearly disagrees with 
the present data. This model predicts, within 10\%, the same $\pt$ dependence 
of $p/\pi$ ($\pbar/\pi$) for all centrality bins.

Recently two new models have been proposed to explain 
the experimental results on the $\pt$ dependence of $p/\pi$ and $\pbar/\pi$ 
ratios. One model is the parton recombination and fragmentation 
model~\cite{recombination} and the other model is the baryon 
junction model~\cite{bjunctions}. Both models explain qualitatively 
the observed feature of $p/\pi$ enhancement in central collisions, 
and their centrality dependencies. Furthermore, both theoretical models 
predict that this baryon enhancement is limited to $\pt < $ 5 -- 6~GeV/$c$. 
This will be discussed in Section~\ref{sec:ncoll} in detail. 

%%%%%%%%%%%%%%%%%%%%%%%%%%%%%%%%%%%%%%%% Figure 17.
%%%%%%%%%%%%%%%%%%%%%%%%%%%%%%%%%%%%%%%% Figure 18.
%%%%%%%%%%%%%%%%%%%%%%%%%%%%%%%%%%%%%%%% Figure 19.
\begin{figure}[ht]
\includegraphics[width=1.0\linewidth]{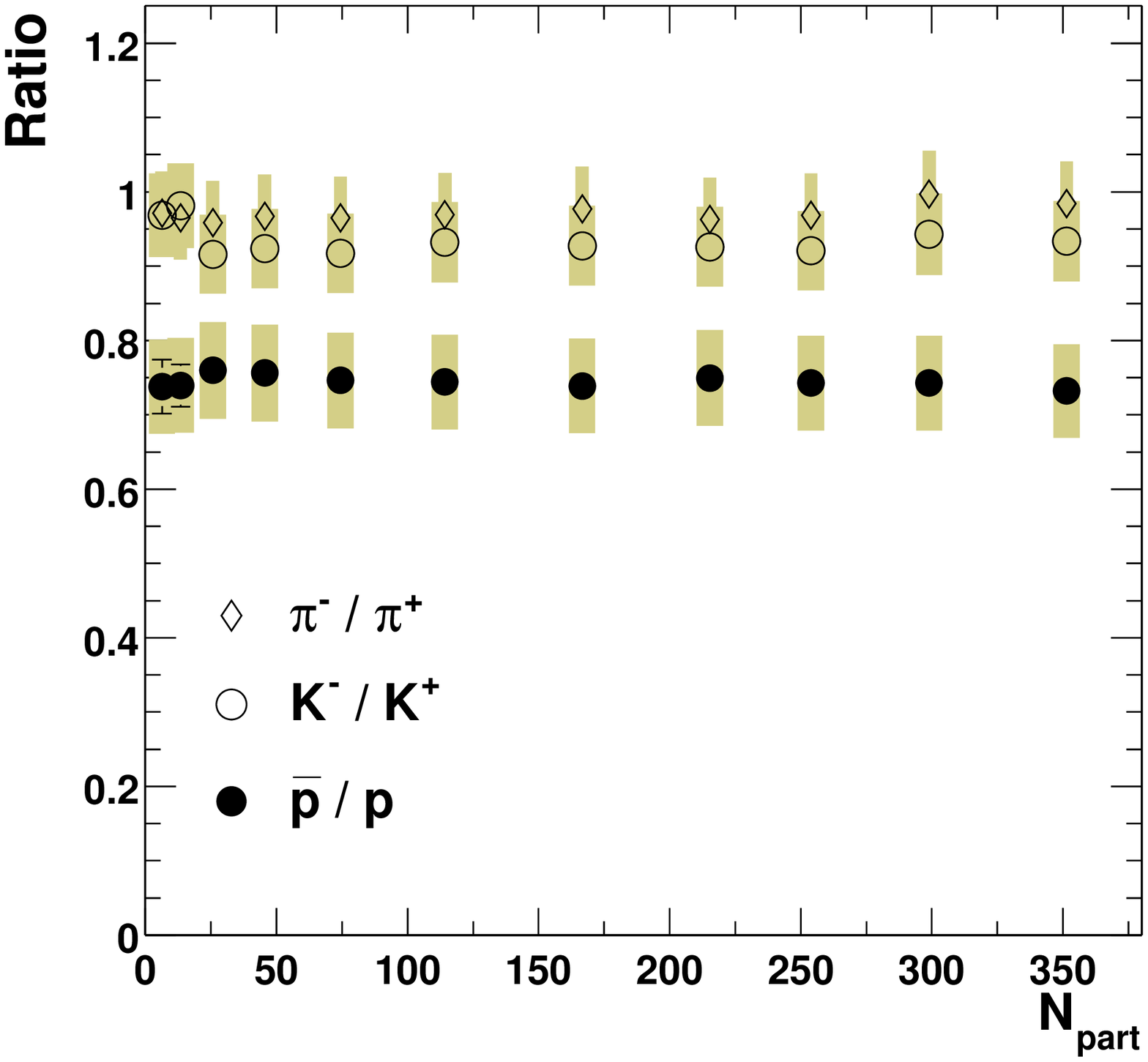}
\caption{Centrality dependence of particle ratios for 
$\pi^{-}/\pi^{+}$, $K^{-}/K^{+}$, and $\pbar/p$ in Au+Au 
collisions at $\sqrt{s_{NN}}$~=~200~GeV. The error bars indicate
the statistical errors. The shaded boxes on each data point are 
the systematic errors.}
\label{fig:ratio_Npart}
\end{figure}   

%%%%%%%%%%%%%%%%%%%%%%%%%%%%%%%%%%%%%%%% Figure 20.
\begin{figure}[ht]
\includegraphics[width=1.0\linewidth]{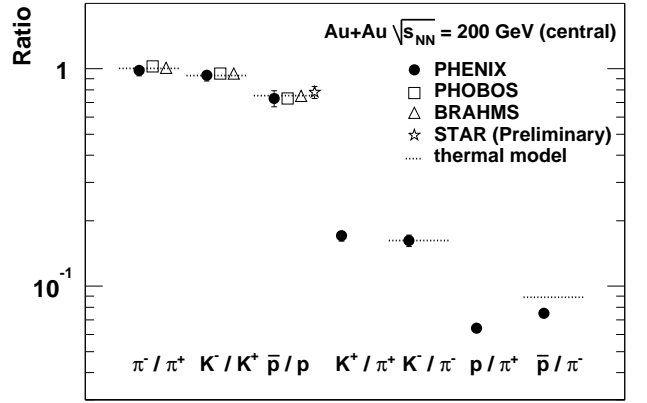}
\caption{Comparison of PHENIX particle ratios with those of 
PHOBOS~\cite{PHOBOS_ratio_200}, BRAHMS~\cite{BRAHMS_ratio_200}, 
and STAR (preliminary)~\cite{STAR_200_QM02} results in Au+Au 
central collisions at $\sqrt{s_{NN}}$~=~200~GeV at mid-rapidity. 
The thermal model prediction~\cite{thermal_1} for 200~GeV Au+Au 
central collisions are also shown as dotted lines. The error
bars on data indicate the systematic errors.}
\label{fig:ratio_all_exp}
\end{figure} 

%%%%%%%%%%%%%%%%%%%%%%%%%%%%%%%%%%%%%%%% Figure 21.
\begin{figure}[ht]
\includegraphics[width=1.0\linewidth]{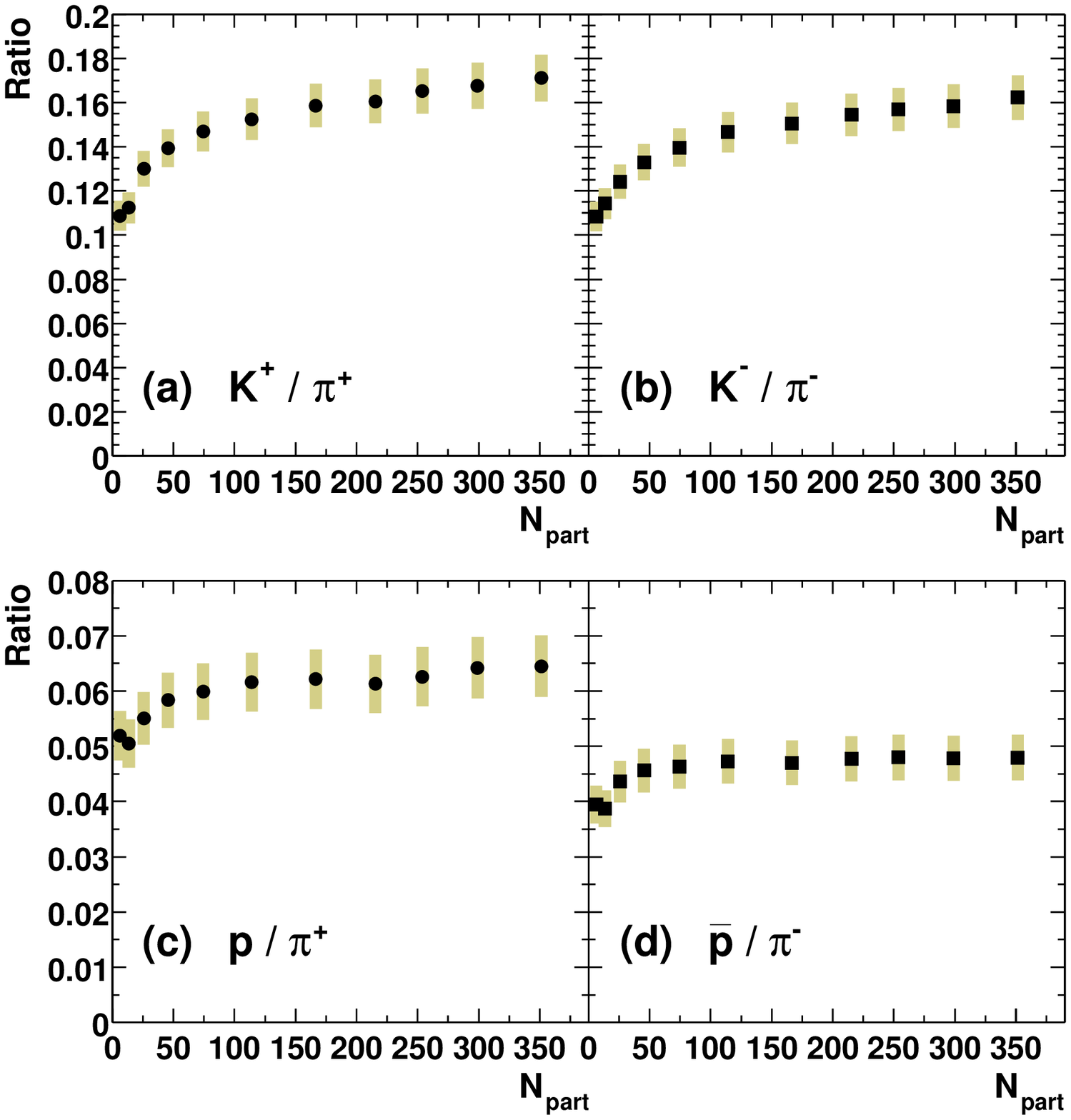}
\caption{Centrality dependence of particle ratios for 
(a) $K^{+}/\pi^{+}$, (b) $K^{-}/\pi^{-}$,
(c) $p/\pi^{+}$, and (d) $\pbar/\pi^{-}$ 
in Au+Au collisions at $\sqrt{s_{NN}}$~=~200~GeV.
The error bars indicate the statistical errors. The shaded 
boxes on each data point are the systematic errors.}
\label{fig:ratio_kpi_ppi_Npart}
\end{figure} 
 
%%%%%%%%%%%%%%%%%%%%%%%%%%%%%%%%%%%%%%%% Figure 22.
\begin{figure*}
\includegraphics[width=1.0\linewidth]{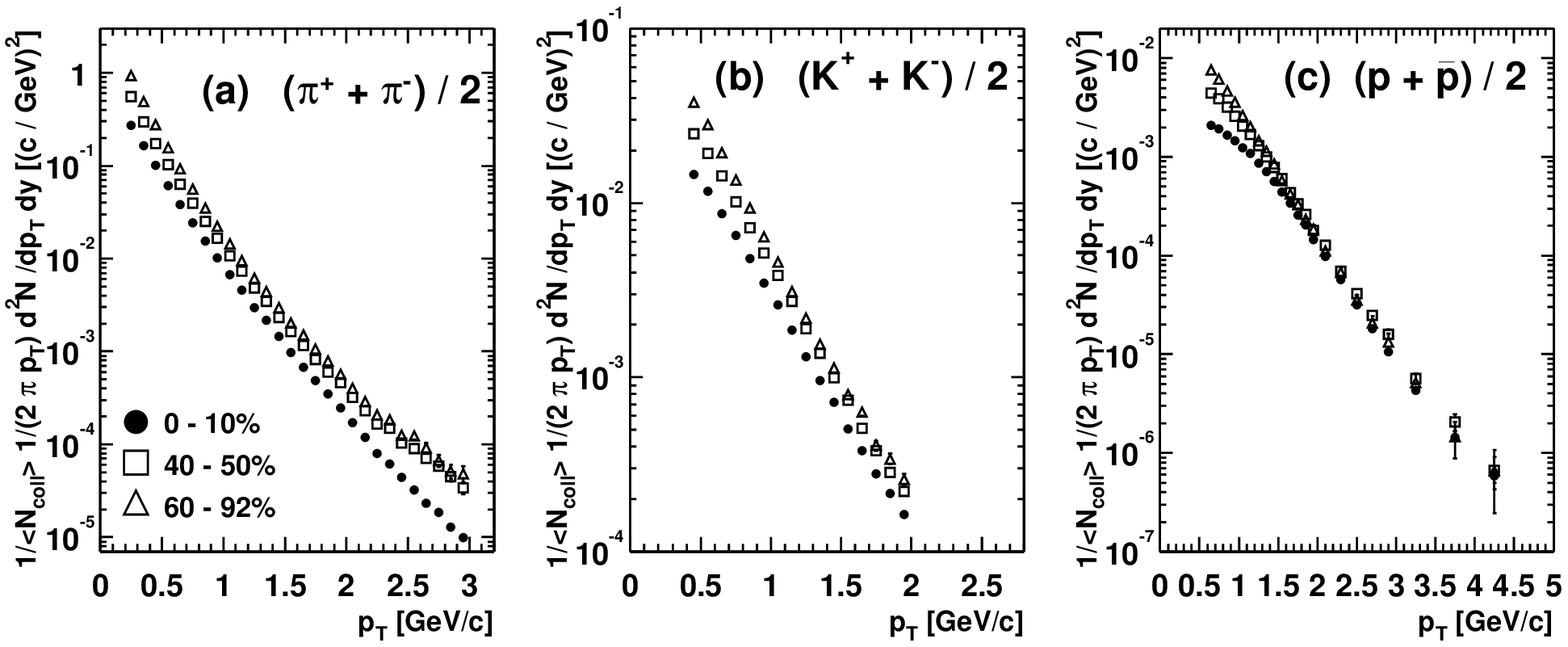}
\caption{$\pt$ spectra scaled by the averaged number
of binary collisions for averaged charged (a) pions, (b) kaons and 
(c) $(p + \pbar)/2$ in three different centrality bins: 
central 0--10 \%, mid-central 40--50\% and peripheral 
60--92 \% in Au+Au collisions at $\sqrt{s_{NN}}$~=~200~GeV. 
The error bars are statistical only. Note the
different horizontal and vertical scales on the three plots.}
\label{fig:pt_Ncoll_all}
\end{figure*}

%%%%%%%%%%%%%%%%%%%%%%%%%%%%%%%%%%%%%
% Ratio vs. Npart AND thermal model %
%%%%%%%%%%%%%%%%%%%%%%%%%%%%%%%%%%%%%
\subsubsection{Particle Ratio versus $\npart$}
\label{sec:ratio_npart}

Figure~\ref{fig:ratio_Npart} shows the centrality dependence of 
particle ratios for $\pi^{-}/\pi^{+}$, $K^{-}/K^{+}$ 
and $\pbar/p$. The ratios presented here are derived from the 
integrated yields over $\pt$ (i.e. $dN/dy$). The shaded boxes on 
each data point indicate the systematic errors. Within uncertainties, 
the ratios are all independent of $\npart$ over the measured range.
% The integrated ratios over the measured $\pt$ range in 
% the most central 0--5\% are:
% 0.98 $\pm$ 0.01 (stat.) $\pm$ 0.05 (syst.) for $\pi^{-}/\pi^{+}$, 
% 0.93 $\pm$ 0.01 (stat.) $\pm$ 0.05 (syst.) for $K^{-}/K^{+}$, and
% 0.73 $\pm$ 0.01 (stat.) $\pm$ 0.06 (syst.) for $\pbar/p$.
Figure~\ref{fig:ratio_all_exp} shows a comparison of the PHENIX 
particle ratios with those from PHOBOS~\cite{PHOBOS_ratio_200}, 
BRAHMS~\cite{BRAHMS_ratio_200}, and STAR (preliminary)~\cite{STAR_200_QM02} 
in Au+Au central collisions at $\sqrt{s_{NN}}$~=~200~GeV at mid-rapidity.
The PHENIX anti-particle to particle ratios are consistent with 
other experimental results within the systematic uncertainties.  

Figure~\ref{fig:ratio_kpi_ppi_Npart} shows the centrality dependence of 
$K/\pi$ and $p/\pi$ ratios. Both $K^{+}/\pi^{+}$ and $K^{-}/\pi^{-}$ ratios 
increase rapidly for peripheral collisions ($\npart < $ 100), and then 
saturate or rise slowly from the mid-central to the most central collisions. 
% For 0--5\% central collisions, we obtain 
% $K^{+}/\pi^{+} = $ 0.17 $\pm$ 0.01 (stat.) $\pm$ 0.01 (syst.) and 
% $K^{-}/\pi^{-} = $ 0.16 $\pm$ 0.01 (stat.) $\pm$ 0.01 (syst.).  
The $p/\pi^{+}$ and $\pbar/\pi^{-}$ ratios increase for
peripheral collisions ($\npart < $ 50) and saturate from
mid-central to central collisions --  similar to the centrality
dependence of $K/\pi$ ratio (but possibly flatter). 
% At 0--5\% central, we obtain:
% $p/\pi^{+} = $0.064 $\pm$ 0.001 (stat.) $\pm$ 0.003 (syst.) and 
% $\pbar/\pi^{-} = $ 0.048 $\pm$ 0.001 (stat.) $\pm$ 0.003 (syst.). 

%%% [thermal model] %%%
Within the framework of the statistical thermal model~\cite{thermal_becattini} 
in a grand canonical ensemble with baryon number, strangeness and charge 
conservation~\cite{thermal_1}, particle ratios measured at $\snn = $130~GeV at 
mid-rapidity have been analyzed with the extracted chemical freeze-out 
temperature $T_{ch} = 174 \pm 7$ MeV and baryon chemical potential 
$\mu_{B} = 46 \pm 5$ MeV.
A set of chemical parameters at $\snn = $200~GeV in Au+Au were also predicted
by using a phenomenological parameterization 
of the energy dependence of $\mu_{B}$. The predictions were $\mu_{B} = $ 29 
$\pm$ 8 MeV and  $T_{ch} = 177 \pm 7$ MeV at $\snn = $200~GeV. The comparison 
between the PHENIX data at 200~GeV for 0--5\% central and the thermal 
model prediction is shown in Table~\ref{tab:thermal} and Figure~\ref{fig:ratio_all_exp}. 
There is a good agreement between data and the model. The thermal model calculation 
was performed by assuming a 50\% reconstruction efficiency of all weakly decaying 
baryons in reference~\cite{thermal_1}. However, our results have been corrected to remove
these contributions. Therefore, Table~\ref{tab:thermal} includes 
$\pbar/p$ and $\pbar/\pi^{-}$ ratios with and without $\Lambda$ ($\lbar$) feed-down 
corrections to the proton and anti-proton spectra. The ratios without the
$\Lambda$ ($\lbar$) feed-down correction are labeled ``inclusive''. 
The small $\mu_{B}$ is qualitatively consistent with our measurement of the number of net 
protons ($\approx 5$) in central Au+Au collisions at $\snn~=~200$~GeV at mid-rapidity. 

%%%%%%%%%%%%%%%%%%%%%%%%
% Particle Ratio Table %
%%%%%%%%%%%%%%%%%%%%%%%%
\begin{table}
\caption{Comparison between the data for the 0--5\% central collisions and the thermal model 
prediction at $\snn$~=~200~GeV with $T_{ch}$~=~177~MeV and $\mu_{B}$~=~29~MeV~\cite{thermal_1}.}
\begin{ruledtabular}\begin{tabular}{lcc}
Particles  &  Ratio $\pm$ stat. $\pm$ sys. & Thermal Model \\ \hline 
$\pi^{-}/\pi^{+}$           &  0.984 $\pm$ 0.004 $\pm$ 0.057     & 1.004   \\
$K^{-}/K^{+}$               &  0.933 $\pm$ 0.007 $\pm$ 0.054     & 0.932   \\
$\pbar/p$                   &  0.731 $\pm$ 0.011 $\pm$ 0.062     &         \\
$\pbar/p$ (inclusive)       &  0.747 $\pm$ 0.007 $\pm$ 0.046     & 0.752   \\
$K^{+}/\pi^{+}$             &  0.171 $\pm$ 0.001 $\pm$ 0.010     &         \\
$K^{-}/\pi^{-}$             &  0.162 $\pm$ 0.001 $\pm$ 0.010     & 0.147   \\
$p/\pi^{+}$                 &  0.064 $\pm$ 0.001 $\pm$ 0.003     &         \\
$p/\pi^{+}$ (inclusive)     &  0.099 $\pm$ 0.001 $\pm$ 0.006     &         \\
$\pbar/\pi^{-}$             &  0.047 $\pm$ 0.001 $\pm$ 0.002     &         \\ 
$\pbar/\pi^{-}$ (inclusive) &  0.075 $\pm$ 0.001 $\pm$ 0.004     & 0.089   \\ 
\end{tabular}\end{ruledtabular}
\label{tab:thermal}
\end{table}

%%%%%%%%%%%%%%%%%%%%%
% 4.5 Ncoll scaling %
%%%%%%%%%%%%%%%%%%%%%
\subsection{Binary Collision Scaling of $\pt$ Spectra}
\label{sec:ncoll}
One of the most striking features in Au+Au collisions
at RHIC is that $\piz$ and non-identified hadron yields at $p_T >$ 
2~GeV/$c$ in central collisions are suppressed with respect 
to the number of nucleon-nucleon binary collisions ($\ncoll$) 
scaled by $p+p$ and peripheral Au+Au results~\cite{PPG003,PPG013,
PPG014}. Moreover, the suppression of $\piz$ is stronger than 
than that for non-identified charged hadrons~\cite{PPG003},
and the yields of protons and anti-protons in central collisions
are comparable to that of pions around $\pt =$2~GeV/$c$~\cite{PPG006}.
The enhancement of the
$p/\pi$ ($\pbar/\pi$) ratio in central collisions at intermediate 
$\pt$ (2.0 -- 4.5~GeV/$c$), which was presented in the previous section,
is consistent with the above observations.
These results show the significant contributions of proton and anti-proton 
yields to the total particle composition at this intermediate $\pt$ region. 
We present here the $\ncoll$ scaling behavior for charged pions, kaons, and 
protons (anti-protons) in order to quantify the particle composition at 
intermediate $\pt$. 

%%%%%%%%%%%%%%%%%%%%%%%%%%%%%%%%%
% 4.5.1 Ncoll scaling (spectra) %
%%%%%%%%%%%%%%%%%%%%%%%%%%%%%%%%%
Figure~\ref{fig:pt_Ncoll_all} shows the $\pt$ spectra scaled by the 
averaged number of binary collisions, $\ncollav$, for 
$(\pi^{+} + \pi^{-})/2$, $(K^{+} + K^{-})/2$, and $(p + \pbar)/2$ 
in three centrality bins: central 0--10\%, mid-central 40--50\% 
and peripheral 60--92\%. For $(p + \pbar)/2$ 
in the range of $\pt=$ 1.5 -- 4.5~GeV/$c$, it is clearly seen that 
the spectra are on top of each other. This indicates that proton and
anti-proton production at high $\pt$ scales with the number of binary
collisions. On the other hand, at $\pt$ below 1.5~GeV/$c$, different 
shapes for different centrality bins are observed, which indicates a 
strong contribution from radial flow. The scaling behavior of the kaons seems
to be similar to protons, but this is not conclusive due to our PID
limitations. For pions, the $\ncoll$ scaled yield in central events 
is suppressed compared to that for peripheral events at $\pt > $ 2 
GeV/$c$, which is consistent with the results in the $\piz$ 
spectra~\cite{PPG003,PPG014}. 

\begin{figure}[t]
\includegraphics[width=1.0\linewidth]{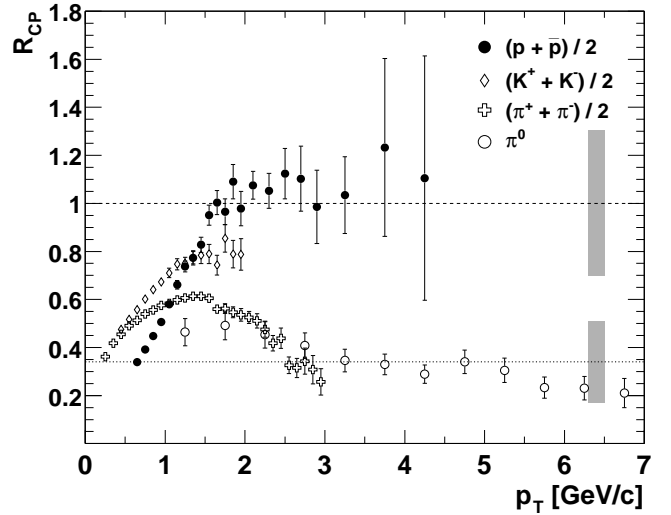}
\caption{Central (0--10\%) to peripheral (60--92\%) ratios of 
binary-collision-scaled $\pt$ spectra, $R_{CP}$, as a function 
of $\pt$ for $(\pbar + p)/2$, charged kaons, charged pions, 
and $\piz$~\cite{PPG014} in Au+Au collisions at $\sqrt{s_{NN}}$ 
= 200~GeV. The lines indicate the expectations of $\npart$ (dotted) 
and $\ncoll$ (dashed) scaling, the shaded bars represent the systematic
errors on these quantities.}
\label{fig:raa_all_60-92}
\end{figure}

%%%%%%%%%%%%%%
% 4.5.2 R_CP %
%%%%%%%%%%%%%%
Figure~\ref{fig:raa_all_60-92} shows the central (0--10\%) to peripheral 
(60--92\%) ratio for $\ncoll$ scaled $\pt$ spectra ($R_{\rm CP}$: 
the nuclear modification factor) of $(\pbar + p)/2$, kaons, charged pions, 
and $\piz$. In this paper we define $R_{\rm CP}$ as: 

\begin{equation}
 R_{\rm CP} =\frac{{\rm Yield^{ 0-10\%}}/\langle \ncoll^{0-10\%} \rangle}
                  {{\rm Yield^{ 60-92\%}}/\langle \ncoll^{60-92\%} \rangle}.
\label{eq:rcp}
\end{equation}
The peripheral 60--92\% Au+Au spectrum is used as an approximation of 
the yields in $\pp$ collisions, based on the experimental fact that 
the peripheral spectra scale with $\ncoll$ by using the yields in $\pp$ 
collisions measured by PHENIX~\cite{PPG014, PPG024}. Thus the meaning
of the $R_{\rm CP}$ is expected to be the same as $R_{\rm AA}$ used in our previous 
publications~\cite{PPG003,PPG013,PPG014}. The lines in 
Figure~\ref{fig:raa_all_60-92} indicate the expectations of $\npart$ 
(dotted) and $\ncoll$ (dashed) scaling. The shaded bars at the end of 
each line represent the systematic error associated with 
the determination of these quantities for central and peripheral events.
The error bars on charged particles are statistical errors only, and those
for $\piz$ are the quadratic sum of the statistical errors and 
the point-to-point systematic errors.  The data show that $(\pbar + p)/2$ 
reaches unity for $\pt \simge$ 1.5~GeV/$c$, consistent with
$\ncoll$ scaling. The data for kaons also show the $\ncoll$ scaling 
behavior around 1.5 -- 2.0~GeV/$c$, but the behavior is weaker than for protons. 
As with neutral pions~\cite{PPG014}, charged pions are also suppressed at 
2 -- 3~GeV/$c$ with respect to peripheral Au+Au collisions. 

\begin{figure}[t]
\includegraphics[width=1.0\linewidth]{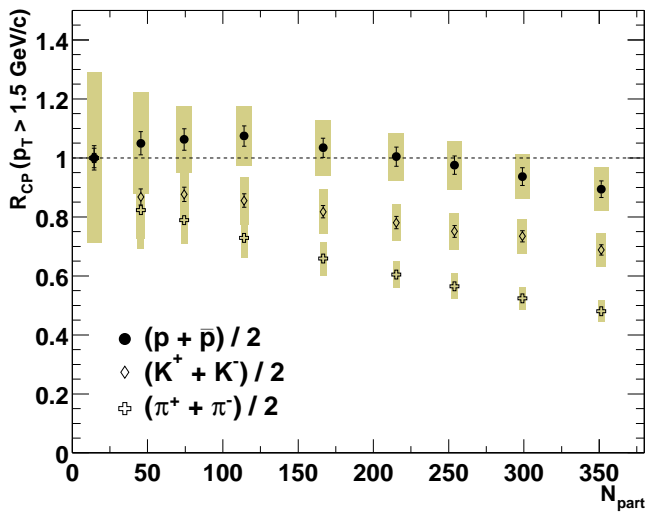}
\caption{Centrality dependence of integrated $R_{CP}$ above 1.5~GeV/$c$
normalized to the most peripheral 60--92\% value.  The data
shows $R_{CP}$ for $(\pbar + p)/2$, charged kaons, and charged pions 
in Au+Au collisions at $\sqrt{s_{NN}}$~=~200~GeV. 
The error bars are statistical only. The shaded boxes represent
the systematic errors (see text for details).}
\label{fig:raa_integ_60-92}
\end{figure}

%%%%%%%%%%%%%%%%%%%%%%%%%
% 4.5.3 Integrated R_CP %
%%%%%%%%%%%%%%%%%%%%%%%%%
Motivated by the observation that the $(\pbar + p)/2$ spectra
scale with $N_{coll}$ above $\pt = $1.5~GeV/$c$, the ratio of 
the integrated yield between central and peripheral events (scaled by 
the corresponding $\ncoll$) above $\pt = $1.5~GeV/$c$ are shown in 
Figure~\ref{fig:raa_integ_60-92} as a function of $\npart$.
The $\pt$ ranges for the integration are, 1.5 -- 4.5~GeV/$c$ 
for $(\pbar + p)/2$, 1.5 -- 2.0~GeV/$c$ for kaons, and 1.5 -- 3.0~GeV/$c$ 
for charged pions. The data points are normalized to 
the most peripheral data point. The shaded boxes in the figure indicate the 
systematic errors, which include the normalization errors on the $\pt$ spectra, 
the errors on the detector occupancy corrections, and the 
uncertainties of the $\langle T_{\rm AuAu}\rangle$ determination for the numerator only. 
Only at the most peripheral data point, the uncertainty on the denominator
$\langle T^{60-92\%}_{\rm AuAu}\rangle$ is also added. The figure shows 
that $(\pbar + p)/2$ scales with $\ncoll$ for all centrality bins, 
while the data for charged pions show a decrease with $\npart$. 
The kaon data points are between the charged pions and the $(\pbar + p)/2$ spectra. 

%%%%%%%%%%%%%%%%%%%%%%%%%%%%%%%%%%%%%%%% Figure 23.
%%%%%%%%%%%%%%%%%%%%%%%%%%%%%%%%%%%%%%%% Figure 24.

%%%%%%%%%%%%%%%%%%%%%%%%%
% 4.5.4 Interpretations %
%%%%%%%%%%%%%%%%%%%%%%%%%
The standard picture of hadron production at high momentum is 
the fragmentation of energetic partons. While the observed 
suppression of the $\piz$ yield at high $\pt$ in central collisions
may be attributed to the energy loss of partons during their 
propagation through the hot and dense matter created in the collisions, i.e. 
jet quenching~\cite{quench_effect,quenching_theory}, it is a
theoretical challenge to explain the absence of suppression for 
baryons up to ~4.5~GeV/$c$ for all centralities along with the
enhancement of the $p/\pi$ ratio at 
$\pt = $ 2 -- 4~GeV/$c$ for central collisions. 

%% Parton Recombination %%
It has been recently proposed that such observations can 
be explained by the dominance of parton recombination
at intermediate $\pt$, rather than by fragmentation~\cite{recombination}.
The competition between recombination and fragmentation of partons
may explain the observed features. The model predicts that
the effect is limited to $\pt < 5$~GeV/$c$, beyond which 
fragmentation becomes the dominant production mechanism for all 
particle species.

%% Baryon Junction %%
Another possible explanation is the baryon junction model~\cite{bjunctions}. 
It invokes a topological gluon configuration with jet quenching. 
With pion production above 2~GeV/$c$ suppressed by jet
quenching, gluon junctions produce copious baryons at intermediate
$\pt$, thus lead to the enhancement of baryons in this $\pt$ region. 
The model reproduces the baryon-to-meson ratio and its 
centrality dependence qualitatively~\cite{ptopi_bjunctions}.

Both theoretical models predict that baryon enhancement is limited to 
$\pt < $ 5 -- 6~GeV/$c$, which is unfortunately beyond our current PID
capability. However, it is possible to test the two predictions indirectly 
by using the non-identified charged hadrons to neutral pion ratio ($h/\piz$) 
as a measure of the baryon content at high $\pt$, as published in~\cite{PPG015}. 
The results support the limited behavior of baryon enhancement up to 
5~GeV/$c$ in $\pt$. Similar trends are observed in $\Lambda$, $K^{0}_{S}$ and 
$K^{\pm}$ measurements by the STAR collaboration~\cite{STAR_k0}. 

%% Cronin Effect %%
On the other hand, it is also possible that nuclear effects, such as 
the ``Cronin effect''~\cite{cronin1975, antreasyan1979}, attributed 
to initial state multiple scattering ($\pt$-broadening)~\cite{pt_broadening},
contribute to the observed species dependence.  At center-of-mass
energies up to $\sqrt{s} = 38.8$~GeV, a nuclear enhancement beyond
$\ncoll$ scaling has been observed for $\pi, K, p$ and their
anti-particles in $p+A$ collisions. The effect is stronger for protons
and anti-protons than for pions which leads to an enhancement of the $p/\pi$
and $\pbar/\pi$ ratios compared to $\pp$ collisions.  In proton-tungsten reactions, 
the increase is a factor of $\sim 2$ in the range 3 $< \pt <$ 6~GeV. 
For pions, theoretical calculations at RHIC energies~\cite{Cronin_RHIC}
predict a reduced strength of the Cronin effect compared to lower energies,
although no prediction exists for protons.  New data from d+Au
collisions at $\snn$~=~200~GeV will help to clarify this issue.

%%%%%%%%%%%%%%%%%%%%%%%%%%%%%%
% (5) Summary and Conclusion %
%%%%%%%%%%%%%%%%%%%%%%%%%%%%%%
\section{SUMMARY AND CONCLUSION}
\label{sec:summary}
In summary, we present the centrality dependence of identified charged 
hadron spectra and yields for $\pi^{\pm}$, $K^{\pm}$, $p$ and $\pbar$ in 
Au+Au collisions at $\snn$~=~200~GeV at mid-rapidity. In central events,
the low $\pt$ region ($\le$ 2.0~GeV/$c$) of the $\pt$ spectra 
show a clear particle mass dependence in their shapes, namely, $p$ and $\pbar$ 
spectra have a shoulder-arm shape while the pion spectra have a concave 
shape. The spectra can be well fit with an exponential function in $\mt$
at the region below 1.0~GeV/$c^{2}$ in $\mt-m_{0}$. The resulting inverse slope parameters 
show clear particle mass and centrality dependences, that increase with 
particle mass and centrality. These observations are consistent with the 
hydrodynamic radial flow picture. Moreover, at around $\pt = $2.0~GeV/$c$ in 
central events, the $p$ and $\pbar$ yields are comparable to the pion 
yields. Here, baryons comprise a significant fraction of the hadron yield in this 
intermediate $\pt$ range.  The $\meanpt$ and $dN/dy$ per participant pair 
increase from peripheral to mid-central collisions and saturate for the 
most central collisions for all particle species. The net proton number 
in Au+Au central collisions at $\snn$~=~200~GeV is $\sim$ 5 at mid-rapidity. 

The particle ratios of $\pi^{-}/\pi^{+}$, $K^{-}/K^{+}$, 
$p/\pbar$, $K/\pi$, $p/\pi$ and $\pbar/\pi$ as a function of $\pt$ and 
centrality have been measured. Particle ratios in central Au+Au collisions are well 
reproduced by the statistical thermal model with a baryon chemical
potential of $\mu_{B}$ = 29 MeV and a chemical freeze-out temperature of 
$T_{ch}$ = 177 MeV. Regardless of the particle species and centrality, 
it is found that ratios for equal mass particles are constant as a function of 
$\pt$, within the systematic uncertainties in the measured $\pt$ range. 
On the other hand, both $K/\pi$ and $p/\pi$ ($\pbar/\pi$) ratios 
increase as a function of $\pt$. This increase with $\pt$ is stronger 
for central than for peripheral events. The $p/\pi$ and $\pbar/\pi$ ratios 
in central events both increase with $\pt$ up to 3~GeV/$c$ 
and approach unity at $\pt \approx$ 2~GeV/$c$. However, in peripheral 
collisions these ratios saturate at the value of 0.3 -- 0.4 around 
$\pt = $ 1.5~GeV/$c$. The observed centrality dependence of $p/\pi$ and 
$\pbar/p$ ratios in intermediate $\pt$ region is not explained 
by the hydrodynamic model alone, but both the parton recombination 
model and the baryon junction model qualitatively agree with data. 

The scaling behavior of identified charged hadrons 
is compared with results for neutral pions. In the $\ncoll$ 
scaled $\pt$ spectra for $(p + \pbar)/2$, the spectra scale with
$\ncoll$ from $\pt = $1.5 -- 4.5~GeV/$c$. The central-to-peripheral 
ratio, $R_{CP}$, approaches unity for $(\pbar + p)/2$ from $\pt =$ 
1.5 up to 4.5~GeV/$c$. Meanwhile, charged and neutral pions are
suppressed. The ratio of integrated $R_{CP}$ from $\pt = $1.5 to 4.5~GeV/$c$ 
exhibits an $\ncoll$ scaling behavior for all centrality bins in
the $(\pbar + p)/2$ data, which is in contrast to the stronger pion 
suppression, that increases with centrality. \\
\\

%%%%%%%%%%%%
% Appendix %
%%%%%%%%%%%%
\begin{appendix}
\section{Table of Invariant Yields}
\label{app}
The invariant yields for $\pi^{\pm}$, $K^{\pm}$, $p$ and $\pbar$ in 
Au+Au collisions at $\snn$~=~200~GeV at mid-rapidity are tabulated
in Tables~\ref{tab:pp_table1} -- ~\ref{tab:peri_table_pr}. 
The data presented here are for the the minimum bias events
and each centrality bin (0--5\%, 5--10\%, 10--15\%, 15--20\%, 
20--30\%, ..., 70--80\%, 80--92\%, and 60--92\%).  
Errors are statistical only. 
\end{appendix}

%%%%%%%%%%%%%%%%%%%%
% Acknowledgments %
%%%%%%%%%%%%%%%%%%%%
\begin{acknowledgments}
%\section{Acknowledgements}   % Run-2 long from for PRC, PLB, etc.
We thank the staff of the Collider-Accelerator and Physics
Departments at Brookhaven National Laboratory and the staff
of the other PHENIX participating institutions for their
vital contributions.  We acknowledge support from the
Department of Energy, Office of Science, Nuclear Physics
Division, the National Science Foundation, Abilene Christian
University Research Council, Research Foundation of SUNY,
and Dean of the College of Arts and Sciences, Vanderbilt
University (U.S.A), Ministry of Education, Culture, Sports,
Science, and Technology and the Japan Society for the
Promotion of Science (Japan), Conselho Nacional de
Desenvolvimento Cient\'{\i}fico e Tecnol{\'o}gico and
Funda\c c{\~a}o de Amparo {\`a} Pesquisa do Estado de
S{\~a}o Paulo (Brazil), Natural Science Foundation of China
(People's Republic of China), Centre National de la
Recherche Scientifique, Commissariat {\`a} l'{\'E}nergie
Atomique, Institut National de Physique Nucl{\'e}aire et
de Physique des Particules, and Institut National
de Physique Nucl{\'e}aire et de Physique des Particules,
(France), Bundesministerium fuer Bildung und
Forschung, Deutscher Akademischer Austausch Dienst, and
Alexander von Humboldt Stiftung (Germany), Hungarian
National Science Fund, OTKA (Hungary), Department of Atomic
Energy and Department of Science and Technology (India),
Israel Science Foundation (Israel), Korea Research
Foundation and Center for High Energy Physics (Korea),
Russian Ministry of Industry, Science and Tekhnologies,
Russian Academy of Science, Russian Ministry of Atomic Energy
(Russia), VR and the Wallenberg Foundation (Sweden), the U.S.
Civilian Research and Development Foundation for the Independent
States of the Former Soviet Union, the US-Hungarian 
NSF-OTKA-MTA,
the US-Israel Binational Science Foundation, and the 5th
European Union TMR Marie-Curie Programme.
\end{acknowledgments}

%%%%%%%%%%%%%%
% REFERENCES %
%%%%%%%%%%%%%%

%%%%%%%%%%%%%%%%%%
% SPECTRA TABLES %
%%%%%%%%%%%%%%%%%%
%\newpage   % This is needed only for the "multicols" style
%TABLES:  If you have a table or two, they should follow the figures:
%         We must use this format with no vertical or horizontal lines
%         in the body of the Table.
%
% 
\begingroup \squeezetable
\begin{table*}
\caption{Invariant yields for $\pi^{+}$ at mid-rapidity in the minimum bias, 
0--5\%, 5--10\%, and 10--15\% centrality bins, normalized to one unit rapidity.
Errors are statistical only.}
\begin{ruledtabular}
\begin{tabular}{ccccc}
$p_T$ [GeV/$c$] & Minimum bias &  0--5\% &  5--10\%  &  10--15\%  \\ \hline
0.25  &  1.07e+02  $\pm$  8.8e-01  &  3.29e+02  $\pm$  2.7e+00  &  2.76e+02  $\pm$  2.3e+00  &  2.39e+02  $\pm$  2.0e+00  \\  
0.35  &  6.06e+01  $\pm$  5.0e-01  &  1.97e+02  $\pm$  1.6e+00  &  1.64e+02  $\pm$  1.4e+00  &  1.39e+02  $\pm$  1.2e+00  \\  
0.45  &  3.63e+01  $\pm$  3.1e-01  &  1.20e+02  $\pm$  1.1e+00  &  9.93e+01  $\pm$  8.7e-01  &  8.41e+01  $\pm$  7.4e-01  \\  
0.55  &  2.18e+01  $\pm$  2.0e-01  &  7.26e+01  $\pm$  6.7e-01  &  6.02e+01  $\pm$  5.6e-01  &  5.08e+01  $\pm$  4.7e-01  \\  
0.65  &  1.34e+01  $\pm$  1.3e-01  &  4.49e+01  $\pm$  4.5e-01  &  3.74e+01  $\pm$  3.8e-01  &  3.16e+01  $\pm$  3.2e-01  \\  
0.75  &  8.71e+00  $\pm$  9.5e-02  &  2.93e+01  $\pm$  3.3e-01  &  2.43e+01  $\pm$  2.7e-01  &  2.05e+01  $\pm$  2.3e-01  \\  
0.85  &  5.41e+00  $\pm$  6.3e-02  &  1.82e+01  $\pm$  2.2e-01  &  1.53e+01  $\pm$  1.8e-01  &  1.29e+01  $\pm$  1.6e-01  \\
\end{tabular}\end{ruledtabular}
\label{tab:pp_table1}
\end{table*} \endgroup
%\cleardoublepage
 
\begingroup \squeezetable
\begin{table*}
\caption{Invariant yields for $\pi^{+}$ at mid-rapidity in 15--20\%, 
20--30\%, 30--40\%, and 40-50\% centrality bins, normalized to one unit rapidity.
Errors are statistical only.}
\begin{ruledtabular}\begin{tabular}{ccccc}
$p_T$ [GeV/$c$] & 15--20\%  &  20--30\% &  30--40\%  &  40--50\%  \\ \hline
0.25  &  2.04e+02  $\pm$  1.7e+00  &  1.57e+02  $\pm$  1.3e+00  &  1.07e+02  $\pm$  8.9e-01  &  6.84e+01  $\pm$  5.7e-01  \\  
0.35  &  1.18e+02  $\pm$  9.9e-01  &  8.82e+01  $\pm$  7.4e-01  &  5.86e+01  $\pm$  4.9e-01  &  3.67e+01  $\pm$  3.1e-01  \\  
0.45  &  7.09e+01  $\pm$  6.2e-01  &  5.27e+01  $\pm$  4.6e-01  &  3.46e+01  $\pm$  3.0e-01  &  2.15e+01  $\pm$  1.9e-01  \\  
0.55  &  4.28e+01  $\pm$  4.0e-01  &  3.17e+01  $\pm$  2.9e-01  &  2.06e+01  $\pm$  1.9e-01  &  1.26e+01  $\pm$  1.2e-01  \\  
0.65  &  2.65e+01  $\pm$  2.7e-01  &  1.95e+01  $\pm$  2.0e-01  &  1.26e+01  $\pm$  1.3e-01  &  7.66e+00  $\pm$  8.0e-02  \\  
0.75  &  1.73e+01  $\pm$  2.0e-01  &  1.27e+01  $\pm$  1.4e-01  &  8.29e+00  $\pm$  9.4e-02  &  4.99e+00  $\pm$  5.8e-02  \\  
0.85  &  1.07e+01  $\pm$  1.3e-01  &  7.94e+00  $\pm$  9.5e-02  &  5.10e+00  $\pm$  6.3e-02  &  3.04e+00  $\pm$  3.9e-02  \\
\end{tabular}\end{ruledtabular}
\label{tab:pp_table2}
\end{table*} \endgroup
%\cleardoublepage
 
\begingroup \squeezetable
\begin{table*}
\caption{Invariant yields for $\pi^{+}$ at mid-rapidity in 50--60\%, 
60--70\%, 70--80\%, and 80-92\% centrality bins, normalized to one unit rapidity.
Errors are statistical only.}
\begin{ruledtabular}\begin{tabular}{ccccc}
$p_T$ [GeV/$c$] & 50--60\%  &  60--70\% &  70--80\%  &  80--92\%  \\ \hline
0.25  &  4.10e+01  $\pm$  3.4e-01  &  2.19e+01  $\pm$  1.9e-01  &  1.03e+01  $\pm$  9.2e-02  &  5.20e+00  $\pm$  5.0e-02  \\  
0.35  &  2.17e+01  $\pm$  1.9e-01  &  1.13e+01  $\pm$  1.0e-01  &  5.27e+00  $\pm$  5.0e-02  &  2.75e+00  $\pm$  2.8e-02  \\  
0.45  &  1.24e+01  $\pm$  1.1e-01  &  6.37e+00  $\pm$  6.0e-02  &  2.95e+00  $\pm$  3.1e-02  &  1.49e+00  $\pm$  1.8e-02  \\  
0.55  &  7.20e+00  $\pm$  7.0e-02  &  3.65e+00  $\pm$  3.8e-02  &  1.62e+00  $\pm$  1.9e-02  &  8.20e-01  $\pm$  1.1e-02  \\  
0.65  &  4.33e+00  $\pm$  4.7e-02  &  2.18e+00  $\pm$  2.6e-02  &  9.63e-01  $\pm$  1.3e-02  &  4.72e-01  $\pm$  8.1e-03  \\  
0.75  &  2.78e+00  $\pm$  3.4e-02  &  1.36e+00  $\pm$  1.9e-02  &  5.91e-01  $\pm$  9.9e-03  &  2.69e-01  $\pm$  5.9e-03  \\  
0.85  &  1.67e+00  $\pm$  2.3e-02  &  8.36e-01  $\pm$  1.3e-02  &  3.53e-01  $\pm$  7.1e-03  &  1.63e-01  $\pm$  4.4e-03  \\
\end{tabular}\end{ruledtabular}
\label{tab:pp_table3}
\end{table*} \endgroup
%\cleardoublepage

\begingroup \squeezetable
\begin{table*}
\caption{Invariant yields for $\pi^{-}$ at mid-rapidity in the minimum bias, 
0--5\%, 5--10\%, and 10--15\% centrality bins, normalized to one unit rapidity.
Errors are statistical only.}
\begin{ruledtabular}\begin{tabular}{ccccc}
$p_T$ [GeV/$c$] & Minimum bias &  0--5\% &  5--10\%  &  10--15\%  \\ \hline
0.25  &  1.02e+02  $\pm$  7.9e-01  &  3.15e+02  $\pm$  2.4e+00  &  2.71e+02  $\pm$  2.1e+00  &  2.27e+02  $\pm$  1.8e+00  \\  
0.35  &  5.92e+01  $\pm$  4.6e-01  &  1.94e+02  $\pm$  1.5e+00  &  1.64e+02  $\pm$  1.3e+00  &  1.35e+02  $\pm$  1.1e+00  \\  
0.45  &  3.56e+01  $\pm$  2.9e-01  &  1.19e+02  $\pm$  9.8e-01  &  9.93e+01  $\pm$  8.2e-01  &  8.18e+01  $\pm$  6.8e-01  \\  
0.55  &  2.18e+01  $\pm$  1.9e-01  &  7.37e+01  $\pm$  6.5e-01  &  6.17e+01  $\pm$  5.4e-01  &  5.04e+01  $\pm$  4.5e-01  \\  
0.65  &  1.34e+01  $\pm$  1.2e-01  &  4.57e+01  $\pm$  4.3e-01  &  3.82e+01  $\pm$  3.6e-01  &  3.15e+01  $\pm$  3.0e-01  \\  
0.75  &  8.36e+00  $\pm$  8.2e-02  &  2.86e+01  $\pm$  2.9e-01  &  2.40e+01  $\pm$  2.4e-01  &  1.96e+01  $\pm$  2.0e-01  \\  
0.85  &  5.44e+00  $\pm$  5.7e-02  &  1.86e+01  $\pm$  2.0e-01  &  1.56e+01  $\pm$  1.7e-01  &  1.28e+01  $\pm$  1.4e-01  \\
\end{tabular}\end{ruledtabular}
\label{tab:pm_table1}
\end{table*} \endgroup
%\cleardoublepage
 
\begingroup \squeezetable
\begin{table*}
\caption{Invariant yields for $\pi^{-}$ at mid-rapidity in 15--20\%, 
20--30\%, 30--40\%, and 40-50\% centrality bins, normalized to one unit rapidity.
Errors are statistical only.}
\begin{ruledtabular}\begin{tabular}{ccccc}
$p_T$ [GeV/$c$] & 15--20\%  &  20--30\% &  30--40\%  &  40--50\%  \\ \hline
0.25  &  1.95e+02  $\pm$  1.5e+00  &  1.51e+02  $\pm$  1.2e+00  &  1.02e+02  $\pm$  7.9e-01  &  6.53e+01  $\pm$  5.1e-01  \\  
0.35  &  1.13e+02  $\pm$  9.0e-01  &  8.62e+01  $\pm$  6.8e-01  &  5.68e+01  $\pm$  4.5e-01  &  3.56e+01  $\pm$  2.8e-01  \\  
0.45  &  6.86e+01  $\pm$  5.7e-01  &  5.18e+01  $\pm$  4.3e-01  &  3.36e+01  $\pm$  2.8e-01  &  2.08e+01  $\pm$  1.7e-01  \\  
0.55  &  4.22e+01  $\pm$  3.7e-01  &  3.17e+01  $\pm$  2.8e-01  &  2.04e+01  $\pm$  1.8e-01  &  1.24e+01  $\pm$  1.1e-01  \\  
0.65  &  2.61e+01  $\pm$  2.5e-01  &  1.95e+01  $\pm$  1.8e-01  &  1.26e+01  $\pm$  1.2e-01  &  7.57e+00  $\pm$  7.4e-02  \\  
0.75  &  1.63e+01  $\pm$  1.7e-01  &  1.22e+01  $\pm$  1.2e-01  &  7.81e+00  $\pm$  8.0e-02  &  4.67e+00  $\pm$  4.9e-02  \\  
0.85  &  1.06e+01  $\pm$  1.2e-01  &  7.96e+00  $\pm$  8.7e-02  &  5.06e+00  $\pm$  5.7e-02  &  3.04e+00  $\pm$  3.5e-02  \\
\end{tabular}\end{ruledtabular}
\label{tab:pm_table2}
\end{table*} \endgroup
%\cleardoublepage

\begingroup \squeezetable
\begin{table*}
\caption{Invariant yields for $\pi^{-}$ at mid-rapidity in 50--60\%, 
60--70\%, 70--80\%, and 80-92\% centrality bins, normalized to one unit rapidity.
Errors are statistical only.}
\begin{ruledtabular}\begin{tabular}{ccccc}
$p_T$ [GeV/$c$] & 50--60\%  &  60--70\% &  70--80\%  &  80--92\%  \\ \hline
0.25  &  3.92e+01  $\pm$  3.1e-01  &  2.07e+01  $\pm$  1.7e-01  &  9.77e+00  $\pm$  8.2e-02  &  5.03e+00  $\pm$  4.5e-02  \\  
0.35  &  2.10e+01  $\pm$  1.7e-01  &  1.09e+01  $\pm$  9.0e-02  &  5.19e+00  $\pm$  4.6e-02  &  2.67e+00  $\pm$  2.6e-02  \\  
0.45  &  1.21e+01  $\pm$  1.0e-01  &  6.21e+00  $\pm$  5.5e-02  &  2.84e+00  $\pm$  2.8e-02  &  1.45e+00  $\pm$  1.6e-02  \\  
0.55  &  7.13e+00  $\pm$  6.6e-02  &  3.59e+00  $\pm$  3.5e-02  &  1.62e+00  $\pm$  1.8e-02  &  8.13e-01  $\pm$  1.1e-02  \\  
0.65  &  4.30e+00  $\pm$  4.4e-02  &  2.16e+00  $\pm$  2.4e-02  &  9.32e-01  $\pm$  1.2e-02  &  4.54e-01  $\pm$  7.3e-03  \\  
0.75  &  2.61e+00  $\pm$  2.9e-02  &  1.30e+00  $\pm$  1.6e-02  &  5.61e-01  $\pm$  8.6e-03  &  2.70e-01  $\pm$  5.3e-03  \\  
0.85  &  1.68e+00  $\pm$  2.1e-02  &  8.30e-01  $\pm$  1.2e-02  &  3.52e-01  $\pm$  6.4e-03  &  1.59e-01  $\pm$  3.9e-03  \\
\end{tabular}\end{ruledtabular}
\label{tab:pm_table3}
\end{table*} \endgroup
\cleardoublepage
%\clearpage
 
\begingroup \squeezetable
\begin{table*}
\caption{Invariant yields for $K^{+}$ at mid-rapidity in the minimum bias, 
0--5\%, 5--10\%, and 10--15\% centrality bins, normalized to one unit rapidity.
Errors are statistical only.}
\begin{ruledtabular}\begin{tabular}{ccccc}
$p_T$ [GeV/$c$] & Minimum bias &  0--5\% &  5--10\%  &  10--15\%  \\ \hline
0.45  &  5.46e+00  $\pm$  1.1e-01  &  1.83e+01  $\pm$  3.9e-01  &  1.50e+01  $\pm$  3.3e-01  &  1.29e+01  $\pm$  2.8e-01  \\  
0.55  &  4.28e+00  $\pm$  7.8e-02  &  1.48e+01  $\pm$  2.9e-01  &  1.20e+01  $\pm$  2.4e-01  &  9.88e+00  $\pm$  2.0e-01  \\  
0.65  &  3.11e+00  $\pm$  5.4e-02  &  1.05e+01  $\pm$  2.0e-01  &  8.75e+00  $\pm$  1.7e-01  &  7.38e+00  $\pm$  1.4e-01  \\  
0.75  &  2.27e+00  $\pm$  3.9e-02  &  7.97e+00  $\pm$  1.5e-01  &  6.48e+00  $\pm$  1.2e-01  &  5.39e+00  $\pm$  1.0e-01  \\  
0.85  &  1.69e+00  $\pm$  3.0e-02  &  5.96e+00  $\pm$  1.2e-01  &  4.81e+00  $\pm$  9.5e-02  &  4.02e+00  $\pm$  8.1e-02  \\  
0.95  &  1.20e+00  $\pm$  2.2e-02  &  4.19e+00  $\pm$  8.5e-02  &  3.47e+00  $\pm$  7.2e-02  &  2.91e+00  $\pm$  6.1e-02  \\  
1.05  &  9.06e-01  $\pm$  1.7e-02  &  3.20e+00  $\pm$  6.8e-02  &  2.61e+00  $\pm$  5.7e-02  &  2.21e+00  $\pm$  5.0e-02  \\
\end{tabular}\end{ruledtabular}
\label{tab:kp_table1}
\end{table*} \endgroup
%\cleardoublepage

\begingroup \squeezetable
\begin{table*}
\caption{Invariant yields for $K^{+}$ at mid-rapidity in 15--20\%, 
20--30\%, 30--40\%, and 40-50\% centrality bins, normalized to one unit rapidity.
Errors are statistical only.}
\begin{ruledtabular}\begin{tabular}{ccccc}
$p_T$ [GeV/$c$] & 15--20\%  &  20--30\% &  30--40\%  &  40--50\%  \\ \hline
0.45  &  1.04e+01  $\pm$  2.3e-01  &  7.81e+00  $\pm$  1.7e-01  &  5.11e+00  $\pm$  1.1e-01  &  3.28e+00  $\pm$  7.8e-02  \\  
0.55  &  8.30e+00  $\pm$  1.7e-01  &  6.22e+00  $\pm$  1.2e-01  &  4.06e+00  $\pm$  8.3e-02  &  2.43e+00  $\pm$  5.3e-02  \\  
0.65  &  6.20e+00  $\pm$  1.2e-01  &  4.51e+00  $\pm$  8.5e-02  &  2.89e+00  $\pm$  5.7e-02  &  1.78e+00  $\pm$  3.8e-02  \\  
0.75  &  4.46e+00  $\pm$  8.8e-02  &  3.31e+00  $\pm$  6.2e-02  &  2.07e+00  $\pm$  4.1e-02  &  1.26e+00  $\pm$  2.7e-02  \\  
0.85  &  3.36e+00  $\pm$  7.0e-02  &  2.50e+00  $\pm$  4.9e-02  &  1.60e+00  $\pm$  3.3e-02  &  9.00e-01  $\pm$  2.1e-02  \\  
0.95  &  2.40e+00  $\pm$  5.2e-02  &  1.74e+00  $\pm$  3.6e-02  &  1.08e+00  $\pm$  2.4e-02  &  6.46e-01  $\pm$  1.6e-02  \\  
1.05  &  1.81e+00  $\pm$  4.2e-02  &  1.31e+00  $\pm$  2.8e-02  &  8.42e-01  $\pm$  2.0e-02  &  4.82e-01  $\pm$  1.3e-02  \\
\end{tabular}\end{ruledtabular}
\label{tab:kp_table2}
\end{table*} \endgroup
%\cleardoublepage

\begingroup \squeezetable
\begin{table*}
\caption{Invariant yields for $K^{+}$ at mid-rapidity in 50--60\%, 
60--70\%, 70--80\%, and 80-92\% centrality bins, normalized to one unit rapidity.
Errors are statistical only.}
\begin{ruledtabular}\begin{tabular}{ccccc}
$p_T$ [GeV/$c$] & 50--60\%  &  60--70\% &  70--80\%  &  80--92\%  \\ \hline
0.45  &  1.93e+00  $\pm$  5.0e-02  &  9.56e-01  $\pm$  2.9e-02  &  4.06e-01  $\pm$  1.7e-02  &  1.88e-01  $\pm$  1.1e-02  \\  
0.55  &  1.36e+00  $\pm$  3.3e-02  &  6.72e-01  $\pm$  2.0e-02  &  2.89e-01  $\pm$  1.2e-02  &  1.48e-01  $\pm$  7.8e-03  \\  
0.65  &  1.01e+00  $\pm$  2.4e-02  &  4.81e-01  $\pm$  1.4e-02  &  1.88e-01  $\pm$  8.0e-03  &  1.02e-01  $\pm$  5.6e-03  \\  
0.75  &  6.82e-01  $\pm$  1.7e-02  &  3.40e-01  $\pm$  1.1e-02  &  1.24e-01  $\pm$  5.8e-03  &  5.88e-02  $\pm$  3.9e-03  \\  
0.85  &  4.77e-01  $\pm$  1.3e-02  &  2.33e-01  $\pm$  8.1e-03  &  9.39e-02  $\pm$  4.8e-03  &  3.87e-02  $\pm$  3.0e-03  \\  
0.95  &  3.51e-01  $\pm$  1.0e-02  &  1.69e-01  $\pm$  6.4e-03  &  5.66e-02  $\pm$  3.5e-03  &  2.99e-02  $\pm$  2.5e-03  \\  
1.05  &  2.54e-01  $\pm$  8.2e-03  &  1.19e-01  $\pm$  5.1e-03  &  4.40e-02  $\pm$  3.0e-03  &  2.07e-02  $\pm$  2.0e-03  \\
\end{tabular}\end{ruledtabular}
\label{tab:kp_table3}
\end{table*} \endgroup
%\cleardoublepage
 
\begingroup \squeezetable
\begin{table*}
\caption{Invariant yields for $K^{-}$ at mid-rapidity in the minimum bias, 
0--5\%, 5--10\%, and 10--15\% centrality bins, normalized to one unit rapidity.
Errors are statistical only.}
\begin{ruledtabular}\begin{tabular}{ccccc}
$p_T$ [GeV/$c$] & Minimum bias &  0--5\% &  5--10\%  &  10--15\%  \\ \hline
0.45  &  4.87e+00  $\pm$  9.3e-02  &  1.64e+01  $\pm$  3.4e-01  &  1.36e+01  $\pm$  2.8e-01  &  1.12e+01  $\pm$  2.4e-01  \\  
0.55  &  3.88e+00  $\pm$  6.7e-02  &  1.31e+01  $\pm$  2.4e-01  &  1.09e+01  $\pm$  2.0e-01  &  8.91e+00  $\pm$  1.7e-01  \\  
0.65  &  2.96e+00  $\pm$  4.9e-02  &  1.01e+01  $\pm$  1.8e-01  &  8.57e+00  $\pm$  1.5e-01  &  6.94e+00  $\pm$  1.3e-01  \\  
0.75  &  2.20e+00  $\pm$  3.6e-02  &  7.69e+00  $\pm$  1.4e-01  &  6.27e+00  $\pm$  1.1e-01  &  5.14e+00  $\pm$  9.5e-02  \\  
0.85  &  1.59e+00  $\pm$  2.6e-02  &  5.61e+00  $\pm$  1.0e-01  &  4.55e+00  $\pm$  8.4e-02  &  3.82e+00  $\pm$  7.2e-02  \\  
0.95  &  1.14e+00  $\pm$  1.9e-02  &  4.11e+00  $\pm$  7.7e-02  &  3.36e+00  $\pm$  6.5e-02  &  2.76e+00  $\pm$  5.4e-02  \\  
1.05  &  8.50e-01  $\pm$  1.5e-02  &  3.03e+00  $\pm$  6.0e-02  &  2.53e+00  $\pm$  5.2e-02  &  2.05e+00  $\pm$  4.3e-02  \\
\end{tabular}\end{ruledtabular}
\label{tab:km_table1}
\end{table*} \endgroup
%\cleardoublepage
 
\begingroup \squeezetable
\begin{table*}
\caption{Invariant yields for $K^{-}$ at mid-rapidity in 15--20\%, 
20--30\%, 30--40\%, and 40-50\% centrality bins, normalized to one unit rapidity.
Errors are statistical only.}
\begin{ruledtabular}\begin{tabular}{ccccc}
$p_T$ [GeV/$c$] & 15--20\%  &  20--30\% &  30--40\%  &  40--50\%  \\ \hline
0.45  &  9.24e+00  $\pm$  2.0e-01  &  7.05e+00  $\pm$  1.5e-01  &  4.60e+00  $\pm$  9.9e-02  &  2.79e+00  $\pm$  6.4e-02  \\  
0.55  &  7.61e+00  $\pm$  1.5e-01  &  5.62e+00  $\pm$  1.0e-01  &  3.68e+00  $\pm$  7.1e-02  &  2.25e+00  $\pm$  4.7e-02  \\  
0.65  &  5.78e+00  $\pm$  1.1e-01  &  4.29e+00  $\pm$  7.7e-02  &  2.74e+00  $\pm$  5.1e-02  &  1.69e+00  $\pm$  3.4e-02  \\  
0.75  &  4.33e+00  $\pm$  8.1e-02  &  3.22e+00  $\pm$  5.8e-02  &  2.04e+00  $\pm$  3.8e-02  &  1.19e+00  $\pm$  2.5e-02  \\  
0.85  &  3.13e+00  $\pm$  6.0e-02  &  2.29e+00  $\pm$  4.2e-02  &  1.49e+00  $\pm$  2.9e-02  &  8.47e-01  $\pm$  1.8e-02  \\  
0.95  &  2.23e+00  $\pm$  4.5e-02  &  1.61e+00  $\pm$  3.1e-02  &  1.04e+00  $\pm$  2.1e-02  &  6.04e-01  $\pm$  1.4e-02  \\  
1.05  &  1.70e+00  $\pm$  3.7e-02  &  1.21e+00  $\pm$  2.5e-02  &  7.74e-01  $\pm$  1.7e-02  &  4.49e-01  $\pm$  1.1e-02  \\
\end{tabular}\end{ruledtabular}
\label{tab:km_table2}
\end{table*} \endgroup
%\cleardoublepage

\begingroup \squeezetable
\begin{table*}
\caption{Invariant yields for $K^{-}$ at mid-rapidity in 50--60\%, 
60--70\%, 70--80\%, and 80-92\% centrality bins, normalized to one unit rapidity.
Errors are statistical only.}
\begin{ruledtabular}\begin{tabular}{ccccc}
$p_T$ [GeV/$c$] & 50--60\%  &  60--70\% &  70--80\%  &  80--92\%  \\ \hline
0.45  &  1.73e+00  $\pm$  4.3e-02  &  8.11e-01  $\pm$  2.5e-02  &  3.89e-01  $\pm$  1.6e-02  &  1.82e-01  $\pm$  9.9e-03  \\  
0.55  &  1.25e+00  $\pm$  2.9e-02  &  6.37e-01  $\pm$  1.8e-02  &  2.80e-01  $\pm$  1.1e-02  &  1.37e-01  $\pm$  7.1e-03  \\  
0.65  &  9.30e-01  $\pm$  2.1e-02  &  4.43e-01  $\pm$  1.3e-02  &  1.83e-01  $\pm$  7.5e-03  &  1.02e-01  $\pm$  5.4e-03  \\  
0.75  &  6.59e-01  $\pm$  1.6e-02  &  3.16e-01  $\pm$  9.5e-03  &  1.40e-01  $\pm$  5.9e-03  &  6.21e-02  $\pm$  3.8e-03  \\  
0.85  &  4.65e-01  $\pm$  1.2e-02  &  2.31e-01  $\pm$  7.4e-03  &  8.42e-02  $\pm$  4.2e-03  &  3.81e-02  $\pm$  2.7e-03  \\  
0.95  &  3.22e-01  $\pm$  9.0e-03  &  1.56e-01  $\pm$  5.7e-03  &  5.67e-02  $\pm$  3.2e-03  &  2.57e-02  $\pm$  2.1e-03  \\  
1.05  &  2.32e-01  $\pm$  7.2e-03  &  1.09e-01  $\pm$  4.5e-03  &  4.26e-02  $\pm$  2.7e-03  &  1.73e-02  $\pm$  1.7e-03  \\
\end{tabular}\end{ruledtabular}
\label{tab:km_table3}
\end{table*} \endgroup
%\cleardoublepage

\begingroup \squeezetable
\begin{table*}
\caption{Invariant yields for protons at mid-rapidity in the minimum bias, 
0--5\%, 5--10\%, and 10--15\% centrality bins, normalized to one unit rapidity.
Errors are statistical only.}
\begin{ruledtabular}\begin{tabular}{ccccc}
$p_T$ [GeV/$c$] & Minimum bias &  0--5\% &  5--10\%  &  10--15\%  \\ \hline
0.65  &  9.51e-01  $\pm$  2.7e-02  &  2.90e+00  $\pm$  9.3e-02  &  2.44e+00  $\pm$  8.0e-02  &  2.09e+00  $\pm$  6.9e-02  \\  
0.75  &  8.47e-01  $\pm$  2.4e-02  &  2.65e+00  $\pm$  8.5e-02  &  2.24e+00  $\pm$  7.3e-02  &  1.87e+00  $\pm$  6.2e-02  \\  
0.85  &  7.08e-01  $\pm$  2.0e-02  &  2.28e+00  $\pm$  7.3e-02  &  1.91e+00  $\pm$  6.3e-02  &  1.60e+00  $\pm$  5.3e-02  \\  
0.95  &  6.06e-01  $\pm$  1.8e-02  &  2.00e+00  $\pm$  6.6e-02  &  1.66e+00  $\pm$  5.5e-02  &  1.41e+00  $\pm$  4.8e-02  \\  
1.05  &  5.05e-01  $\pm$  1.5e-02  &  1.68e+00  $\pm$  5.7e-02  &  1.43e+00  $\pm$  4.9e-02  &  1.16e+00  $\pm$  4.1e-02  \\  
1.15  &  4.23e-01  $\pm$  1.3e-02  &  1.46e+00  $\pm$  5.1e-02  &  1.22e+00  $\pm$  4.3e-02  &  9.85e-01  $\pm$  3.6e-02  \\  
1.25  &  3.30e-01  $\pm$  1.0e-02  &  1.16e+00  $\pm$  4.2e-02  &  9.51e-01  $\pm$  3.5e-02  &  7.92e-01  $\pm$  3.0e-02  \\
\end{tabular}\end{ruledtabular}
\label{tab:pr_table1}
\end{table*} \endgroup
%\cleardoublepage

\begingroup \squeezetable
\begin{table*}
\caption{Invariant yields for protons at mid-rapidity in 15--20\%, 
20--30\%, 30--40\%, and 40-50\% centrality bins, normalized to one unit rapidity.
Errors are statistical only.}
\begin{ruledtabular}\begin{tabular}{ccccc}
$p_T$ [GeV/$c$] & 15--20\%  &  20--30\% &  30--40\%  &  40--50\%  \\ \hline
0.65  &  1.76e+00  $\pm$  6.0e-02  &  1.37e+00  $\pm$  4.4e-02  &  9.68e-01  $\pm$  3.2e-02  &  6.31e-01  $\pm$  2.2e-02  \\  
0.75  &  1.59e+00  $\pm$  5.4e-02  &  1.24e+00  $\pm$  4.0e-02  &  8.52e-01  $\pm$  2.9e-02  &  5.39e-01  $\pm$  1.9e-02  \\  
0.85  &  1.34e+00  $\pm$  4.6e-02  &  1.02e+00  $\pm$  3.3e-02  &  7.06e-01  $\pm$  2.4e-02  &  4.33e-01  $\pm$  1.6e-02  \\  
0.95  &  1.16e+00  $\pm$  4.1e-02  &  8.90e-01  $\pm$  2.9e-02  &  5.79e-01  $\pm$  2.0e-02  &  3.60e-01  $\pm$  1.4e-02  \\  
1.05  &  9.75e-01  $\pm$  3.5e-02  &  7.41e-01  $\pm$  2.5e-02  &  4.83e-01  $\pm$  1.7e-02  &  2.96e-01  $\pm$  1.2e-02  \\  
1.15  &  8.38e-01  $\pm$  3.1e-02  &  6.27e-01  $\pm$  2.2e-02  &  3.93e-01  $\pm$  1.5e-02  &  2.33e-01  $\pm$  9.7e-03  \\  
1.25  &  6.47e-01  $\pm$  2.5e-02  &  4.83e-01  $\pm$  1.8e-02  &  3.09e-01  $\pm$  1.2e-02  &  1.77e-01  $\pm$  7.9e-03  \\
\end{tabular}\end{ruledtabular}
\label{tab:pr_table2}
\end{table*} \endgroup
%\cleardoublepage

\begingroup \squeezetable
\begin{table*}
\caption{Invariant yields for protons at mid-rapidity in 50--60\%, 
60--70\%, 70--80\%, and 80-92\% centrality bins, normalized to one unit rapidity.
Errors are statistical only.}
\begin{ruledtabular}\begin{tabular}{ccccc}
$p_T$ [GeV/$c$] & 50--60\%  &  60--70\% &  70--80\%  &  80--92\%  \\ \hline
0.65  &  3.82e-01  $\pm$  1.5e-02  &  2.04e-01  $\pm$  9.7e-03  &  9.09e-02  $\pm$  5.9e-03  &  4.96e-02  $\pm$  4.2e-03  \\  
0.75  &  3.25e-01  $\pm$  1.3e-02  &  1.65e-01  $\pm$  8.1e-03  &  7.04e-02  $\pm$  4.9e-03  &  3.79e-02  $\pm$  3.4e-03  \\  
0.85  &  2.60e-01  $\pm$  1.1e-02  &  1.27e-01  $\pm$  6.5e-03  &  5.41e-02  $\pm$  4.0e-03  &  2.62e-02  $\pm$  2.7e-03  \\  
0.95  &  2.08e-01  $\pm$  9.1e-03  &  1.00e-01  $\pm$  5.5e-03  &  4.11e-02  $\pm$  3.3e-03  &  2.06e-02  $\pm$  2.3e-03  \\  
1.05  &  1.61e-01  $\pm$  7.5e-03  &  7.43e-02  $\pm$  4.5e-03  &  3.14e-02  $\pm$  2.8e-03  &  1.54e-02  $\pm$  1.9e-03  \\  
1.15  &  1.24e-01  $\pm$  6.2e-03  &  5.88e-02  $\pm$  3.8e-03  &  2.40e-02  $\pm$  2.3e-03  &  8.08e-03  $\pm$  1.3e-03  \\  
1.25  &  9.20e-02  $\pm$  5.0e-03  &  3.98e-02  $\pm$  3.0e-03  &  1.68e-02  $\pm$  1.9e-03  &  6.94e-03  $\pm$  1.2e-03  \\
\end{tabular}\end{ruledtabular}
\label{tab:pr_table3}
\end{table*} \endgroup
%\cleardoublepage

\begingroup \squeezetable
\begin{table*}
\caption{Invariant yields for anti-protons at mid-rapidity in the minimum bias, 
0--5\%, 5--10\%, and 10--15\% centrality bins, normalized to one unit rapidity.
Errors are statistical only.}
\begin{ruledtabular}\begin{tabular}{ccccc}
$p_T$ [GeV/$c$] & Minimum bias &  0--5\% &  5--10\%  &  10--15\%  \\ \hline
0.65  &  6.73e-01  $\pm$  2.0e-02  &  2.00e+00  $\pm$  6.8e-02  &  1.73e+00  $\pm$  6.0e-02  &  1.48e+00  $\pm$  5.2e-02  \\  
0.75  &  6.16e-01  $\pm$  1.8e-02  &  1.89e+00  $\pm$  6.2e-02  &  1.61e+00  $\pm$  5.4e-02  &  1.34e+00  $\pm$  4.6e-02  \\  
0.85  &  5.28e-01  $\pm$  1.5e-02  &  1.67e+00  $\pm$  5.4e-02  &  1.42e+00  $\pm$  4.7e-02  &  1.19e+00  $\pm$  4.1e-02  \\  
0.95  &  4.52e-01  $\pm$  1.3e-02  &  1.47e+00  $\pm$  4.8e-02  &  1.25e+00  $\pm$  4.2e-02  &  1.05e+00  $\pm$  3.6e-02  \\  
1.05  &  3.65e-01  $\pm$  1.1e-02  &  1.21e+00  $\pm$  4.1e-02  &  1.04e+00  $\pm$  3.6e-02  &  8.82e-01  $\pm$  3.1e-02  \\  
1.15  &  3.19e-01  $\pm$  9.7e-03  &  1.10e+00  $\pm$  3.9e-02  &  9.28e-01  $\pm$  3.4e-02  &  7.39e-01  $\pm$  2.8e-02  \\  
1.25  &  2.53e-01  $\pm$  7.9e-03  &  8.90e-01  $\pm$  3.3e-02  &  7.47e-01  $\pm$  2.8e-02  &  6.15e-01  $\pm$  2.4e-02  \\
\end{tabular}\end{ruledtabular}
\label{tab:pb_table1}
\end{table*} \endgroup
%\cleardoublepage

\begingroup \squeezetable
\begin{table*}
\caption{Invariant yields for anti-protons at mid-rapidity in 15--20\%, 
20--30\%, 30--40\%, and 40-50\% centrality bins, normalized to one unit rapidity.
Errors are statistical only.}
\begin{ruledtabular}\begin{tabular}{ccccc}
$p_T$ [GeV/$c$] & 15--20\%  &  20--30\% &  30--40\%  &  40--50\%  \\ \hline
0.65  &  1.25e+00  $\pm$  4.5e-02  &  9.68e-01  $\pm$  3.2e-02  &  6.98e-01  $\pm$  2.4e-02  &  4.51e-01  $\pm$  1.7e-02  \\  
0.75  &  1.16e+00  $\pm$  4.1e-02  &  8.94e-01  $\pm$  2.9e-02  &  6.35e-01  $\pm$  2.2e-02  &  4.06e-01  $\pm$  1.5e-02  \\  
0.85  &  1.02e+00  $\pm$  3.5e-02  &  7.83e-01  $\pm$  2.5e-02  &  5.21e-01  $\pm$  1.8e-02  &  3.37e-01  $\pm$  1.3e-02  \\  
0.95  &  8.85e-01  $\pm$  3.1e-02  &  6.61e-01  $\pm$  2.2e-02  &  4.42e-01  $\pm$  1.5e-02  &  2.70e-01  $\pm$  1.0e-02  \\  
1.05  &  7.26e-01  $\pm$  2.6e-02  &  5.25e-01  $\pm$  1.8e-02  &  3.54e-01  $\pm$  1.3e-02  &  2.05e-01  $\pm$  8.4e-03  \\  
1.15  &  6.43e-01  $\pm$  2.5e-02  &  4.63e-01  $\pm$  1.6e-02  &  2.99e-01  $\pm$  1.2e-02  &  1.79e-01  $\pm$  7.7e-03  \\  
1.25  &  4.99e-01  $\pm$  2.0e-02  &  3.65e-01  $\pm$  1.4e-02  &  2.33e-01  $\pm$  9.5e-03  &  1.37e-01  $\pm$  6.4e-03  \\
\end{tabular}\end{ruledtabular}
\label{tab:pb_table2}
\end{table*} \endgroup
%\cleardoublepage
 
\begingroup \squeezetable
\begin{table*}
\caption{Invariant yields for anti-protons at mid-rapidity in 50--60\%, 
60--70\%, 70--80\%, and 80-92\% centrality bins, normalized to one unit rapidity.
Errors are statistical only.}
\begin{ruledtabular}\begin{tabular}{ccccc}
$p_T$ [GeV/$c$] & 50--60\%  &  60--70\% &  70--80\%  &  80--92\%  \\ \hline
0.65  &  2.84e-01  $\pm$  1.2e-02  &  1.58e-01  $\pm$  8.1e-03  &  6.22e-02  $\pm$  4.7e-03  &  3.55e-02  $\pm$  3.4e-03  \\  
0.75  &  2.50e-01  $\pm$  1.1e-02  &  1.25e-01  $\pm$  6.6e-03  &  5.43e-02  $\pm$  4.0e-03  &  2.77e-02  $\pm$  2.8e-03  \\  
0.85  &  1.89e-01  $\pm$  8.3e-03  &  9.50e-02  $\pm$  5.2e-03  &  4.16e-02  $\pm$  3.3e-03  &  2.06e-02  $\pm$  2.2e-03  \\  
0.95  &  1.58e-01  $\pm$  7.1e-03  &  7.38e-02  $\pm$  4.3e-03  &  3.13e-02  $\pm$  2.7e-03  &  1.56e-02  $\pm$  1.8e-03  \\  
1.05  &  1.19e-01  $\pm$  5.8e-03  &  5.50e-02  $\pm$  3.5e-03  &  2.12e-02  $\pm$  2.1e-03  &  1.01e-02  $\pm$  1.4e-03  \\  
1.15  &  9.60e-02  $\pm$  5.1e-03  &  4.34e-02  $\pm$  3.1e-03  &  1.73e-02  $\pm$  1.9e-03  &  7.94e-03  $\pm$  1.2e-03  \\  
1.25  &  7.11e-02  $\pm$  4.1e-03  &  3.19e-02  $\pm$  2.5e-03  &  1.22e-02  $\pm$  1.5e-03  &  6.05e-03  $\pm$  1.1e-03  \\
\end{tabular}\end{ruledtabular}
\label{tab:pb_table3}
\end{table*} \endgroup
%\cleardoublepage

\begingroup \squeezetable
\begin{table*}
\caption{Invariant yields for $\pi^{\pm}$ and $K^{\pm}$ at mid-rapidity 
in 60--92\% centrality bin, normalized to one unit rapidity.
Errors are statistical only.}
\begin{ruledtabular}\begin{tabular}{ccccc}
$p_T$ [GeV/$c$] & $\pi^{+}$  &  $\pi^{-}$  &  $K^{+}$  &  $K^{-}$ \\ \hline
0.25  &  1.28e+01  $\pm$  1.1e-01  &  1.21e+01  $\pm$  9.5e-02  &  & \\
0.35  &  6.61e+00  $\pm$  5.7e-02  &  6.42e+00  $\pm$  5.2e-02  &  & \\ 
0.45  &  3.71e+00  $\pm$  3.4e-02  &  3.59e+00  $\pm$  3.1e-02  &  5.35e-01  $\pm$  1.5e-02  &  4.74e-01  $\pm$  1.3e-02  \\  
0.55  &  2.09e+00  $\pm$  2.1e-02  &  2.06e+00  $\pm$  1.9e-02  &  3.83e-01  $\pm$  9.7e-03  &  3.62e-01  $\pm$  8.8e-03  \\  
0.65  &  1.24e+00  $\pm$  1.4e-02  &  1.21e+00  $\pm$  1.3e-02  &  2.66e-01  $\pm$  6.8e-03  &  2.50e-01  $\pm$  6.2e-03  \\  
0.75  &  7.63e-01  $\pm$  9.6e-03  &  7.31e-01  $\pm$  8.4e-03  &  1.81e-01  $\pm$  4.9e-03  &  1.78e-01  $\pm$  4.6e-03  \\  
0.85  &  4.64e-01  $\pm$  6.6e-03  &  4.60e-01  $\pm$  5.9e-03  &  1.26e-01  $\pm$  3.7e-03  &  1.21e-01  $\pm$  3.4e-03  \\
\end{tabular}\end{ruledtabular}
\label{tab:peri_table_pik}
\end{table*} \endgroup
%\cleardoublepage

\begin{table}
\caption{Invariant yields for protons and anti-protons at mid-rapidity 
in 60--92\% centrality bin, normalized to one unit rapidity.
Errors are statistical only.}
\begin{ruledtabular}\begin{tabular}{ccc}
$p_T$ [GeV/$c$] & $p$  & $\pbar$  \\ \hline
0.65  &  1.17e-01  $\pm$  4.8e-03  &  8.63e-02  $\pm$  3.8e-03  \\  
0.75  &  9.26e-02  $\pm$  3.9e-03  &  7.00e-02  $\pm$  3.1e-03  \\  
0.85  &  7.01e-02  $\pm$  3.1e-03  &  5.31e-02  $\pm$  2.5e-03  \\  
0.95  &  5.48e-02  $\pm$  2.6e-03  &  4.07e-02  $\pm$  2.0e-03  \\  
1.05  &  4.10e-02  $\pm$  2.1e-03  &  2.92e-02  $\pm$  1.6e-03  \\  
1.15  &  3.09e-02  $\pm$  1.7e-03  &  2.32e-02  $\pm$  1.4e-03  \\  
1.25  &  2.16e-02  $\pm$  1.3e-03  &  1.70e-02  $\pm$  1.1e-03  \\  
1.35  &  1.77e-02  $\pm$  1.2e-03  &  1.27e-02  $\pm$  9.4e-04  \\  
1.45  &  1.25e-02  $\pm$  9.4e-04  &  1.02e-02  $\pm$  8.3e-04  \\  
1.55  &  8.85e-03  $\pm$  7.8e-04  &  6.51e-03  $\pm$  6.2e-04  \\  
1.65  &  6.42e-03  $\pm$  6.3e-04  &  4.76e-03  $\pm$  5.2e-04  \\  
1.75  &  5.08e-03  $\pm$  5.5e-04  &  3.69e-03  $\pm$  4.5e-04  \\  
1.85  &  3.58e-03  $\pm$  4.6e-04  &  2.60e-03  $\pm$  3.7e-04  \\  
1.95  &  2.79e-03  $\pm$  3.9e-04  &  2.11e-03  $\pm$  3.4e-04  \\  
2.10  &  1.77e-03  $\pm$  1.8e-04  &  1.23e-03  $\pm$  1.5e-04  \\  
2.30  &  1.08e-03  $\pm$  1.4e-04  &  7.22e-04  $\pm$  1.2e-04  \\  
2.50  &  5.42e-04  $\pm$  9.5e-05  &  3.97e-04  $\pm$  8.5e-05  \\  
2.70  &  3.32e-04  $\pm$  7.4e-05  &  2.17e-04  $\pm$  6.2e-05  \\  
2.90  &  2.04e-04  $\pm$  5.8e-05  &  1.49e-04  $\pm$  5.2e-05  \\  
3.25  &  8.58e-05  $\pm$  2.3e-05  &  5.24e-05  $\pm$  1.9e-05  \\  
3.75  &  2.76e-05  $\pm$  1.3e-05  &  1.14e-05  $\pm$  8.7e-06  \\  
4.25  &  1.24e-05  $\pm$  9.1e-06  &  5.08e-06  $\pm$  6.1e-06  \\  
\end{tabular}\end{ruledtabular}
\label{tab:peri_table_pr}
\end{table} 

%%%%%%%
% END %
%%%%%%%
\end{document}